\documentclass[lettersize,journal]{IEEEtran}
\usepackage{graphicx}
\usepackage{subfigure}

\usepackage{fancyhdr}
\usepackage{longtable}

\usepackage{amsmath}
\usepackage{amsthm}
\newtheorem{definition}{\textbf{Definition}}[section]

\usepackage{float}
\usepackage{amssymb}
\usepackage{multirow}
\usepackage{mathrsfs}
\usepackage{algorithmic}
\usepackage[ruled,linesnumbered]{algorithm2e}
\SetKwRepeat{Do}{do}{while}

\usepackage{rotating}
\usepackage{balance}
\usepackage{threeparttable}
\usepackage{color}
\usepackage{colortbl}
\usepackage{makecell}
\usepackage{pdflscape}

\usepackage[pagewise]{lineno}

\usepackage{stfloats}
\usepackage{verbatim}
\usepackage{setspace}
\usepackage{threeparttable}
\usepackage{supertabular,booktabs}
\usepackage{rotating}

\usepackage{supertabular}
\usepackage{pifont}

\usepackage{xcolor,pict2e}

\usepackage{lettrine}
\usepackage{hyperref}
\usepackage[hyphenbreaks]{breakurl}
\usepackage{fancyhdr}

\allowdisplaybreaks[4]

\usepackage{tcolorbox}
\tcbuselibrary{breakable}

\renewcommand{\raggedright}{\leftskip=0pt \rightskip=0pt plus 0cm}

\begin{document}

\title{Toward Cost-effective Adaptive Random Testing: An Approximate Nearest Neighbor Approach}

\author{Rubing Huang,~\IEEEmembership{Senior Member,~IEEE,} Chenhui Cui, Junlong Lian, Dave Towey,~\IEEEmembership{Senior Member,~IEEE,} Weifeng Sun,~\IEEEmembership{Graduate Student Member,~IEEE,} Haibo Chen
\thanks{Rubing Huang and Chenhui Cui are with the School of Computer Science and Engineering, Macau University of Science and Technology, Taipa, Macau 999078, China. E-mail: rbhuang@must.edu.mo, 3230002105@student.must.edu.mo}
\thanks{Junlong Lian and Haibo Chen are with the School of Computer Science and Communication Engineering, Jiangsu University, Zhenjiang, Jiangsu 212013, China. E-mail: \{2211908018, 2112108002\}@stmail.ujs.edu.cn.}
\thanks{Dave Towey is with the School of Computer Science, University of Nottingham Ningbo China, Ningbo, Zhejiang 315100, China. E-mail: dave.towey@nottingham.edu.cn.}
\thanks{Weifeng Sun is with the School of Big Data and Software Engineering, Chongqing University, Chongqing 401331, China. E-mail: weifeng.sun@cqu.edu.cn.}
}

\markboth{Journal of \LaTeX\ Class Files,~Vol.~14, No.~8, August~2021}
{Shell \MakeLowercase{\textit{et al.}}: A Sample Article Using IEEEtran.cls for IEEE Journals}


\maketitle

\begin{abstract}
\textit{Adaptive Random Testing} (ART)  enhances the testing effectiveness (including fault-detection capability) of \textit{Random Testing} (RT) by increasing the diversity of the random test cases throughout the input domain.
Many ART algorithms have been investigated such as \textit{Fixed-Size-Candidate-Set ART} (FSCS) and \textit{Restricted Random Testing} (RRT), and have been widely used in many practical applications.
Despite its popularity, ART suffers from the problem of high computational costs during test-case generation, especially as the number of test cases increases.
Although several strategies have been proposed to enhance the ART testing efficiency,  such as the \textit{forgetting strategy} and the \textit{$k$-dimensional tree strategy}, these algorithms still face some challenges, including:
(1) Although these algorithms can reduce the computation time, their execution costs are still very high, especially when the number of test cases is large; and
(2) To achieve low computational costs, they may sacrifice some fault-detection capability.
In this paper, we propose an approach based on \textit{Approximate Nearest Neighbors} (ANNs), called \textit{Locality-Sensitive Hashing ART} (LSH-ART).
When calculating distances among different test inputs, LSH-ART identifies the approximate (not necessarily exact) nearest neighbors for candidates in an efficient way.
LSH-ART attempts to balance ART testing effectiveness and efficiency.
\end{abstract}


\section{Introduction}

\IEEEPARstart{S}{oftware} testing is an essential software quality assurance activity in the software development life-cycle \cite{Orso2014,Anand2013}.
Among the many software testing techniques, one fundamental approach is \textit{Random Testing} (RT) \cite{Hamlet2002}, which simply selects test cases in a random manner from the \textit{input domain} (the set of all possible test inputs).
RT is considered popular due to the usual ease of implementation, its efficient generation of random test cases, and
its ability to provide quantitative estimation of the reliability of software \cite{Frankl1998}.
RT has been widely applied to different software applications, including:
SQL database systems \cite{Slutz1998,Bati2007};
embedded software systems \cite{Regehr2005};
Java Just-In-Time (JIT) compilers \cite{Yoshikawa2003};
security assessment \cite{Godefroid2008}; and
.NET error detection \cite{Pacheco2008}.
As Arcuri et al. \cite{Arcuri2012} have pointed out, RT should be recommended as the first choice for testing many practical software scenarios.

Despite its popularity, though, RT has been criticized for generally adopting little, or none, of the available information to support its test-case generation.
Myers et al. \cite{Myers2011}, for example, described RT as perhaps the ``least effective'' testing approach.
Many approaches have been proposed to enhance RT fault detection, including one of the most popular, \textit{Adaptive Random Testing} (ART) \cite{Chen2010,Huang2019}.
ART is a family of enhanced RT methods that are motivated by observations of clustering of failure regions (regions of test inputs that can identify software failures) \cite{White1980,Ammann1988,Finelli1991,Bishop1993,Schneckenburger2007}.
ART aims to achieve an even spread of random test cases over the input domain, which may deliver better test case diversity than RT \cite{chen2015revisit}.
Previous studies \cite{Huang2019} have shown ART to be more effective than RT, in terms of various evaluation criteria, including:
the number of test cases required to detect the first failure (\textit{F-measure}) \cite{Chen2007a};
code coverage \cite{Chen2013}; and
test case distribution \cite{Chen2007}.

\begin{figure*}[!b]
\centering
\resizebox{\textwidth}{!}{
\graphicspath{{Graphs/failurePattern/}}
    \subfigure[Point pattern]
    {
        \includegraphics[width=0.295\textwidth]{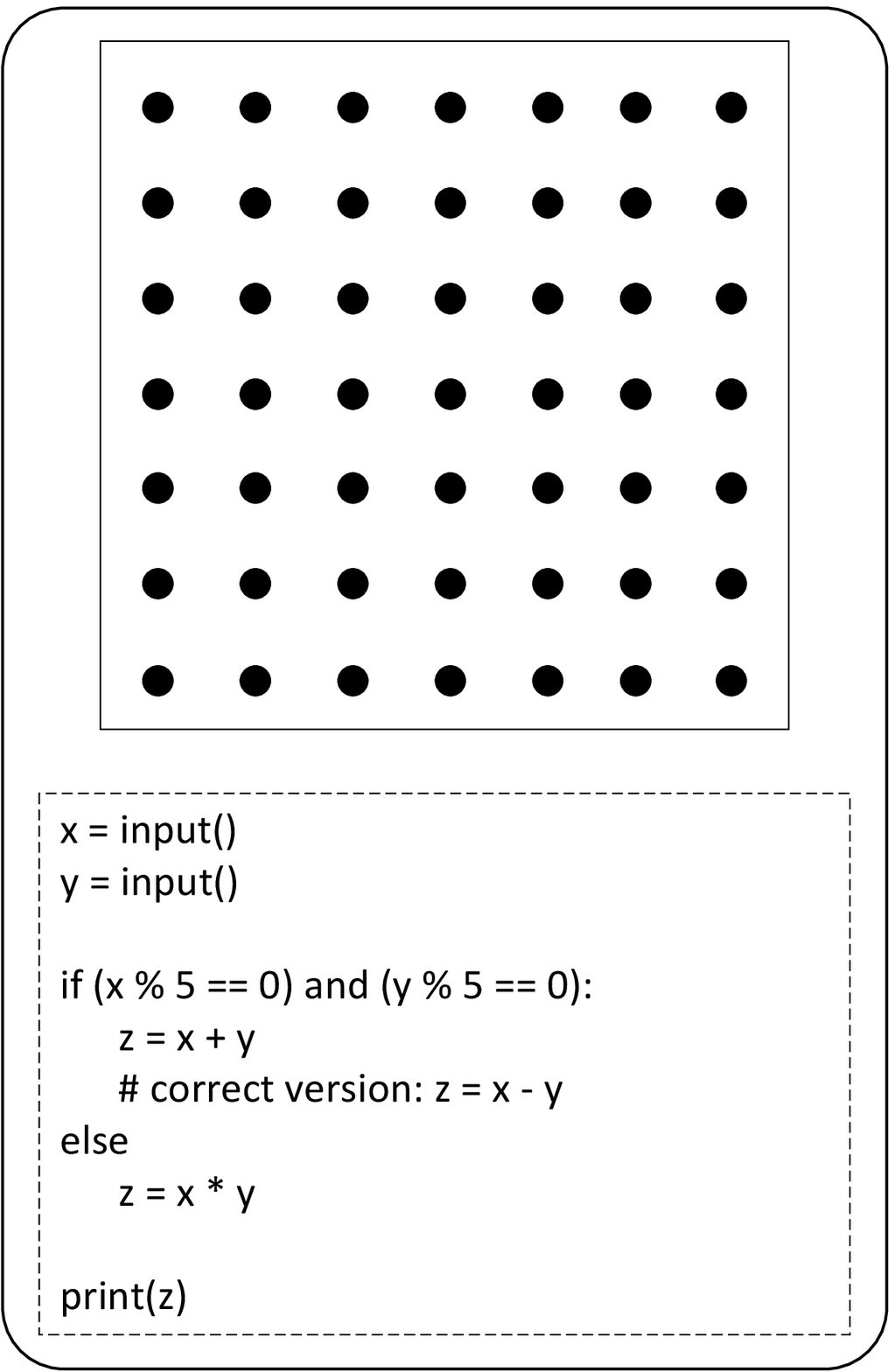}
        \label{figure:1.1}
    }
    \hspace*{12pt}
    \subfigure[Strip pattern]
    {
        \includegraphics[width=0.295\textwidth]{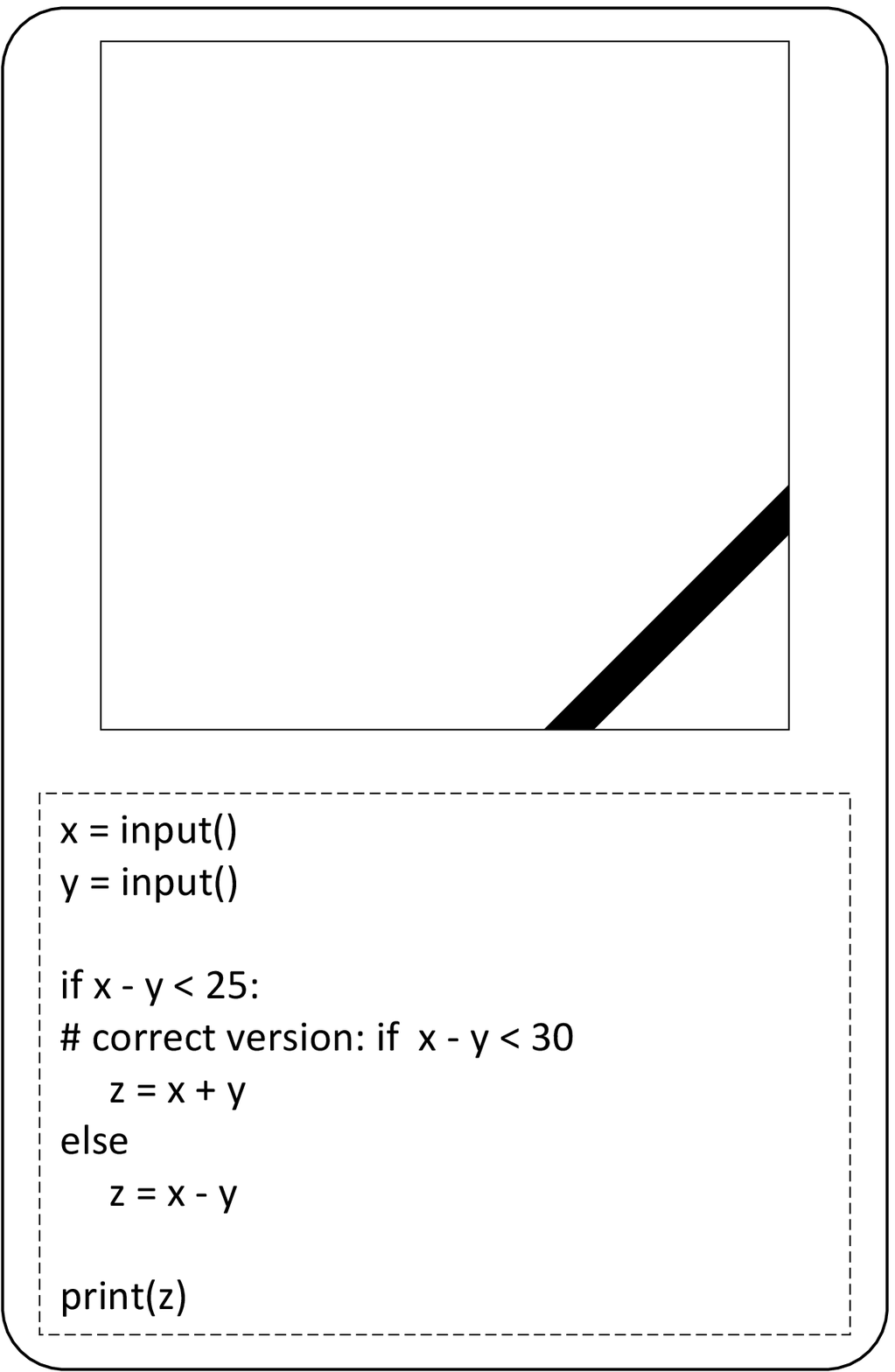}
        \label{figure:1.2}
    }
    \hspace*{12pt}
    \subfigure[Block pattern]
    {
        \includegraphics[width=0.295\textwidth]{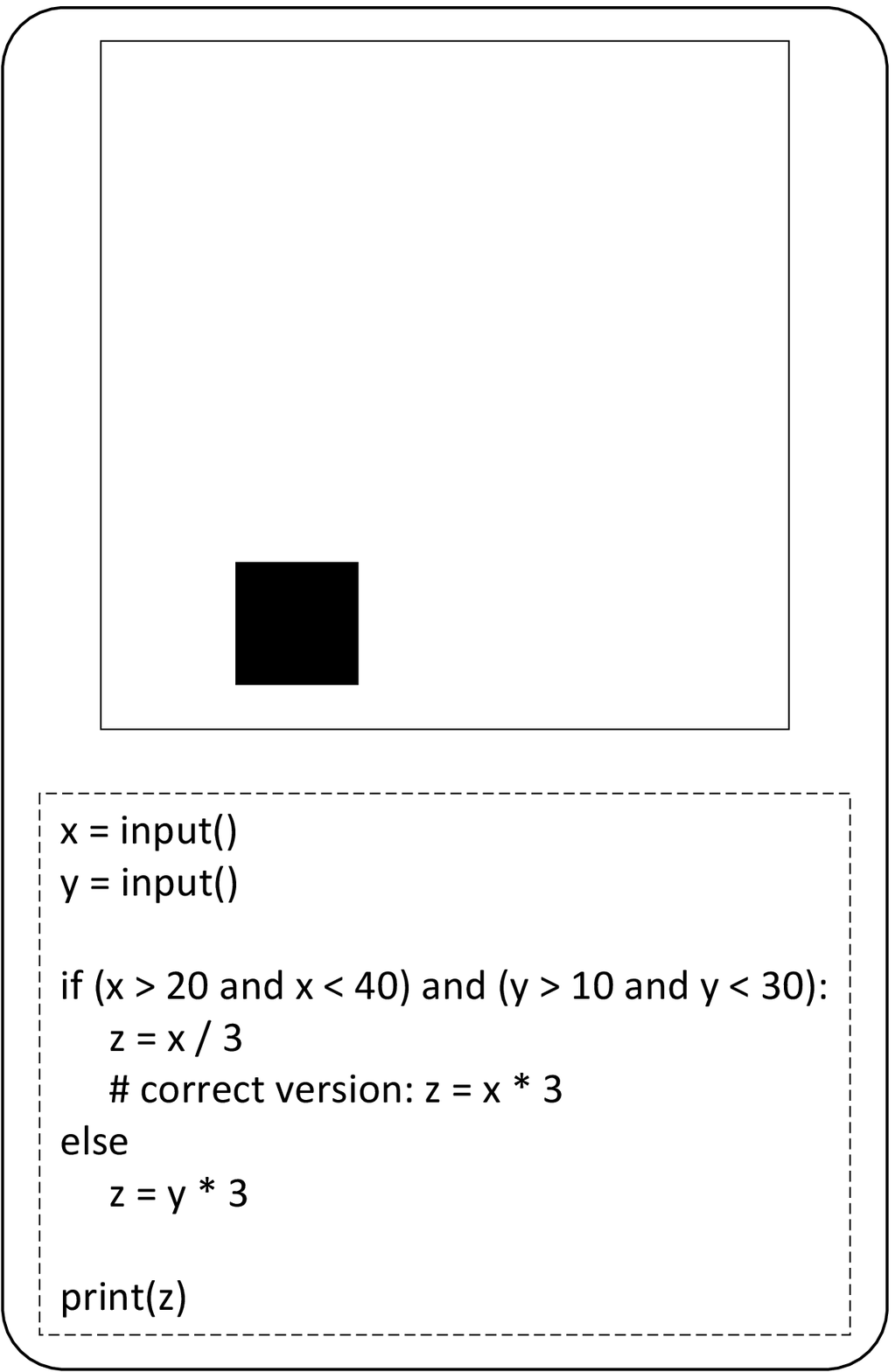}
        \label{figure:1.3}
    }}
    \caption{Examples of three failure patterns and sample faulty programs in two-dimensional input domains.}
    \label{FIG:failurePattern}
\end{figure*}

Many different strategies have been adopted for ART, resulting in multiple ART implementations \cite{Huang2019}, the most well-known and widely used of which is the \textit{Select-Test-From-Candidates Strategy} (STFCS).
STFCS chooses the next test case from a set of random candidates based on some criteria or evaluation of the previously-executed test cases~\cite{Huang2019}:
Each random candidate is compared (in some way) with each previously-executed test case, with the ``\textit{best}'' candidate being selected as the next test case.
Two basic (and popular) STFCS implementations are
\textit{Fixed-Size-Candidate-Set ART} (FSCS) \cite{Chen2010} and \textit{Restricted Random Testing} (RRT) \cite{Chan2006}.
FSCS-ART chooses one of the candidates as the next test case such that it is farthest away from the already-executed test cases;
RRT selects the next test case such that its distance from all previously-executed test cases is greater than a given threshold value.
Generating each new test case requires calculation of the distance between each candidate and each executed test case, which represents a high computational overhead, especially when the number of executed test cases becomes large.
STFCS involves identifying the nearest neighboring executed test case for each candidate, as part of the decision process to choose the ``\textit{best}'' one as the next test case.
A key part of STFCS, therefore, can be viewed as an instance of finding a \textit{Nearest Neighbor} (NN)~\cite{Wilfong1991}.

Various enhanced ART implementations have been proposed to address the STFCS overhead problem, including \textit{forgetting} \cite{Chan2006a}, and a \textit{precise NN} approach \cite{Mao2019}.
The forgetting approach involves ignoring (``forgetting'')  some of the previously-executed test cases to reduce the number of distance computations for each candidate.
The precise NN approach, in contrast, often uses a tree-index structure (such as a \textit{$k$-dimensional tree}, KD-tree \cite{Bentley1975}) to speed up the search process.
Because forgetting does not use all the information of the already-executed test cases, it can be considered to sacrifice some fault-detection capability.
The precise NN approach still needs to precisely identify the nearest neighbor for each candidate, which can still be computationally expensive, especially when the input domain dimensionality is high, or the number of executed test cases is large \cite{Wilfong1991}.

In this paper, we propose an \textit{Approximate Nearest Neighbor} (ANN) ART approach using the well-known ANN algorithm \textit{Locality-Sensitive Hashing}
(LSH)\footnote{Because the set of executed test cases dynamically increases with ART, either the \textit{dynamic LSH} \cite{Jagadish2005,Bawa2005} or \textit{Scalable LSH} (SLSH) \cite{Hu2015,Li2017} approaches need to be used in place of the static LSH: We adopted SLSH.} \cite{Indyk1998}:
We call this \textit{LSH-ART}.
LSH-ART attempts to balance testing effectiveness and efficiency:
It makes use of all previously-executed test cases to maintain fault-detection performance, but retrieves the approximate (not exact) nearest neighbor of each candidate, thus reducing the search cost.

To evaluate our proposed approach, we conducted a series of simulations and empirical studies with more than 30 real-life programs, comparing them with the original ART algorithms and their corresponding enhanced variants.
The main contributions and findings of this paper are:
\begin{itemize}
    \item
    To the best of our knowledge, this is the first paper that proposes an LSH-based approach to enhance ART performance.
    \item
    We propose a framework to support LSH-ART, and present two implementations: LSH-FSCS and LSH-RRT.
    \item
    LSH-ART maintains comparable testing effectiveness (including fault-detection effectiveness) to the original ART and its variants,
    sometimes obtaining better results, especially for high dimensions.
    \item
    LSH-ART has better testing efficiency, generally incurring lower computational costs (including for test-case generation) than the original ART and its variants,  especially when the input domain dimensionality is high.
    \item
    LSH-ART is much more cost-effective than the original ART, in all scenarios, and outperforms the variants in most scenarios.
\end{itemize}

The rest of this paper is organized as follows:
Section \ref{SECTION:background} presents some background information.
Section \ref{SECTION:method} introduces the proposed approach, including a motivating example, the framework, algorithms, and complexity analyses.
Section \ref{SECTION:design} explains the experimental set-up to examine the performance of our proposed approach.
Section \ref{SECTION:results} presents and analyzes the experimental results, and examines some potential threats to validity.
Section \ref{SECTION:configurable} explores the extension of our proposed method from numerical to non-numerical domains.
Section \ref{SECTION:discussions} provides some discussions regarding software failures and faults.
Section \ref{SECTION:relatedWork} discusses some related work about ART.
Section \ref{SECTION:conclusions} concludes this paper, and identifies some potential future work.

\section{Background}
\label{SECTION:background}
In this section, we present some of the preliminary concepts, including an introduction to both ART and ANN.

\subsection{Preliminaries
\label{SECTION:Preliminaries}}

Given a faulty software-system under test (SUT), a test case $t$ is called a \textit{failure-causing input} if the output or behavior of the SUT, when executed with $t$, is not as expected, as determined by the \textit{test oracle} \cite{Weyuker1982,barr2014oracle,zhou2018introduction}.
Generally speaking, two fundamental features can be used to describe the properties of the faulty SUT:
the \textit{failure rate}; and the \textit{failure pattern} \cite{Chan1996}.
The failure rate is the number of failure-causing inputs as a proportion of all possible inputs; and the failure pattern is the distribution of failure-causing inputs across the input domain, including both the locations and geometric shapes.
Before testing, these two features are fixed, but unknown.

Figure~\ref{FIG:failurePattern} shows examples of the three broad categories of {\em point}, {\em strip}, and {\em block} failure patterns, as originally identified by Chan et al.~\cite{Chan1996}:
The squares in the upper part of each subfigure represent the two-dimensional input domain boundary;
and the black dots, a strip, and a block represent the failure-causing inputs (the failure pattern).
The code snippets in each dashed box in the subfigures represent sample faulty programs that can produce the corresponding failure pattern.
Previous studies have noted that block and strip patterns are more commonly encountered than point patterns \cite{White1980,Ammann1988,Finelli1991,Bishop1993,Schneckenburger2007}.
Although many different geometric shapes of failure patterns exist, the three broad categories (of point, strip, and block) can generally represent them all:
For example, a two-dimensional (2D) failure region may be a circle-like shape, which can be categorized as a block pattern, because a circle-like failure region is also a block; similarly, a narrow (2D) eclipse can be classified as a strip pattern;
and so on.

\subsection{Adaptive Random Testing (ART)}

\begin{figure}[!t]
\centering
\graphicspath{{Graphs/}}
\resizebox{0.48\textwidth}{!}{
    \subfigure[]
    {
        \includegraphics[width=0.225\textwidth]{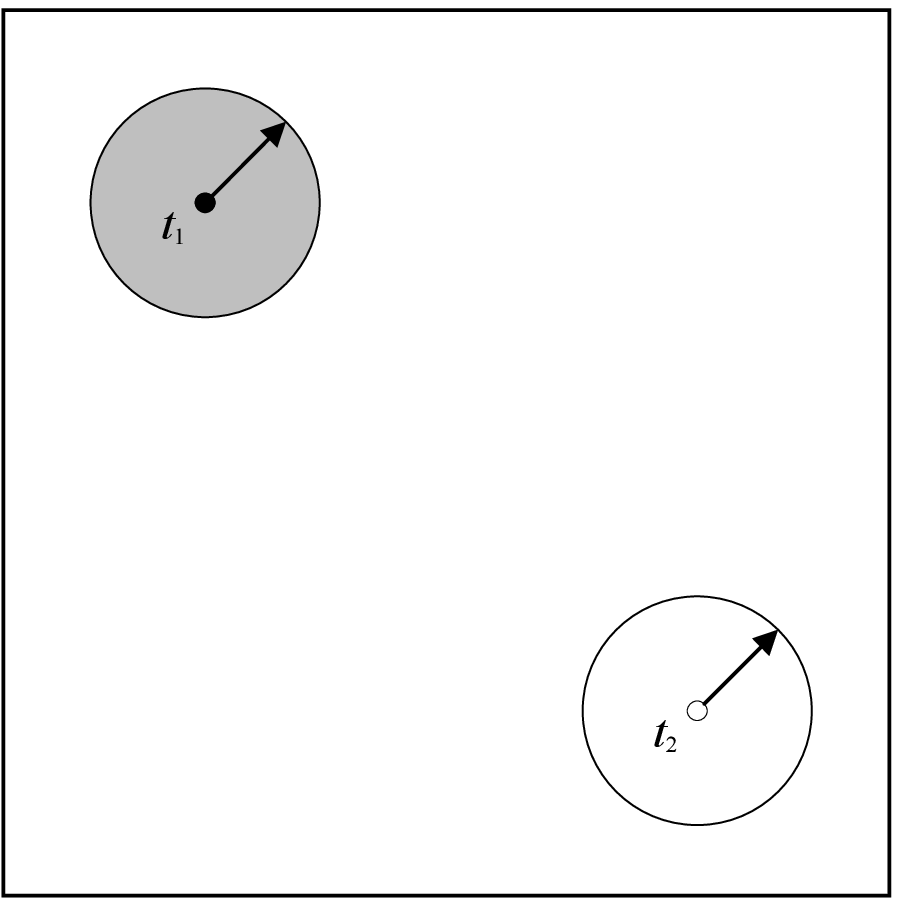}
        \label{figure:2.1}
    }
    \subfigure[]
    {
        \includegraphics[width=0.225\textwidth]{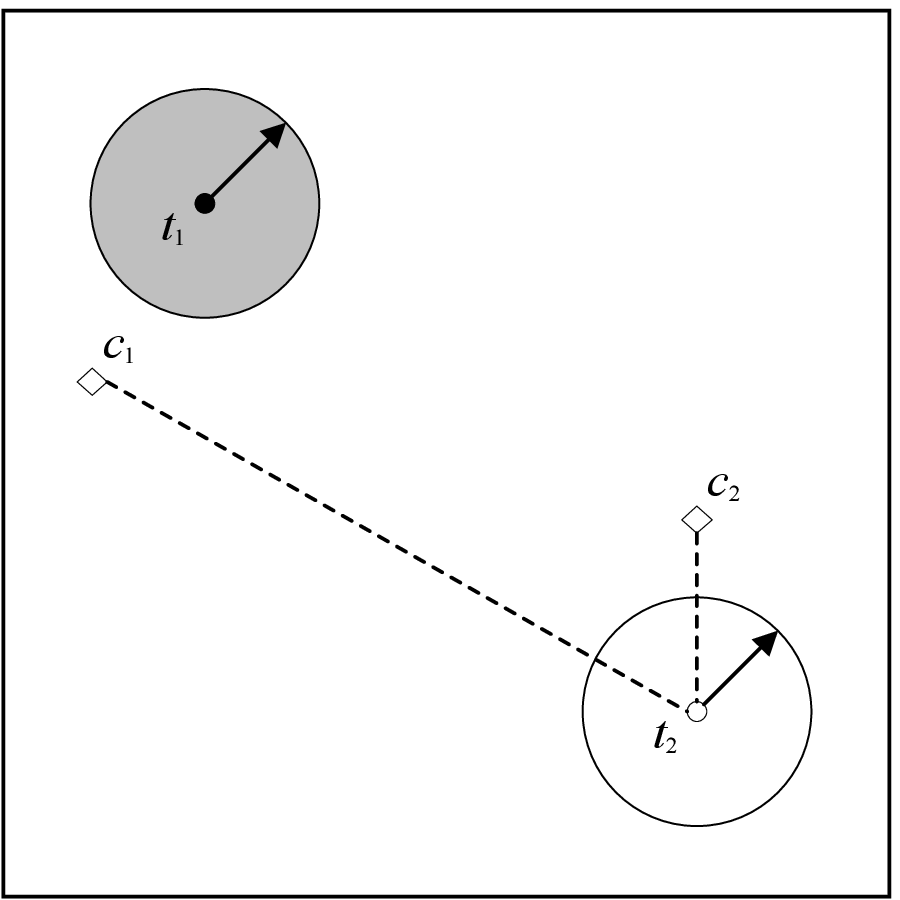}
        \label{figure:2.2}
    }
    }
    \caption{Motivation of ART.}
    \label{FIG:ARTmotivation}
\end{figure}

An important observation made, independently, by multiple researchers in different areas, is that failure-causing inputs tend to cluster into connected regions:
\textit{failure regions} \cite{White1980,Ammann1988,Finelli1991,Bishop1993,Schneckenburger2007}.
Accordingly, if a test case $t_1$ is a failure-causing input, then it is highly probable that its neighbors are also failure-causing inputs; similarly, if $t_2$ is not a failure-causing input, then its neighbors are also likely to not be failure-causing.
This is shown in Figure \ref{FIG:ARTmotivation}.
Consider the two candidate test cases $c_1$ and $c_2$ in Figure \ref{figure:2.2}.
Because $c_1$ is close to the failure-causing input $t_1$, it is considered to have a higher probability to be a failure-causing input.
Similarly, because $c_2$ is close to the non-failure-causing input $t_2$, it has a higher probability of being a non-failure-causing input.
This leads to the heuristic that a program input far away from non-failure-causing inputs may have a higher probability of causing failure than neighboring test inputs.
This has inspired \textit{Adaptive Random Testing} (ART), which aims to achieve a more diverse, or even, spread of test cases across the input domain~\cite{Chen2010,chen2015revisit}.

\begin{figure}[!b]
    \centering
    \fbox{
        \centering
        \parbox{0.95\linewidth}{
            \begin{algorithmic}[1]
                \renewcommand{\algorithmicrequire}{\textbf{Input:}}
                \renewcommand{\algorithmicensure}{\textbf{Output:}}
                \renewcommand{\algorithmicelsif}{\algorithmicelse}
                \renewcommand{\algorithmicthen}{}
            \STATE Set $C\leftarrow\{\}$, $E\leftarrow\{\}$
            \STATE Randomly generate a test case ${t}$ from the input domain, according to uniform distribution
            \WHILE {The stopping condition is not satisfied}
            \STATE Add ${t}$ into $E$, i.e., $E\leftarrow E\bigcup\{{t}\}$
            \STATE Randomly choose a specific number of elements from $\mathcal{D}$ to form $C$ according to the specific criterion
            \STATE Find a ${t}\in C$ as the next test case satisfying the specific criterion
            \ENDWHILE
            \STATE Report the result and exit
            \end{algorithmic}
        }
    }
    \caption{Framework pseudocode for STFCS ART \cite{Huang2019}.}
    \label{FIG:STFCS}
\end{figure}

ART refers to a family of testing approaches that randomly generate test cases that are evenly spread over the entire input domain \cite{Anand2013,Huang2019}.
Because there are many approaches to generating these test cases, there are many different ART implementations \cite{Huang2019}.
In this paper, we focus on the \textit{Select-Test-From-Candidates Strategy} (STFCS)
ART category \cite{Huang2019}, which has been the most popular and extensively-studied category.
STFCS selects the next test case from a set of randomly-generated candidates based on some criteria or evaluation involving the previously-executed test cases.
Figure \ref{FIG:STFCS} shows the STFCS framework \cite{Huang2019}, which contains two components:
\textit{random-candidate-set-construction}; and
\textit{test-case-selection}.
STFCS makes use of two sets:
one to store the random candidates (the \textit{candidate set}, $C$); and
one to store the previously-executed test cases (the \textit{executed set}, $E$).
Two of the most popular ART STFCS algorithms are \textit{Fixed-Size-Candidate-Set ART} (FSCS) \cite{Chen2010} and \textit{Restricted Random Testing} (RRT) \cite{Chan2006}, both of which are explored in this paper.
\subsubsection{Fixed-Size-Candidate-Set ART (FSCS)}
\label{SECTION:FSCS}

When generating the next test case, FSCS randomly generates $k$ candidates to form the candidate set $C$.
Each candidate is then evaluated against all the previously-executed test cases
---
all elements in $E$.
An element $c$ from $C$ will be the ``\textit{best}" choice as the next test case if it satisfies the following criterion:
\begin{equation}
\label{EQ:FSCS}
    \forall c' \in C,~\min_{e \in E} dist(c, e) \geq \min_{e \in E}dist(c', e),
\end{equation}
where $dist(c,e)$ is a function that measures the distance between two test inputs $c$ and $e$.
In other words, assessment of each random candidate $c'$ in $C$ involves identification of the closest element in $E$
---
the \textit{Nearest Neighbor} (NN) of $c'$ in $E$.
Because $k$ is usually a fixed constant ($10$ in most cases \cite{Chen2004}), the generation of each next test case only requires the identification of this fixed number of nearest neighbors.
However, as the size of $E$ increases, identification of each nearest neighbor incurs an increasingly large amount of overhead.
Therefore, the FSCS time overheads mainly depend on the NN search process.

\subsubsection{Restricted Random Testing (RRT)
\label{SECTION:RRT}}

In contrast to FSCS, RRT continues to generate and check random candidates, one by one, until the first suitable one is identified as the next test case.
This means that the number of random candidates is flexible, not fixed, during test-case generation.
The RRT criterion to determine whether or not a random candidate $c$ can be selected as the next test case, against $E$, is defined as:
\begin{equation}
    \label{EQ:RRT}
    \min_{e \in E} dist(c, e) > \sqrt[d]{\frac{\sigma_{d} \times A \times R}{\pi^{\lfloor d/2\rfloor} \times |E|}},
\end{equation}
where
$d$ is the dimension of the input domain $\mathcal{D}$;
$A$ is the size of $\mathcal{D}$;
$R$ is a constant parameter (generally called the \textit{exclusion ratio} \cite{chan2007controlling}); and
$\sigma_{d}$ is the formula coefficient, which can be written as:
\begin{equation}
    \sigma_{d} =
    \left\{
        \begin{array}{rl}
            {(\sigma_{d-2} \times d)}/{2}, &d > 2, \\
            1,&d = 2,  \\
            1/2,&d = 1.  \\
         \end{array}
     \right.
\end{equation}
Similar to FSCS, the RRT computational overheads are strongly connected to the speed of identification of each NN, and can increase as the size of $E$ increases.

\subsection{Approximate Nearest Neighbor (ANN)}

In this section, we first introduce the \textit{Nearest Neighbor} (NN) problem, and then focus on the \textit{Approximate Nearest Neighbor} (ANN).
A definition of NN is \cite{Arya1993,Arya1998}:
\begin{definition}[\textbf{\textit{Nearest Neighbor, NN}}]
    \textit{Given a set $V$ of $m$ data points in a $d$-dimensional space $\mathbb{R}^d$
    (i.e., $V \subset \mathbb{R}^d$),
    and given a query point $q \in \mathbb{R}^d$, the point $v$ of $T$ is the nearest neighbor to $q$ such that:}
    \begin{equation}
        \forall v' \in V,~dist(v, q) \leq dist(v', q),
    \end{equation}
    \textit{where $dist(v,q)$ gives the distance between points $v$ and $q$.}
\end{definition}

Although the NN problem can find the exact nearest neighbors, its query efficiency (for both space and query time) can be low, especially when the dimensionality is high.
An alternative to NN is to query \textit{approximate} (not exact) nearest neighbors, as in the Approximate NN (ANN) problem:\,
ANN may enable a more efficient performance.
The concept of ANN can be defined as \cite{Arya1993,Arya1998}:
\begin{definition}[\textbf{\textit{Approximate Nearest Neighbor, ANN}}]
    \textit{Given a set $V$ of data points in a metric space $\mathbb{R}^d$ (i.e., $V \subset \mathbb{R}^d$), and
    given a query point $q \in \mathbb{R}^d$,
    a point $v$ from $V$ is an approximate nearest neighbor satisfying the following condition:}
    \begin{equation}
        dist(v, q) \leq (1+\epsilon)dist(v^*, q),
    \end{equation}
    \textit{where $v^*$ is the exact or real nearest neighbor to $q$,
    and $\epsilon$ is an \textit{approximation factor} parameter  ($\epsilon > 0$).}
\end{definition}

ANN algorithms aim to find a point whose distance from the query point is at most $(1+\epsilon)$ times the distance from the query to the actual nearest neighbor, where $(1+\epsilon)$ is generally called the \textit{approximation factor} \cite{Andoni2006}.

\subsubsection{Locality-Sensitive Hashing (LSH)}

There are many techniques to solve the ANN problem, one of which is \textit{Locality-Sensitive Hashing} (LSH) \cite{Indyk1998}.
The basic principle of LSH is to make use of a {hash function} from the same {hash family} for placing each data point into a {bucket} such that the probability of {collision} is much higher for points that are close to each other than for those that are far apart.
An ANN search can be carried out as follows:
(1) Find the bucket to which a query point is hashed and choose the data points in these buckets as candidates; and
(2) Rank candidates according to their actual distance to the query point to find the nearest neighbor.

LSH relies on the existence of LSH functions.
Given $\mathcal{H}$, a family of hash functions mapping $\mathbb{R}^d$ to the integer set $\mathbb{N}$
(i.e., $\mathcal{H} = \{h: \mathbb{R}^d \rightarrow \mathbb{N}\}$),
then a definition of LSH \cite{Indyk1998} is as follows:
\begin{definition}[\textbf{\textit{Locality-Sensitive Hashing, LSH}}]
    \textit{A family $\mathcal{H}$ is called $(r_1, r_2, p_1, p_2,)$-sensitive if, for any two points $v,q \in \mathbb{R}^d$},
    \begin{equation}
            \left\{
        \begin{array}{rl}
            \textrm{if~} dist(v,q) \leq r_1 \textrm{,~then~} \textrm{Pr}_{\mathcal{H}}[h(q) = h(v)] \geq p_1, \\
            \textrm{if~} dist(v,q) \geq r_2 \textrm{,~then~} \textrm{Pr}_{\mathcal{H}}[h(q) = h(v)] \leq p_2, \\
         \end{array}
         \right.
    \end{equation}
    where $h$ is a function selected randomly (with uniform distribution) from $\mathcal{H}$;
    $\textrm{Pr}_{\mathcal{H}}[h(q) = h(v)]$ represents the probability that $h(q) = h(v)$; and
    $r_1$, $r_2$, $p_1$, and $p_2$ are parameters satisfying $p_1 > p_2$ and $r_1 < r_2$.
\end{definition}

In this study, we adopted an LSH family based on $p$-stable
distributions\footnote{For example,
a \textit{Cauchy distribution}, defined by the density function $g(x)=\frac{1}{\pi}\frac{1}{1+x^2}$, is 1-stable; 
and
a \textit{Gaussian (normal) distribution}, defined by the density function $g(x)=\frac{1}{\sqrt{2\pi}}e^{-x^2/2}$, is 2-stable \cite{Datar2004}.}
(also called E$^2$LSH) \cite{Datar2004}, that works for all $p \in (0,2]$.
Formally, each hash function $h_{\boldsymbol{a},b}(\boldsymbol{v})$: $\mathbb{R}^d \rightarrow \mathbb{N}$ maps a $d$-dimensional vector $\boldsymbol{v}$ onto the set of integers, which is defined as:
\begin{equation}
    \label{EQ:LSH}
    h_{\boldsymbol{a},b}(\boldsymbol{v}) = \Bigg\lfloor \frac{\boldsymbol{a} \cdot \boldsymbol{v} + b}{w} \Bigg\rfloor,
\end{equation}
where $\boldsymbol{a}$ is a random $d$-dimensional vector each entry in which is selected independently from a $p$-stable distribution;
$w$ is a fixed parameter for the entire family ($w$  acts as a quantization width); and
$b$ is a real number selected uniformly from the range $[0,w)$.
The $h_{\boldsymbol{a},b}(\boldsymbol{v})$ corresponds to the composition of a projection to a random direction with a random offset and a quantization by a constant.
According to Eq. (\ref{EQ:LSH}), the collision probability between two points $\boldsymbol{q}$ and $\boldsymbol{v}$ can be calculated as:
\begin{equation}
    \label{EQ:PRO}
    \textrm{Pr}_{\mathcal{H}}[h(\boldsymbol{q}) = h(\boldsymbol{v})] = \int_0^w\frac{1}{\alpha} \cdot g\bigg(\frac{x}{\alpha}\bigg)\cdot\bigg(1-\frac{x}{w}\bigg)dx,
\end{equation}
where $\alpha = dist(\boldsymbol{q},\boldsymbol{v})$, and $g(x)$ is the density function describing the probability density distribution of the variable $x$.
It can be seen that with the increase of $\alpha$, the collision probability decreases monotonically.

To make locality-sensitiveness actually useful, LSH employs a function family $\mathcal{F}$ (a hash table) by concatenating $M$ hash functions from $\mathcal{H}$,
i.e., $\mathcal{F}=\{f(\boldsymbol{v}):\mathbb{R}^d \rightarrow \mathbb{N}^M\}$,
such that each $f \in \mathcal{F}$ can be represented as:
$f(\boldsymbol{v})=(h_1(\boldsymbol{v}), h_2(\boldsymbol{v}), \cdots, h_M(\boldsymbol{v}))$,
where $h_i \in \mathcal{H}~(1\leq i \leq M)$.

\subsubsection{Scalable Locality-Sensitive Hashing}

\begin{figure}[!b]
    \centering
    \graphicspath{{Graphs/}}
    \includegraphics[width=0.45\textwidth]{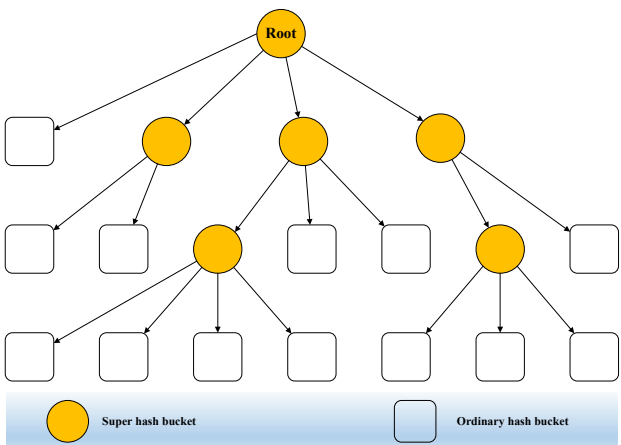}
    \caption{An example of an SLSH indexing structure.}
    \label{FIG:SLSH}
\end{figure}

Because traditional LSH (including E$^2$LSH) deals with ANN for static data sets, \textit{Scalable Locality-Sensitive Hashing} (SLSH) \cite{Hu2015,Li2017} has been proposed for dynamic data sets.
In this study, we adopted the SLSH from Hu \& Jiang \cite{Hu2015}, which adjusts the original E$^2$LSH in the following two ways:
\begin{itemize}
    \item
    Scalable hash family:
    SLSH extends the hash family $\mathcal{H}$ of E$^2$LSH by dynamically adjusting the value of $w$:
    If $\mathcal{H}(w)$ is a scalable hash family, and
    $\forall \boldsymbol{v}\in \mathbb{R}^d$, $\mathcal{H}(w)=\{h(\boldsymbol{v},w):\mathbb{R}^d \rightarrow \mathbb{N}\}$,
    then $\mathcal{H}(w) = \{h_1(\boldsymbol{v},w_1), h_2(\boldsymbol{v},w_2), \cdots, h_M(\boldsymbol{v},w_M)\}$.

    \item
    Scalable indexing structure:
    According to the scalable hash family, a scalable hash table is designed as $\mathcal{T}(\boldsymbol{w}) = \{t(\boldsymbol{v},\boldsymbol{w}):
    \mathbb{R}^d \rightarrow \mathbb{N}^M\}$,
    from which an individual hash table $t(\boldsymbol{v},\boldsymbol{w})$ can be described as
    $t(\boldsymbol{v},\boldsymbol{w}) = (h_1(\boldsymbol{v},w_1), h_2(\boldsymbol{v},w_2), \cdots, h_M(\boldsymbol{v},w_M))$,
    where $w_1 > w_2 > \cdots > w_M$.
    Unlike LSH, SLSH adopts a hierarchical indexing structure to organize $M$ hash functions.
    SLSH introduces a new parameter $m$ to configure the upper bound of the hash bucket capacity.
    If the number of collision elements in a hash bucket exceeds the limit $m$
    (this \textit{ordinary hash bucket} becomes a \textit{super hash bucket}),
    then these elements will be re-hashed to the next level of hash buckets.
    In other words, each ordinary hash bucket stores data points,
    and each super hash bucket stores some pointers to other hash buckets.
    Figure \ref{FIG:SLSH} presents an example of the scalable indexing structure for SLSH,
    where it can be seen that different hash tables may have different numbers of hash functions.
\end{itemize}

ART generates the new test cases based on a dynamic set of previously executed test cases \cite{Chen2010,Huang2019}.
Because SLSH is a scalable version of LSH for dynamic data sets, it is more suitable for ART.
For ease of description, unless explicitly stated otherwise, LSH in the following sections refers to SLSH.

\section{Proposed LSH-ART Approach}
\label{SECTION:method}

In this section, we propose a novel approach for ART STFCS based on the LSH version of ANN:
\textit{LSH-based ART} (LSH-ART).
We first introduce the motivation of this work, and then provide a framework to support LSH-ART.
We also present two detailed algorithms to implement the framework, and present their complexity analyses.

\begin{figure}[!t]
    \centering
    \graphicspath{{Graphs/}}
    \includegraphics[width=0.32\textwidth]{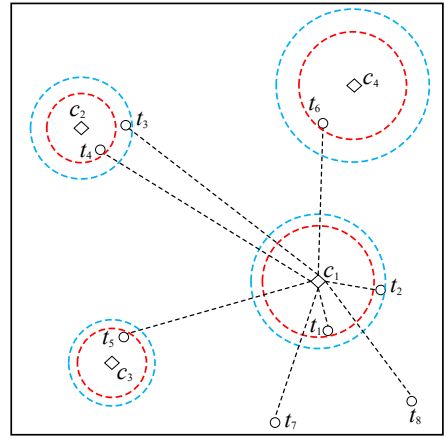}
    \caption{A motivating example.}
    \label{FIG:motivation}
\end{figure}

\begin{figure*}[!t]
    \centering
    \graphicspath{{Graphs/}}
    \includegraphics[width=\textwidth]{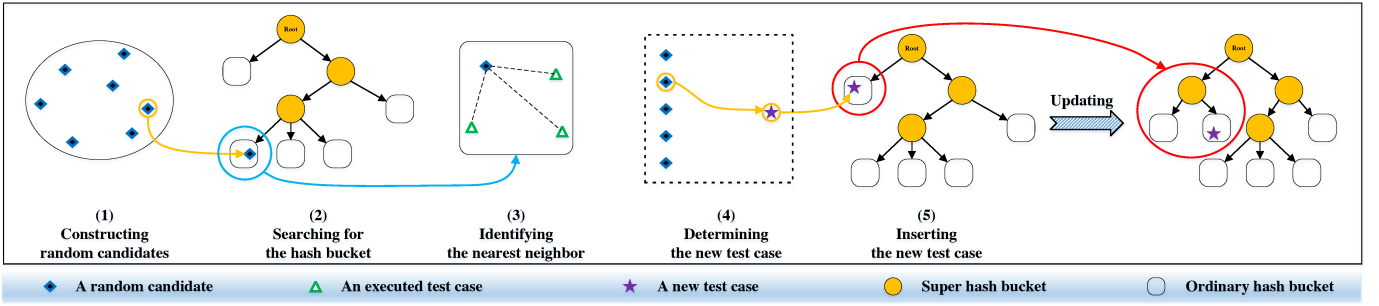}
    \caption{LSH-ART Framework.}
    \label{FIG:Framework}
\end{figure*}

\subsection{Motivating Example}

As discussed in Sections \ref{SECTION:FSCS} and \ref{SECTION:RRT}, the computational overheads of both FSCS and RRT depend strongly on the NN search process, especially when the number of already-executed test cases is large.
This is because it is necessary to consider all executed test cases in the NN search process.
To reduce the computational overheads, therefore, it is necessary to reduce the number of executed test cases that need to be processed.
One way to do this is to sacrifice an amount of accuracy in the NN search, by adopting the ANN search instead.
We next present a motivating example to illustrate this.

Figure \ref{FIG:motivation} shows a situation where eight test cases have already been executed, $t_1, t_2,\cdots,t_8$.
In FSCS (and somewhat similarly for RRT), the nearest (executed test case) neighbors for each of the four candidate test cases ($c_1$, $c_2$, $c_3$, and $c_4$) must be identified.
To calculate the NN of $c_1$, for example, the distance from $c_1$ to each executed test case $t_i~(i=1,2,\cdots,8)$ is calculated, producing eight distances.
With an ANN  (instead of exact NN) search, both $t_1$ and $t_2$ may be selected as the (approximate) NN of $c_1$.
(If $t_1$ is selected as the NN for $c_1$, then the ANN search process is equivalent to the (exact) NN search process.)
Although $t_2$ is slightly farther away from $c_1$ than $t_1$, it is still close.
The $c_2$ candidate has a similar situation with $t_3$ and $t_4$, but the $c_3$ and $c_4$ candidates do not face this:
$c_3$ and $c_4$ may have only one element for both NN and ANN search processes
($t_5$ for $c_3$; and $t_6$ for $c_4$),
resulting in that element being selected as the NN/ANN.

\subsection{Framework}

Figure \ref{FIG:Framework} shows the proposed LSH-ART framework for generating each test case.
The framework consists of five components:
(1) Constructing random candidates;
(2) Searching for the hash bucket;
(3) Identifying the nearest neighbor;
(4) Determining the new test case; and
(5) Inserting the new test case into the LSH tree.
We next explain each component.

Component (1) involves constructing/selecting a random candidate (or some random candidates) from the input domain.
Component (2) consists of hashing each random candidate in the corresponding ordinary hash bucket.
The hashing process of each candidate $c$ starts with the root of the LSH indexing structure, and then checks its hash bucket type:
If the current hash bucket $\mathcal{B}$ is an ordinary hash bucket, $\mathcal{B}$ will be returned for $c$ (i.e., $c$ should be included in $\mathcal{B}$);
however, if $\mathcal{B}$ is a super hash bucket, then the hashing process is repeated using the respective hash function of each level until an ordinary hash bucket is found for $c$.

For each candidate $c$, when an ordinary hash bucket $\mathcal{B}$ is found, it is believed that all elements in $\mathcal{B}$ are considered as potential ANNs for $c$.
Then, Component (3) identifies the NN within this hash bucket $\mathcal{B}$ (compared with all executed test cases, however, this NN is approximate).
Component (4) then determines which candidate should be selected as the next test case (i.e., the new test case), according to some criterion.
Once a new test case is determined, Component (5) inserts this test case in the LSH indexing tree:
If the corresponding ordinary hash bucket has not exceeded its limit, then this new test case is directly inserted into the hash bucket;
otherwise, this ordinary hash bucket is promoted into a super hash bucket (involving an \textit{updating} process), and its elements are rehashed in the ordinary hash buckets in the next level.

\subsection{Algorithms\label{SECTION:algorithm}}

In this study, we focus on the STFCS category of ART, proposing an STFCS version of LSH-ART, \textit{LSH-based STFCS} (LSH-STFCS).
We first present the LSH-STFCS algorithm, and then discuss two well-known implementations of STFCS ART (FSCS \cite{Chen2010} and RRT \cite{Chan2006}).
We then present the LSH-STFCS versions of FSCS and RRT:
\textit{LSH-based FSCS} (LSH-FSCS); and
\textit{LSH-based RRT} (LSH-RRT).

\subsubsection{LSH-STFCS}

\begin{algorithm}[!b]
    \DontPrintSemicolon
    \footnotesize
    \caption{LSH-STFCS Pseudocode}
    \label{ALG:LSH-STFCS}
      \SetKwData{Left}{left}\SetKwData{This}{this}\SetKwData{Up}{up}
      \SetKwInOut{Input}{Input}\SetKwInOut{Output}{Output}
      \renewcommand{\algorithmicendwhile}{\algorithmicend~\algorithmicwhile}

    \Input{Input domain $\mathcal{D}$.}
    \Output{Executed test cases $E$.}

    $C\leftarrow \texttt{\textbf{EmptySet()}}$\;
    $E\leftarrow \texttt{\textbf{EmptySet()}}$\;
    $\mathcal{T} \leftarrow \texttt{\textbf{CreateHashBucket()}}$\;
    $t \leftarrow \texttt{\textbf{ConstructCandidate(}}\mathcal{D}\texttt{\textbf{)}}$\;

    \While{The stopping condition is not satisfied}
    {
    $E \leftarrow \texttt{\textbf{Append(}}E,t\texttt{\textbf{)}}$\;
    $\mathcal{T} \leftarrow \texttt{\textbf{InsertTest(}}\mathcal{T},t\texttt{\textbf{)}}$~~~~~~~~~~~~~~~~~~~~~~~$\triangleright$ Component (5)\;
    $C \leftarrow \bigcup\{\texttt{\textbf{ConstructCandidate(}}\mathcal{D}\texttt{\textbf{)}}\}$~~~~~~$\triangleright$ Component (1)\;
    $S \leftarrow \texttt{\textbf{EmptySet()}}$\;

    \For{$\forall c \in C$}
    {
    $\mathcal{B} \leftarrow \texttt{\textbf{SearchForHashBucket(}} \mathcal{T},c\texttt{\textbf{)}}$~~~$\triangleright$ Component (2)\;
    $s \leftarrow \texttt{\textbf{IdentifyNN(}}\mathcal{B}, c\texttt{\textbf{)}}$~~~~~~~~~~~~~~~~~~~~$\triangleright$ Component (3)\;
    $S \leftarrow S \bigcup\{{s}\}$\;
    }

    $t \leftarrow \texttt{\textbf{DetermineNewTest(}}C,S\texttt{\textbf{)}}$~~~~~~~~~~~~~~$\triangleright$ Component (4)\;
    }
    \textbf{return} $E$\;
\end{algorithm}

Algorithm \ref{ALG:LSH-STFCS} presents a pseudocode description of LSH-STFCS.
In the initialization stage (Lines 1 to 3), $C$ is used to store random candidates,  $E$ is used to store the already-executed test cases, and $\mathcal{T}$ is an LSH table (or an LSH indexing tree):
$C$ and $E$ are initialized as empty sets;
and $\mathcal{T}$ is initialized as a null tree.
The algorithm randomly generates a candidate $t$ from the input domain $\mathcal{D}$ as the first test case (line 4).
If the stopping condition is not satisfied, then $t$ is appended to $E$ (Lines 5 and 6).
The algorithm repeats the generation process until the stopping condition is satisfied (Lines 5 to 16).
When a candidate $t$ is selected as the next test case, it is inserted into the LSH indexing tree $\mathcal{T}$ using the function $\texttt{\textbf{InsertTest(}}\mathcal{T},t\texttt{\textbf{)}}$ (Component (5) of the framework) (Line 7).
A set of random candidates is constructed from the input domain $\mathcal{D}$ using the function $\texttt{\textbf{ConstructCandidate(}}\mathcal{D}\texttt{\textbf{)}}$ (Component (1) of the framework).
The number of random candidates can be either fixed or flexible.
For each candidate $c$, the algorithm tries to find a potential ordinary hash bucket $\mathcal{B}$ in which to store $c$, by searching the hash table $\mathcal{T}$ (the function $\texttt{\textbf{SearchForHashBucket(}} \mathcal{T},c\texttt{\textbf{)}}$; 
Component (2)).
Then, the function $\texttt{\textbf{IdentifyNN(}}\mathcal{B}, c\texttt{\textbf{)}}$ identifies the NN of $c$ in $\mathcal{B}$ (Component (3)).
A new test case is determined,
according to the NN $s$ in $S$ of each candidate $c$ in $C$,
using the function $\texttt{\textbf{DetermineNewTest(}}C,S\texttt{\textbf{)}}$ (Component (4)).

\begin{algorithm}[!t]
    \DontPrintSemicolon
    \footnotesize
    \caption{$\texttt{\textbf{SearchForHashBucket(}}\mathcal{T},c\texttt{\textbf{)}}$}
    \label{ALG:Searching}
      \SetKwData{Left}{left}\SetKwData{This}{this}\SetKwData{Up}{up}
      \SetKwInOut{Input}{Input}\SetKwInOut{Output}{Output}
      \renewcommand{\algorithmicendwhile}{\algorithmicend~\algorithmicwhile}

    \Input{LSH indexing tree $\mathcal{T}$, candidate test case $c$.}
    \Output{Hash bucket $\mathcal{B}$.}

    $\mathcal{B} \leftarrow \mathcal{T}$\;
    $flag \leftarrow \texttt{true}$\;

    \While{$flag$}
    {
    $flag\leftarrow \texttt{\textbf{IsSuperBucket(}}\mathcal{B}\texttt{\textbf{)}}$\;

    \eIf{$flag$}
    {
    $\mathcal{B} \leftarrow \texttt{\textbf{GetHashBucket(}}\mathcal{B},c\texttt{\textbf{)}}$\;
    }
    {
    $flag \leftarrow \texttt{false}$\;
    }
    }
    \textbf{return} $\mathcal{B}$\;
\end{algorithm}

\begin{algorithm}[!b]
    \DontPrintSemicolon
    \footnotesize
    \caption{$\texttt{\textbf{IdentifyNN(}}\mathcal{B}, c\texttt{\textbf{)}}$}
    \label{ALG:IdentifyingNN}
      \SetKwData{Left}{left}\SetKwData{This}{this}\SetKwData{Up}{up}
      \SetKwInOut{Input}{Input}\SetKwInOut{Output}{Output}
      \renewcommand{\algorithmicendwhile}{\algorithmicend~\algorithmicwhile}

    \Input{Ordinary hash bucket $\mathcal{B}$, random candidate $c$.}
    \Output{Nearest neighbor $s$.}

    $\delta \leftarrow MaxValue$\;

    \ForEach{$e \in \mathcal{B}$}
    {
     $l \leftarrow dist(e,c)$\;
     \If{$l < \delta$}
     {
     $\delta \leftarrow l$\;
     $s \leftarrow e$\;
     }
    }
    \textbf{return} $s$\;
\end{algorithm}

\begin{algorithm}[!t]
    \DontPrintSemicolon
    \footnotesize
    \caption{$\texttt{\textbf{InsertTest(}}\mathcal{T},t\texttt{\textbf{)}}$}
    \label{ALG:Inserting}
      \SetKwData{Left}{left}\SetKwData{This}{this}\SetKwData{Up}{up}
      \SetKwInOut{Input}{Input}\SetKwInOut{Output}{Output}
      \renewcommand{\algorithmicendwhile}{\algorithmicend~\algorithmicwhile}

    \Input{LSH indexing tree $\mathcal{T}$, test case $t$.}
    \Output{Updated LSH indexing tree $\mathcal{T}$.}

    $\mathcal{B} \leftarrow \texttt{\textbf{SearchForHashBucket(}}\mathcal{T},t\texttt{\textbf{)}}$\;

    $\mathcal{B}' \leftarrow \mathcal{B}$\;
    \eIf{$|\mathcal{B}'| + 1 < m $}
    {
    /* $m$ is the capacity of the ordinary hash bucket */\;
    $\mathcal{B}'\leftarrow \texttt{\textbf{Append(}}\mathcal{B}',t\texttt{\textbf{)}}$\;
    }
    {
    $\mathcal{B}' \leftarrow \texttt{\textbf{CreateHashBucket()}}$\;
    $\mathcal{B}' \leftarrow \texttt{\textbf{HashElements(}}\mathcal{B}',\mathcal{B}\texttt{\textbf{)}}$\;
    }
    $\mathcal{T} \leftarrow \texttt{\textbf{ReplaceHashBucket(}} \mathcal{T}, \mathcal{B}', \mathcal{B} \texttt{\textbf{)}}$\;

    \textbf{return} $\mathcal{T}$\;
\end{algorithm}

Because Components (1) and (4) can be implemented using different strategies, we discuss them separately in the following sections.
Here, we discuss the common components, Components (2), (3), and (5)
(the functions $\texttt{\textbf{SearchForHashBucket(}} \mathcal{T},c\texttt{\textbf{)}}$, $\texttt{\textbf{IdentifyNN(}}\mathcal{B}, c\texttt{\textbf{)}}$, and $\texttt{\textbf{InsertTest(}}\mathcal{T},t\texttt{\textbf{)}}$).
Algorithm \ref{ALG:Searching} contains the pseudocode for the function $\texttt{\textbf{SearchForHashBucket(}}\mathcal{T},c\texttt{\textbf{)}}$:
The algorithm repeatedly checks the hash bucket by hashing the candidate test case $c$ ($\texttt{\textbf{GetHashBucket(}} \mathcal{B},c\texttt{\textbf{)}}$) until an ordinary hash bucket is found ($\texttt{\textbf{IsSuperBucket(}} \mathcal{B}\texttt{\textbf{)}}$ is used to determine the bucket type).
The identified ordinary hash bucket is then used to store $c$.
Algorithm \ref{ALG:IdentifyingNN} shows the implementation of the function $\texttt{\textbf{IdentifyNN(}}\mathcal{B}, c\texttt{\textbf{)}}$, which identifies the nearest neighbor of a candidate $c$ in an ordinary hash bucket $\mathcal{B}$, by calculating the distances between $c$ and each element $e\in \mathcal{B}$.
Algorithm \ref{ALG:Inserting} lists the pseudocode for the function $\texttt{\textbf{InsertTest(}}\mathcal{T},t\texttt{\textbf{)}}$, which starts by finding the ordinary hash bucket $\mathcal{B}$ for $t$, using the function $\texttt{\textbf{SearchForHashBucket(}}\mathcal{T},t\texttt{\textbf{)}}$.
The algorithm then checks whether or not $\mathcal{B}$ needs to be updated, according to its capacity limit threshold $m$ (defined in advance).
If $\mathcal{B}$ is not already at capacity, then $t$ is inserted directly, using the $\texttt{\textbf{Append(}} \mathcal{B}',t\texttt{\textbf{)}}$ function;
otherwise, $\mathcal{B}$ is promoted into a super hash bucket, and its elements are re-hashed into ordinary hash buckets using the $\texttt{\textbf{CreateHashBucket()}}$ and $\texttt{\textbf{HashElements(}}\mathcal{B}',\mathcal{B}\texttt{\textbf{)}}$ functions.
Finally, the original $\mathcal{B}$ is replaced by $\mathcal{B}'$ in $\mathcal{T}$, using the function $\texttt{\textbf{ReplaceHashBucket(}} \mathcal{T}, \mathcal{B}', \mathcal{B} \texttt{\textbf{)}}$.

\subsubsection{LSH-FSCS}
To maintain the failure-finding benefits of the original FSCS, LSH-FSCS keeps the main FSCS processes, changing only those related to storing previously-executed test cases, and searching for each candidate's NN.
Accordingly, the LSH-FSCS Component (1) involves the construction of the fixed number ($k$) of randomly-generated candidates from the input domain (according to uniform distribution).
The LSH-FSCS Component (4) selects as the next test case that candidate $c$ whose distance to its NN is greatest, in terms of Eq. (\ref{EQ:FSCS}).

\subsubsection{LSH-RRT}
As with LSH-FSCS, LSH-RRT also retains the core original processes of RRT, changing only how executed test cases are stored, and how the NNs for each candidate are identified.
As discussed in Section \ref{SECTION:RRT}, RRT checks successive randomly-generated candidates to determine whether or not they satisfy the selection criteria, in terms of Eq. (\ref{EQ:RRT}).
In other words, RRT's generation of the next test case requires the construction and checking of a flexibly-sized set of random candidates.
More specifically,
LSH-RRT uses Component (1) to construct a random candidate and then Component (4) to check whether or not it is valid.
This process is repeated until a satisfactory new test case is generated.

\subsection{Complexity Analysis \label{SECTION:complexity}}
This section presents a brief investigation of the LSH-ART time and space complexity, through a formal mathematical analysis.
For ease of description, $n$ denotes the number of $d$-dimensional test cases to be generated, and $\mu_n$ is an upper bound of the number of hash functions used to generate the $n$ test cases.
As discussed by Hu \& Jiang \cite{Hu2015}, $\mu_n$ is generally equal to $\log(n)$.
In addition, $\psi_n$ denotes the depth of the LSH indexing tree when generating the $n$ test cases, where $\psi_n \leq \mu_n$. Finally, an analysis of time and space complexity is given for different ART implementations.

\begin{table*}[!t]
\centering
\scriptsize
\caption{Comparisons of Time and Space Complexity Among Different Implementations for ART}
\label{TAB:complexityComparison}
\resizebox{\textwidth}{!}{
\begin{tabular}{@{}cclllll@{}}
\hline
\multirow{2}*{\textbf{Complexity}} &\multirow{2}*{\textbf{Version}} &\multicolumn{5}{c}{\textbf{Different ART Implementations}}\\\cline{3-7}
&&\textit{Original} &\textit{Forgetting} &\textit{Distance-aware Forgetting} &\textit{KD-tree} &\textit{LSH}    \\\hline

\multirow{2}*{\textbf{Time}}
&FSCS  &$O(k \cdot d \cdot n^2)$ &$O(k \cdot d \cdot \lambda \cdot n)$ &$O(k \cdot \tau \cdot 3^d \cdot n)$ &$O(k \cdot d^2 \cdot n \cdot \log n)$* &$O(k \cdot n \cdot \log n)$* \\
&RRT  &$O(d \cdot n^2 \cdot \log n)$ &$O(d \cdot \lambda \cdot \log\lambda \cdot n)$ &$O(\log(\tau \cdot 3^d) \cdot \tau \cdot 3^d \cdot n)$ &$O(d^2 \cdot n \cdot \log n \cdot \log(d^2 \cdot \log n))$* &$O( \log m \cdot n \cdot \log n)$* \\\hline

\multirow{2}*{\textbf{Space}}
&FSCS  &$O(d \cdot n)$ &$O(d \cdot n)$ &$O(d \cdot 2^d \cdot  n)$ &$O(d \cdot n)$ &$O(n \cdot \log n)$* \\
&RRT  &$O(d \cdot n)$ &$O(d \cdot n)$ &$O(d \cdot 2^d \cdot n)$ &$O(d \cdot n)$ &$O(n \cdot \log n)$* \\\hline
\multicolumn{7}{l}{* in the worst-case scenario.}

\end{tabular}}
\end{table*}

\subsubsection{Time Complexity}

The time complexity involved in generating the $i$-th ($1 \leq i \leq n$) test case from $k$ random candidates $C=\{c_1,c_2,\cdots,c_k\}$ using LSH-ART is as follows.
As discussed in Section \ref{SECTION:algorithm}, each candidate $c \in C$ requires the following two steps that incur computational costs:
(1) Searching for the ordinary bucket $\mathcal{B}$ in the LSH indexing tree (or the LSH table) in which to (potentially) store $c$; and
(2) Identifying $c$'s NN in $\mathcal{B}$.
LSH-ART then needs to complete two further tasks:
(3) Determining the next test case $t$; and
(4) Inserting $t$ into the hash table.

The worst case scenario for Step (1) is that the targeted ordinary bucket $\mathcal{B}_i$ is in the $\psi_i$-th level (the deepest level) of the LSH indexing tree, resulting in an $O(\psi_i)$ order of time complexity for each candidate.
For Step (2), because each ordinary bucket $\mathcal{B}$ contains at most $m$ previously-executed test cases, a maximum time complexity of $O(d \times m)$ is needed to calculate the distance between each candidate and the $m$ executed test cases, and a further complexity of $O(m)$ to identify its NN.
Therefore, a time complexity of $O(k \times (\psi_i+d\times m + m))$ is required, for $k$ candidates, for (1) and (2).
Step (3) involves LSH-ART choosing an element from $k$ candidates as the next test case, resulting in a time complexity of $O(k)$.
For Step (4), LSH-ART requires a maximum time complexity of $O(\psi_i)$ to insert the new test case into the ordinary bucket $\mathcal{B}$.
If $\mathcal{B}_i$ exceeds its maximum capacity, then it will be upgraded into a super bucket, and all elements will be re-hashed into different ordinary buckets, resulting in a time complexity of $O(m)$.
In total, the worst time complexity for generating the $i$-th ($1 \leq i \leq n$) test case can be described as $O(k \times (\psi_i+d\times m + m)+k+\psi_i+m)$.

In conclusion, when generating $n$ test cases, the worst order of time complexity of LSH-ART can be written as:
\begin{small}
\begin{align}
    &O(\textrm{LSH-ART}) \nonumber\\
    &= O\Bigg(\sum_{i=1}^n{\Big(k \times (\psi_i+d\times m + m)+k+\psi_i+m\Big)}\Bigg) \nonumber\\
    &\leq O\Bigg(\sum_{i=1}^n{\Big(k \times (\mu_i+d\times m + m)+k+\mu_i+m\Big)}\Bigg) \nonumber\\
    &\leq O\Bigg(\sum_{i=1}^n{\Big(k \times (\log i+d\times m + m)+k+\log i+m\Big)}\Bigg) \nonumber\\
    &= O\Big(k \times (d \times m + m + 1) + m\Big) + O\Bigg((k+1)\times\sum_{i=1}^n{\log i}\Bigg) \nonumber\\
    &= O(k \times d \times m) + O\Big(k\times \log n!\Big) \nonumber\\
    &< O(k \times d \times m) + O\Big(k\times n \times \log n\Big) \nonumber\\
    &= O(k \times d \times m) + O(k \times n \times \log n).
\end{align}
\end{small}

The input domain dimensionality ($d$) and the maximum size of the ordinary bucket ($m$) are constants, set before testing begins.
The number of random candidates ($k$) is also a constant in LSH-FSCS, though not in LSH-RRT.
As explained by Mayer \& Schneckenburger \cite{Mayer2006a}, however, the average number of random candidates for RRT is probably logarithmic to the number of previously executed test cases.
Because the maximum size of each ordinary bucket is equal to $m$, the LSH-RRT $k$ is, at most, approximately $\log m$, 
which can also be considered a constant.
In summary, therefore, the worst order of time complexity of LSH-FSCS is $O(k \cdot n \cdot \log n)$, and that of LSH-RRT is $O(\log m \cdot n \cdot \log n)$.

\subsubsection{Space Complexity}
LSH-ART requires memory space to store
(1) all $n$ executed test cases,
(2) all $k$ candidates, and
(3) all hash buckets (both super and ordinary) of the SLSH indexing structure.
Obviously, the executed test cases need an order of space complexity of $O(n \times d)$; while the candidate test cases need an order of space complexity of $O(k \times d)$.
For the hash buckets, in the worst-case scenario
---
when each ordinary hash bucket contains only one test case, and all buckets are in the $\mu_n$-th level
---
the order of space complexity is $O(n \times \psi_n)$.
Therefore, the worst order of space complexity for LSH-ART can be described as:
\begin{align}
    O(\textrm{LSH-ART})
    &= O(n \times d + k \times d + n \times \psi_n)  \nonumber \\
    &\leq O(n \times d + k \times d + n \times \mu_n)  \nonumber \\
    &= O(n \times d + k \times d + n \times \log n)
\end{align}
Because the input domain dimension $d$ is determined before LSH-ART is applied, $d$ is a constant. 
In addition, $k$ is also effectively a constant in both FSCS and RRT versions of LSH-ART.
Therefore, the worst order of LSH-ART space complexity is $O(n \cdot \log n)$.

\subsubsection{Complexity Comparisons}
The time complexity of some ART implementations has been examined in previous studies \cite{Mao2019,Mayer2006a,Chen2004a,Mao2017}.
Chen et al. \cite{Chen2004a}, for example, reported on the order of time complexity of the original FSCS being equal to $O(n^2)$, when generating $n$ test cases.
Mayer \cite{Mayer2006a} showed the time complexity of the original RRT to be of the order of $n^2\log n$.
Mao et al. \cite{Mao2019,Mao2017} gave the order of time complexity for FSCS with distance-aware forgetting as $O(n)$, and for FSCS with KD-tree as $O(n \cdot \log n)$.
To the best of our knowledge, however, there are no time or space complexity analyses available for the other ART implementations.
In this paper, therefore, we provide an analysis of time and space complexity for these ART implementations:
Table \ref{TAB:complexityComparison} presents a detailed comparison of these complexities, for different ART implementations, when generating $n$ test cases.
In the table, $d$ is the dimension of the input domain; and 
$k$, $\lambda$, $\tau$, and $m$ are constant parameters.

\section{Experimental Frameworks
\label{SECTION:design}}
In this section, we first present the research questions related to the performance of our proposed approach, and then examine the experiments we conducted to answer them, from the perspectives of testing effectiveness, efficiency, and cost-effectiveness.

\subsection{Research Questions}

A goal of LSH-ART is to reduce the computational overheads of the original ART (FSCS \cite{Chen2010} and RRT \cite{Chan2006}) methods.
Therefore, an examination of the test-case generation time is necessary.
In addition to reducing computational costs, LSH-ART also aims to maintain comparable fault detection, thus delivering a high testing cost-effectiveness.
The LSH-ART fault-detection performance, thus, also needs to be examined, under different scenarios.

Unlike some other ART enhancements and variants, such as forgetting \cite{Chan2006a}, LSH-ART does not discard any test case information, and may thus achieve better fault detection, while maintaining comparable computational cost reductions.
Similarly, because LSH-ART implements ART with an ANN search, it should have less computational overheads than an NN search (such as with a KD-tree ART, in KDFC-ART \cite{Mao2019}), while hopefully maintaining comparable fault-detection effectiveness.

Based on this, we designed the following three research questions:
\begin{description}
    \item[\textbf{RQ1:}] [Effectiveness] \textit{How effective is LSH-ART at identifying software failures?}
      \begin{itemize}
       \item[\textbf{RQ1.1:}] \textit{How effective is LSH-ART at detecting the first software failure?}
       \item[\textbf{RQ1.2:}] \textit{How effective is LSH-ART at triggering software failures when running a fixed number of test cases?}
      \end{itemize}
    \item[\textbf{RQ2:}] [Efficiency] \textit{How efficient is LSH-ART at generating test cases?}
    \item[\textbf{RQ3:}] [Cost-effectiveness] \textit{How cost-effective is LSH-ART at revealing software failures?}
\end{description}

For each research question, LSH-ART was compared with both the original ART algorithms and some enhanced variants.

\subsection{Independent Variables}
We focused on the test-case generation methods and their parameters as the independent variables.

\subsubsection{Test-Case Generation Methods}

Because LSH-ART aims to enhance two well-known versions of STFCS ART (FSCS  and RRT), both LSH-FSCS and LSH-RRT were chosen for this variable;
as were the original FSCS and RRT (as baselines for comparisons).
We also selected another three types of enhanced variants of both FSCS and RRT \cite{Chan2006a,Mao2019,Mao2017}.
The first enhancement strategy was \textit{forgetting} \cite{Chan2006a}, which attempts to discard some executed test cases during the test-case generation process, to reduce computation time.
In this study, we considered two versions of forgetting, \textit{Random Forgetting} (RF) and \textit{Consecutive Retention} (CR) \cite{Chan2006a}.
When the number of already-executed test cases exceeds a predefined parameter $\lambda$, RF randomly selects $\lambda$ executed test cases for distance comparisons, whereas CR uses only the most recent $\lambda$ executed test cases.

The second enhancement strategy is a \textit{Distance-aware forgetting strategy} (DF) \cite{Mao2017}, which tries to partition the input domain into some disjoint subdomains, and then discard some distance calculations to reduce the computational overheads.
The final enhancement strategy takes advantage of a \textit{KD-tree} structure to store the executed test cases (which corresponds to the \textit{KD-tree strategy}), thus speeding up the precise NN search.

Overall, in addition to the original ART strategy (FSCS and RRT), we also applied five others, resulting in six ART test-case generation methods in the experiments:
We use \textit{ART}, \textit{RF}, \textit{CR}, \textit{DF}, \textit{KD}, and \textit{LSH} to represent these six strategies.
We also compared our proposed method with RT.

\subsubsection{Test-Case Generation Method Parameters}
\label{SECTION:Parameter}

Each ART algorithm has some associated parameters, which are considered constants in the experiments.
We used the recommended values for these parameters as identified in their respective studies.

For all ART methods, because all the input domains in this study were numerical, the Euclidean distance was used to measure the distance between test cases.
FSCS (both the original and its enhanced variants) constructs a fixed number ($k$) of random candidates to generate each test case:
As recommended by Chen et al. \cite{Chen2004}, $k$ was set to 10.
Similarly, RRT and its variants require that the exclusion ratio, $R$ in Eq. (\ref{EQ:RRT}), be set in advance:
Following previous studies \cite{Liu2011,Ackah-Arthur2019}, $R$ was set to 0.75.

LSH-ART also has parameters that need to be set before testing, including the quantization width $w$ and the hash bucket capacity $m$.
The quantization width $w$ should be neither very large nor very small
---
if $w$ is very large, all the data points could be assigned to the same hash bucket;
however, if it is very small, a large number of hash buckets could be created in the hash table, potentially resulting in each bucket having only one data point.
In this study, we set the collision probability as $p=0.9$, and then made use of the inverse of Eq. (\ref{EQ:PRO}) to get the value of $\alpha$.
If $dist(\boldsymbol{q},\boldsymbol{v})$ is given, then the lower bound of $w$ can be calculated.
For the lower bound of $w$ to be as large as possible, $dist(\boldsymbol{q},\boldsymbol{v})$ should also be as large as possible,
which means that $\boldsymbol{q}$ and $\boldsymbol{v}$ are two data points that are as far apart as possible in the input domain $\mathcal{D}$.
In addition, when an ordinary hash bucket is promoted into a super hash bucket, all elements need to be re-hashed with a different $w$.
To make the next level of the hash function more powerful, $w$ should be decreased:
$w_i > w_j$ for $i < j$.
Following previous studies \cite{Hu2015}, we set $w_{i+1} = w_i/2$ for any $i \geq 1$

If the capacity of the ordinary hash bucket, $m$, is very large, then LSH-ART degrades to the original ART;
however, if it is very small, then data points with a high collision probability may be assigned to different hash buckets.
Because of the lack of guidelines for how to choose an appropriate value of $m$ \cite{Hu2015}, we conducted a simulation study to examine its impact on LSH-ART performance, varying $m$ from 10 to 100 in increments of 10.
We observed that larger values of $m$ resulted in fewer test cases being required to reveal the first failure (\textit{the F-measure}), but with little difference for $m \geq 60$.
We therefore set $m=60$ in the experiments.

The forgetting strategy \cite{Chan2006a} also has few guidelines regarding the choice of value for $\lambda$.
However, because $\lambda$ restricts the number of executed test cases used in distance calculations, it is similar to the LSH-ART parameter $m$.
To ensure fair comparisons, therefore, we also set $\lambda=60$.
Two parameters also need to be set for the distance-aware forgetting strategy \cite{Mao2017}:
$p_0$ is the initial partition number in each dimension; and
$\tau$ is the pre-set threshold for the average number of test cases in a cell (a sub-domain after partitioning).
As suggested by Mao et al.~\cite{Mao2017}, we set $p_0$ to 3, and $\tau$ to 5.
Because the DF strategy may not always provide an exact nearest neighbor, due to the parameter configurations, it can be considered an ANN approach.
We choose the most efficient KD-tree strategy version \cite{Mao2019}, which, through parameter design and configuration, can also implement an ANN search.

\subsection{Dependent Variables}

According to the research questions, there are three dependent variables, relating to
testing effectiveness;
efficiency; and
cost-effectiveness.

\subsubsection{Testing Effectiveness}
The dependent variable for \textbf{RQ1} (for both \textbf{RQ1.1} and \textbf{RQ1.2}) is the metric to evaluate the fault-detection effectiveness.

(1) For \textbf{RQ1.1}, we adopted the \textit{F-measure} \cite{Chen2004}, which is defined as the expected number of test case executions required before detecting the first software failure.
When a software failure is identified, as explained by Chen et al. \cite{Chen2004}, it is common to stop testing and commence the debugging process.
Therefore, the F-measure is realistic from a practical testing perspective.
In addition, most ART approaches generally generate test cases one at a time \cite{Huang2019}, which means that the F-measure is an appropriate measure for evaluating ART.
Lower F-measure values indicate better testing performance.

Given a failure rate $\theta$, the F-measure of RT (according to uniform distribution), is theoretically equal to $\theta^{-1}$ \cite{Chen2004}.
In this study, we also use the \textit{F-ratio} to denote the ratio of the F-measure for a method $\mathcal{M}$ to RT's theoretical F-measure, thus showing the F-measure improvement of $\mathcal{M}$ over RT:
An F-ratio of less than 1.0 indicates that $\mathcal{M}$ requires fewer test cases than RT to detect the first failure (i.e., $\mathcal{M}$ has better testing effectiveness);
in contrast, an F-ratio greater than 1.0 indicates that RT performs better than $\mathcal{M}$.

(2) For \textbf{RQ1.2}, we adopted the \textit{P-measure} \cite{Duran1984}, which is defined as the probability of a given test set detecting at least one failure.
Although the P-measure may appear less practical than the F-measure, it has been widely used in many testing scenarios, especially in automated software testing \cite{Shahbazi2012}.
Another evaluation metric is the \textit{E-measure} \cite{Duran1984}, which is defined as the expected number of failures to be detected by a given test set.
Similar to the P-measure, the E-measure has also been applied to many automated software testing scenarios \cite{Shahbazi2012}.
However, because failure regions may tend to cluster \cite{White1980,Ammann1988,Finelli1991,Bishop1993,Schneckenburger2007}, higher E-measure values do not imply more faults or more distinct failures \cite{Shahbazi2012}.
Therefore, the P-measure may be a more appropriate measure than the E-measure.
Higher P-measure values indicate more effective software-failure detection.
If the total number of test sets is $N_t$ and the number of test sets that identify at least one failure is $N_f$, then the P-measure can be estimated as $N_f/N_t$ \cite{Shahbazi2012}.
For RT, the expected P-measure is equal to $1-(1-\theta)^{n}$ \cite{Chan1996}, where $n$ is the size of the test set.

\subsubsection{Testing Efficiency}

The \textbf{RQ2} dependent variable relates to the test-case generation time and includes the time taken for both {\em generation} and {\em execution} of a certain number of test cases.

\subsubsection{Testing Cost-effectiveness}

The \textbf{RQ3} dependent variable needs to measure both the testing {\em effectiveness} and the {\em efficiency}:
As discussed in Huang et al.~\cite{Huang2019}, the \textit{F-time}
--- the time required to detect the first failure
--- is a good measure of the ART's cost-effectiveness.

\begin{table*}[!t]
\centering
\scriptsize
\caption{Subject Programs for the F-measure Evaluation Framework}
\label{TAB:program}
\resizebox{\textwidth}{!}{
\begin{tabular}{@{}ccccccccc@{}}
\hline
\multirow{2}*{\textbf{ID}} &\multirow{2}*{\textbf{Program}} &\textbf{Size} &\textbf{Dimension} &\multicolumn{2}{c}{\textbf{Input Domain} ($\mathcal{D}$)}  &\multirow{2}*{\textbf{Fault Type}} &\textbf{Total} &\textbf{Failure Rate} \\ \cline{5-6}
& &(LOC) &($d$) &\textit{From} &\textit{To} & &\textbf{Faults} &($\theta$) \\\hline	
P1	&airy	&43	&1	&$-5000$	&$5000$	&CR	&1	&0.000716 	\\ \hline
P2	&bessj0	&28	&1	&$-300,000$	&$300,000$  	&AOR; ROR; SVR; CR	&5	&0.001373 	\\ \hline
P3	&erfcc	&14	&1	&$-30,000$	&$30,000$	&AOR; ROR; SVR; CR	&4	&0.000574 	\\ \hline
P4	&probks	&22	&1	&$-50,000$	&$50,000$	&AOR; ROR; SVR; CR	&4	&0.000387 	\\ \hline
P5	&tanh	&18	&1	&$-500$	&$500$	&AOR; ROR; SVR; CR	&4	&0.001817 	\\ \hline
P6	&bessj	&99	&2	&$(2; -1000)$	&$(300; 15,000)$	&AOR; ROR; CR	&4	&0.000716 	\\ \hline
P7	&gammq	&106	&2	&$(0; 0)$	&$(1700; 40)$	&ROR; CR	&4	&0.000830 	\\ \hline
P8	&sncndn	&64	&2	&$(-5000; -5000)$	&$(5000; 5000)$	&SVR; CR	&5	&0.001623 	\\ \hline
P9	&golden	&80	&3	&$(-100; -100; -100)$	&$(60; 60; 60)$	&ROR; SVR; CR	&5	&0.000550 	\\ \hline
P10	&plgndr	&36	&3	&$(10; 0; 0)$	&$(500; 11; 1)$	&AOR; ROR; CR	&5	&0.000368 	\\ \hline
P11	&cel 	&49	&4	&$(0.001; 0.001; 0.001; 0.001)$	&$(1; 300; 10,000; 1000)$	&AOR; ROR; CR	&3	&0.000332 	\\ \hline
P12	&el2 	&78	&4	&$(0; 0; 0; 0)$	&$(250; 250; 250; 250)$	&AOR; ROR; SVR; CR	&9	&0.000690 	\\ \hline
P13	&calDay	&37	&5	&$(1; 1; 1; 1; 1800)$  	&$(12; 31; 12; 31; 2200)$	&STD	&1	&0.000632 	\\ \hline
P14	&complex	&68	&6	&$(-20; -20; -20; -20; -20; -20)$	&$(20; 20; 20; 20; 20; 20)$	&SVR	&1	&0.000901 	\\ \hline
P15	&pntLinePos	&23	&6	&$(-25; -25; -25; -25; -25; -25)$	&$(25; 25; 25; 25; 25; 25)$	&CR	&1	&0.000728 	\\ \hline
P16	&triangle	&21	&6	&$(-25; -25; -25; -25; -25; -25)$	&$(25; 25; 25; 25; 25; 25)$	&CR	&1	&0.000713 	\\ \hline
P17	&line	&86	&8	&$(-10; -10; -10; -10; -10; -10; -10; -10)$	&$(10; 10; 10; 10; 10; 10; 10; 10)$	&ROR	&1	&0.000303 	\\ \hline
P18	&pntTrianglePos	&68	&8	&$(-10; -10; -10; -10;  -10; -10; -10; -10)$	&$(10; 10; 10; 10; 10; 10; 10; 10)$	&CR	&1	&0.000141 	\\ \hline
P19	&twoLinePos	&28	&8	&$(-15; -15; -15; -15; -15; -15; -15; -15)$	&$(15; 15; 15; 15; 15; 15; 15; 15)$	&CR	&1	&0.000133 	\\ \hline

\multirow{2}*{P20}	&\multirow{2}*{calGCD}	&\multirow{2}*{26}	&\multirow{2}*{10}	&\multirow{2}*{$(1; 1; 1; 1; 1; 1; 1; 1; 1; 1)$}	& $(1000; 1000; 1000; 1000; 1000;$ 	&\multirow{2}*{AOR}	&\multirow{2}*{1}	&\multirow{2}*{NR*} 	\\
&&&&&$1000; 1000; 1000; 1000; 1000)$&&&\\\hline

\multirow{2}*{P21}	&\multirow{2}*{nearestDistance}	&\multirow{2}*{24}	&\multirow{2}*{10}	&\multirow{2}*{$(1; 1; 1; 1; 1; 1; 1; 1; 1; 1)$} &$(15; 15; 15; 15; 15;$	&\multirow{2}*{CR}	&\multirow{2}*{1}	&\multirow{2}*{0.000256}	\\
&&&&&$15; 15; 15; 15; 15)$&&& \\\hline

\multirow{2}*{P22}	&\multirow{2}*{select}	&\multirow{2}*{117}	&\multirow{2}*{11}	&\multirow{2}*{$(1; 1; 1; 1; 1; 1; 1; 1; 1; 1; 1)$}	&$(100; 100; 100; 100; 100; 100;$ &\multirow{2}*{RSR; CR}	&\multirow{2}*{2}	&\multirow{2}*{NR*}	\\
&&&&&$100; 100; 100; 100; 100)$&&& \\\hline

\multirow{2}*{P23}	&\multirow{2}*{tcas}	&\multirow{2}*{182}	&\multirow{2}*{12}	&\multirow{2}*{$(0; 0; 0; 0; 0; 0; 0; 0; 0; 0; 0; 0)$}	&$(1000; 1; 1; 50,000; 1000; 50,000;$ 	&\multirow{2}*{CR}	&\multirow{2}*{1}	&\multirow{2}*{NR*}	\\
&&&&&$3; 1000; 1000; 2; 2; 1)$&&& \\\hline
\multicolumn{9}{l}{* NR indicates that the failure rate was not reported in the original paper.}
\end{tabular}}
\end{table*}

\subsection{Simulation Framework}

We created a simulation framework to enable the evaluation of our proposed approach.
This framework contains some configurable features, including:
the failure pattern;
failure rate related to each failure pattern; and
the number of test cases in each test set (used for P-measure evaluations).

\subsubsection{Failure Pattern and Failure Rate}

Within the input domain $\mathcal{D}$, a faulty program was simulated by designating either a continuous region, or some disjoint regions.
When a test case was selected from inside such a region, a software failure was considered to be triggered.
The simulation framework, therefore, involves the setting up of the failure pattern (the shape of the failure region(s)) and failure rate $\theta$ (the size of the failure region(s)).
The input domain dimensionality $d$ must also be set in advance.

As discussed in Section \ref{SECTION:Preliminaries}, Chan et al. \cite{Chan1996} identified three common failure pattern types:
block;
strip; and
point (Figure \ref{FIG:failurePattern}).
Following previous ART studies \cite{Mao2019,Mao2017,Mayer2005,Mayer2005a,Tappenden2009,Ashfaq2020}, we also used these patterns in our simulation framework.
Using a unit input domain ($\mathcal{D}$ was $[0,1.0)^d$), the block pattern was simulated as a single hypercube randomly constructed and located within $\mathcal{D}$.
This was achieved by selecting a random point and then extending the same length for each dimension (with respect to $\theta$), producing, for example, a square in two dimensions, or a cube in three dimensions.
The strip pattern was simulated using a single strip at any angle:
As previously noted \cite{Mao2019},
because strips that are close to the corners of the input domain result in  ``fat'' but unrealistic strips, these strips were excluded from the experiments.
For the point pattern, $25$ disjoint equally-sized hypercubes were randomly generated within $\mathcal{D}$.
Representative values for $d$ and $\theta$ were selected based on previous ART studies \cite{Mao2019, Mayer2005}, as follows:
\begin{itemize}
    \item
        Dimension ($d$): $1$, $2$, $3$, $4$, $5$, and $10$.
    \item
    Failure rate ($\theta$):
        $1.0\times 10^{-2}$, $5.0\times 10^{-3}$,
        $2.0\times 10^{-3}$, $1.0\times 10^{-3}$,
        $5.0\times 10^{-4}$, $2.0\times 10^{-4}$, and
        $1.0\times 10^{-4}$.
\end{itemize}

\subsubsection{Number of Test Cases for P-measure Evaluations}

Calculation of the P-measure requires that the number of test cases (denoted $n$) in each test set $T$ be set in advance (i.e., $n=|T|$).
As discussed by Shahbazi et al. \cite{Shahbazi2012}, a good choice of value for $n$ when using the P-measure to analyze the test-case generation approaches is the worst case according to the standard error, $SE$:
i.e., the maximum $SE$.
This can be estimated as:
\begin{equation}
    SE = \frac{SD}{\sqrt{N_t}},
\end{equation}
where $SD$ is the standard deviation.

Because $N_t$ is a constant, the maximum $SD$ leads to the worst case of $SE$.
As reported by Chen et al. \cite{Chen2006a}, the maximum $SD$ of the P-measure calculation is $0.5$, and the expected P-measure $SD$ for RT can be approximated as:
\begin{equation}
    SD \approx\sqrt{(1-\theta)^{n}-(1-\theta)^{2n}}.
\end{equation}
The value for $n$ can therefore be estimated using the following equation:
\begin{equation}
    n = \frac{\log(0.5)}{\log(1-\theta)}.
\end{equation}

Seven values of $\theta$ were selected for the simulations:
$1.0\times 10^{-2}$;
$5.0\times 10^{-3}$;
$2.0\times 10^{-3}$;
$1.0\times 10^{-3}$;
$5.0\times 10^{-4}$;
$2.0\times 10^{-4}$; and
$1.0\times 10^{-4}$.
The corresponding values of $n$, an integer, were calculated as:
$69$;
$138$;
$346$;
$693$;
$1386$;
$3465$; and
$6931$.

\subsection{Empirical Study Framework}

To further evaluate the proposed LSH approach, we conducted an empirical study based on mutation testing \cite{Jia2011}.
This study made use of many subject programs.

Following previous ART studies \cite{Mao2019,Shahbazi2012,Arcuri2011}, we adopted two different evaluation frameworks, using the two different evaluation metrics (F-measure and P-measure).

\begin{table*}[!b]
\centering
\scriptsize
\caption{Subject Programs for the P-measure Evaluation Framework}
\label{TAB:program-P}
\resizebox{\textwidth}{!}{
\begin{tabular}{@{}cccccccrrrrr@{}}
\hline
\multirow{2}*{\textbf{ID}} &\multirow{2}*{\textbf{Program}} &\textbf{Size} &\textbf{Dimension} &\multicolumn{2}{c}{\textbf{Input Domain} ($\mathcal{D}$)}  & &\multicolumn{5}{c}{\textbf{Mutants}} \\ \cline{5-6} \cline{8-12}
& &(\textit{LOC}) &($d$) &\textit{From} &\textit{To} & &\textit{Total Mutants} &\textit{Equivalent Mutants} &\textit{Mutants with Timeout} &\textit{Easy Mutants} &\textit{Appropriate Mutants}\\\hline	
Q1	&Bessj	&131	&2	&$(0; 0)$	&$(4095;4095)$	&	&1007	&111	&26	&551	&319	\\ \hline
Q2	&Expint	&86	&2	&$(0; 0)$	&$(4095;4095)$	&	&432	&46	&16	&188	&182	\\ \hline
Q3	&Gammq	&89	&2	&$(0; 0)$	&$(4095;4095)$	&	&258	&0	&37	&213	&8	\\ \hline
Q4	&Remainder	&48	&2	&$(0; 0)$	&$(4095;4095)$	&	&382	&290	&42	&49	&1	\\ \hline
Q5	&Fisher	&71	&3	&$(0; 0; 0)$	&$(255;255;255)$	&	&615	&0	&16	&599	&0	\\ \hline
Q6	&Triangle	&26	&3	&$(0; 0; 0)$	&$(255;255;255)$	&	&191	&34	&0	&95	&62	\\ \hline
Q7	&Triangle2	&41	&3	&$(0; 0; 0)$	&$(255;255;255)$	&	&333	&47	&0	&118	&168	\\ \hline
Q8	&BubbleSort	&14	&4	&$(0; 0; 0; 0)$	&$(63; 63; 63; 63)$	&	&66	&3	&10	&53	&0	\\ \hline
Q9	&Encoder	&65	&4	&$(0; 0; 0; 0)$	&$(63; 63; 63; 63)$	&	&283	&123	&0	&160	&0	\\ \hline
Q10	&Median	&20	&4	&$(0; 0; 0; 0)$	&$(63; 63; 63; 63)$	&	&89	&6	&11	&72	&0	\\ \hline
Q11	&Variance	&22	&4	&$(0; 0; 0; 0)$	&$(63; 63; 63; 63)$	&	&71	&10	&2	&59	&0	\\ \hline
\end{tabular}}
\end{table*}

\subsubsection{F-measure Evaluation Framework}

The F-measure evaluation framework involved 23 subject programs, implemented in Java, that have been used in previous studies \cite{Mao2019,Chen2004,Mao2017,Liu2011,Ackah-Arthur2019}.
Table \ref{TAB:program} presents detailed information about these 23 programs.
The programs ranged in size from 14 to 182 lines of code (LOC), with their input domain dimensions ranging from one to 12.
As shown in the ``Fault Types'' column of Table \ref{TAB:program}, each program was seeded with one to nine faults using some common mutation operators \cite{Jia2011}:
\textit{constant replacement} (CR);
\textit{arithmetic operator replacement} (AOR);
\textit{relational operator replacement} (ROR);
\textit{scalar variable replacement} (SVR);
\textit{statement deletion} (STD); and
\textit{return statement replacement} (RSR).
The table also shows the input domain and failure rate data for each program.

As noted by Huang et al. \cite{Huang2019}, the first 12 programs have been used in more than 15 previous ART studies, and are considered to be the classic numeric-input benchmark programs for ART.
The other 11 programs, each of which has a high-dimensional numeric input domain, can be considered new benchmark programs.
Although it may not be possible to collect all subject programs from previous studies (due to a lack of available source code or other reasons), our use of these 23 subject programs seems comprehensive and sufficient to support the evaluation process.
We directly adopted the program's source code and mutations, without any modification, in order to ensure fair comparisons.

\subsubsection{P-measure Evaluation Framework}

The P-measure evaluation framework used 11 Java subject programs that have been used in the ART literature \cite{Shahbazi2012,Arcuri2011}.
In our study, we directly used the original program source code, without any modification.
Table \ref{TAB:program-P} lists the details of these programs.

The input domain for each program was constructed by assigning each dimension as an integer range of $[0,2^{24/d}-1]$, leading to $2^{24}$ possible inputs
---
for example, when $d=3$, the input domain is $[0, 255]^3$.
A total of 3727 mutants of the 11 subject programs were created using the well-known mutation testing tool MuJava \cite{Ma2005}.
Exhaustive testing was conducted for each mutant:
Each mutant was executed with all $2^{24}$ different inputs.
After the exhaustive testing, the following mutants were excluded:
(1) \textit{Equivalent Mutants}, whose outputs or behavior was identical to the original program for all inputs;
(2) \textit{Mutants with Timeout}, whose execution did not terminate within the time limit of 10 minutes; and
(3) \textit{Easy Mutants}, whose failure rates were greater than $0.01$.
This left a total of 780 \textit{Appropriate Mutants} that were used in the original study \cite{Arcuri2011}.

We obtained all 3727 mutant programs used in the original study \cite{Arcuri2011}, and reran the mutant exclusion process in our environment.
Due to differences in our experimental environment compared with the original study \cite{Arcuri2011}, 40 of the 780 mutants did not terminate within the time limit, so we only used 740. These 740 Appropriate Mutants were used, without any modification, to calculate the P-measure of our proposed approach, and compare its performance with the other ART approaches.
The failure rate of the subject programs was assumed to be unknown before testing.
We set the test set sizes in the study to range from 2 to 10, in intervals of 2; and
from 10 to 100, in intervals of 5.
Each approach was evaluated according to the P-measure for each test set size.

\subsection{Number of Runs and Statistical Analysis}

Because ART involves random test case construction (which obviously involves some randomness), it was necessary to repeat the experiments a number of times to obtain a large enough sample size to ensure reliable statistical estimates.
There were different requirements for the F-measure and P-measure, which are discussed in the following.

\subsubsection{F-measure Settings}

Arcuri and Briand \cite{Arcuri2014} have recommended that experiments involving randomization should be run at least 1000 times for each subject program:
In our study, therefore, there were 3000 runs for each approach under each testing scenario (or experiment), resulting in 3000 sets of data points (F-measure or F-time data) for further statistical analysis.

The two-tailed nonparametric \textit{Mann-Whitney U test} (U-test) \cite{Wilcoxon1945} has been recommended
for detecting statistical differences in interval-scale results \cite{Arcuri2014}.
Both the F-measure and F-time generally have interval-scale results.
When comparing two ART approaches, therefore, we used the U-test to collect $p$-values to identify significant differences in F-measure or F-time results (at a significance level of 1\%).
Similarly, standardized \textit{effect size} measures were also used:
Vargha and Delaney's $\hat{\textrm{A}}_{12}$ statistic \cite{Vargha2000} compares two approaches,
$\mathcal{M}_1$ and $\mathcal{M}_2$, as:
\begin{equation}
    \hat{\textrm{A}}_{12}(\mathcal{M}_1,\mathcal{M}_2) = (R_1 /X - (X+1)/2)/Y,
\end{equation}
where $R_1$ is the rank sum of the first data group under comparison;
$X$ is the number of observations in the data sample of $\mathcal{M}_1$; and
$Y$ is the number of observations in the data sample of $\mathcal{M}_2$.
Vargha and Delaney's $\hat{\textrm{A}}_{12}$ statistic shows the magnitude of improvement of one ART approach over another
---
$\hat{\textrm{A}}_{12}(\mathcal{M}_1,\mathcal{M}_2)$ indicates the probability that algorithm $\mathcal{M}_1$ outperforms algorithm $\mathcal{M}_2$.
For example,
$\hat{\textrm{A}}_{12}(\mathcal{M}_1,\mathcal{M}_2)=0.68$ means that $\mathcal{M}_1$ outperforms $\mathcal{M}_2$ with a probability of 68\%.
In general, $\hat{\textrm{A}}_{12}(\mathcal{M}_1,\mathcal{M}_2)=0.5$ means $\mathcal{M}_1$ and $\mathcal{M}_2$ have equal performance;
$\hat{\textrm{A}}_{12}(\mathcal{M}_1,\mathcal{M}_2)>0.5$, means that $\mathcal{M}_1$ is better than $\mathcal{M}_2$; and
$\hat{\textrm{A}}_{12}(\mathcal{M}_1,\mathcal{M}_2)<0.5$, means that $\mathcal{M}_2$ is better than $\mathcal{M}_1$.

\subsubsection{P-measure Settings}

A number of test sets, $N_t$, is needed to calculate the P-measure.
In our P-measure simulations, following previous studies \cite{Shahbazi2012},
we generated $N_t=100$ distinct test sets for RT, ART, RF, CR, DF, KD, and LSH.
Because each failure pattern was randomly constructed within the input domain, we ran the experiments 3000 times (resulting in 3000 different failure patterns).
In the empirical P-measure study, the number of mutants was fixed (740), which means that there were only 740 failure patterns. 
Therefore, more distinct test sets were needed ($N_t=10,000$) for the empirical study than for the simulations, to collect more experimental data for the statistical analysis.

When the result of an experiment is dichotomous
---
i.e., either one finds a solution to the problem (\emph{success}) or one does not (\emph{fail})
---
the \emph{Fisher exact test} \cite{Klein2003} has been recommended \cite{Arcuri2014} to measure whether or not two approaches are significantly different (at a significance level of 1\%).
When calculating the P-measure, each fixed-size test set is executed on the program, yielding a dichotomous result (whether the software failure is triggered or not).
The Fisher exact test was therefore appropriate for measuring the significant differences in the P-measure experiments.
When calculating the standardized effect size measures, the \textit{odds ratio} $\psi$ \cite{Grissom2005} has been recommended for dichotomous results between the two algorithms \cite{Arcuri2014}.
Its definition is as follows:
\begin{equation}
    \psi(\mathcal{M}_1,\mathcal{M}_2) = \frac{a_1+\rho}{n+\rho-a_1}\Big/\frac{a_2+\rho}{n+\rho-a_2},
\end{equation}
where
$a_1$ and $a_2$ are the number of success times for the algorithms $\mathcal{M}_1$ and $\mathcal{M}_2$, respectively;
$n$ is the number of observations; and
$\rho$ is any arbitrary positive constant (usually $\rho=0.5$) used to avoid problems with zero occurrences.
In general,
$\psi(\mathcal{M}_1,\mathcal{M}_2)=1.0$ indicates that there is no difference between $\mathcal{M}_1$ and $\mathcal{M}_2$;
$\psi(\mathcal{M}_1,\mathcal{M}_2)> 1.0$ means that algorithm $\mathcal{M}_1$ has higher chances of success (of identifying a failure) than algorithm $\mathcal{M}_2$; and
$\psi(\mathcal{M}_1,\mathcal{M}_2) < 1.0$ means that $\mathcal{M}_2$ outperforms $\mathcal{M}_1$.

\section{Experimental Results and Discussions}
\label{SECTION:results}

In this section, we provide the experimental results and statistical analyses to answer the three research questions.
In the analyses, when comparing two methods $\mathcal{M}_1$ and $\mathcal{M}_2$, we used
the \ding{109} symbol to indicate that there was no statistical difference between them (their $p$-value was greater than 0.01);
the \ding{52} symbol to indicate that $\mathcal{M}_1$ was significantly better ($p$-value was less than 0.01, and the effect size was greater than 0.50 for the F-measure or 1.0 for the P-measure); and
the \ding{54} symbol to indicate that $\mathcal{M}_2$ was significantly better ($p$-value was less than 0.01, and the effect size was less than 0.50 for the F-measure or 1.0 for the P-measure).
Each effect size value
---
$\hat{\textrm{A}}_{12}(\mathcal{M}_1,\mathcal{M}_2)$ for the F-measure and $\psi(\mathcal{M}_1,\mathcal{M}_2)$ for the P-measure
---
is listed in the parenthesis immediately following the comparison symbol.
All the results are included in the Appendix (see supplementary material); and we have also made them available online \cite{detailLSH}.

\subsection{Answer to RQ1.1: Effectiveness: F-measure Simulations}
\label{SECTION:Fsimulation}

Tables \ref{TAB:FSCS-block} to \ref{TAB:FSCS-point} present the FSCS F-measure simulation results for block, strip, and point patterns, respectively.
Tables \ref{TAB:RRT-block} to \ref{TAB:RRT-point} show the corresponding RRT F-measure simulation results.
Each table presents the mean F-ratio results and statistical pairwise comparisons of LSH against the other methods.

\subsubsection{General LSH F-measure Observations}

Based on all the F-measure simulation results (Tables \ref{TAB:FSCS-block} to \ref{TAB:RRT-point}), we have the following general observations for LSH:
\begin{itemize}
    \item[(a)]
        Similar to the original ART algorithms (both FSCS and RRT),
        the LSH F-measure depends on many factors,
        including the failure pattern, dimensionality $d$, and the failure rate $\theta$.
    \item[(b)]
        For a fixed failure rate $\theta$, the LSH F-ratio increases as $d$ increases,
        irrespective of the failure pattern types.
        This indicates that LSH has poorer fault-detection performance in higher dimensions.
    \item[(c)]
        For a fixed dimensionality $d$, as $\theta$ decreases, the LSH F-ratio decreases for block and point patterns,
        but remains very similar for the strip pattern.
\end{itemize}

\subsubsection{F-measure Simulation Observations: Block Pattern}

Based on the block pattern F-measure simulation results (Tables \ref{TAB:FSCS-block} and \ref{TAB:RRT-block}), we have the following observations:
\begin{itemize}
    \item[\emph{\textbf{(a)}}] \textbf{\textit{LSH vs. RT:}}
    The FSCS version of LSH has much better performance than RT when $d$ is low ($d \leq 3$).
    As $d$ increases, however, LSH becomes inferior to RT, regardless of failure rates.
    The RRT version of LSH also significantly outperforms RT when $d \leq 3$, but becomes similar to RT when $d \geq 4$, even in the case of $d=10$.
    The statistical pairwise comparisons between LSH and RT generally support these observations.

    \item[\emph{\textbf{(b)}}] \textbf{\textit{LSH vs. ART:}}
    The FSCS version of LSH is similar to, or slightly worse than, ART for $d \leq 5$, but significantly better for $d=10$. 
    The RRT version of LSH has very similar performance to ART when $d \leq 5$, and, when $d=10$, is similar to, or slightly better than, ART.
    The $p$-values and effect size values indicate that most comparisons between LSH and ART have no significant differences.

    \item[\emph{\textbf{(c)}}] \textbf{\textit{LSH vs. RF:}}
    The F-ratio of the FSCS version of LSH is much less than that of RF, in all cases, irrespective of failure rates and dimensions.
    The RRT version of LSH, when $d=1$, has very similar performance to RF, and performs significantly better than RF when $d \geq 2$.
    Compared with the FSCS version, however, the F-ratio difference is relatively small.

    \item[\emph{\textbf{(d)}}] \textbf{\textit{LSH vs. CR:}}
    The comparisons between LSH (both FSCS and RRT) and CR are very similar to LSH vs. RF:
    LSH shows significantly better fault-detection capability than CR for FSCS, and similar, or better, for  RRT.

    \item[\emph{\textbf{(e)}}] \textbf{\textit{LSH vs. DF:}}
    The FSCS version of LSH has a similar or slightly worse performance than DF when $d \leq 5$.
    However, for $d=10$, apart from one case with $\theta = 1.0 \times 10^{-2}$, LSH has better performance than DF.
    For the RRT version, when $d=1$ or $d = 10$, the mean F-ratio differences between LSH and DF are large,  with LSH generally performing better.
    For other dimensions, however, the mean F-ratio differences are relatively small, which means that LSH and DF have very similar performances.

    \item[\emph{\textbf{(f)}}] \textbf{\textit{LSH vs. KD:}}
    The mean F-ratio values for the FSCS version of LSH are similar to or slightly greater than those of KD, in most cases.
    However, the differences are relatively small.
    Similarly, the F-ratio differences between LSH-RRT and KD are very small, when $d \leq 5$.
    When $d=10$, LSH is similar to, or slightly better than, KD.
    Overall, the statistical comparisons show no significant differences between LSH and KD in most cases.
\end{itemize}

\begin{table*}[!b]
\centering
\footnotesize
\caption{\textbf{F-measure Simulation Scenarios} where LSH is Significantly Superior (\ding{52}), Indistinguishable (\ding{109}), or Significantly Inferior (\ding{54}), with respect to \textbf{Failure Pattern}}
\label{TAB:SimulationCollection1}
\resizebox{\textwidth}{!}{
\begin{tabular}{@{}cccccccccccccccccccccccccc@{}}
\hline
\multirow{2}*{\textbf{ART Version}} &\multirow{2}*{\textbf{Failure Pattern}} & &\multicolumn{3}{c}{\textit{LSH vs. RT}} & &\multicolumn{3}{c}{\textit{LSH vs. ART}} & &\multicolumn{3}{c}{\textit{LSH vs. RF}} & &\multicolumn{3}{c}{\textit{LSH vs. CR}} & &\multicolumn{3}{c}{\textit{LSH vs. DF}} & &\multicolumn{3}{c}{\textit{LSH vs. KD}}\\\cline{4-6}\cline{8-10}\cline{12-14} \cline{16-18}\cline{20-22} \cline{24-26}
&&&\ding{52} &\ding{109} &\ding{54} & &\ding{52} &\ding{109} &\ding{54} & &\ding{52} &\ding{109} &\ding{54} & &\ding{52} &\ding{109} &\ding{54} & &\ding{52} &\ding{109} &\ding{54} &&\ding{52} &\ding{109} &\ding{54}
\\\hline
\multirow{4}*{FSCS Version}	&\textit{Block Pattern}	&	&19	&7	&16	&	&6	&28	&8	&	&42	&0	&0	&	&41	&1	&0	&	&3	&29	&10	&	&0	&34	&8	\\
	&\textit{Strip Pattern}	&	&8	&34	&0	&	&1	&40	&1	&	&8	&34	&0	&	&6	&36	&0	&	&0	&42	&0	&	&0	&42	&0	\\
	&\textit{Point Pattern}	&	&0	&21	&21	&	&5	&37	&0	&	&21	&21	&0	&	&19	&23	&0	&	&4	&36	&2	&	&5	&37	&0	\\\cline{2-26}
	&\textit{\textbf{Sum}} 	&	&27	&62	&37	&	&12	&105	&9	&	&71	&55	&0	&	&66	&60	&0	&	&7	&107	&12	&	&5	&113	&8	\\\hline
\multirow{4}*{RRT Version}	&\textit{Block Pattern}	&	&22	&17	&3	&	&5	&36	&1	&	&30	&12	&0	&	&30	&12	&0	&	&11	&30	&1	&	&4	&38	&0	\\
	&\textit{Strip Pattern}	&	&7	&35	&0	&	&0	&42	&0	&	&0	&42	&0	&	&0	&39	&3	&	&7	&35	&0	&	&0	&42	&0	\\
	&\textit{Point Pattern}	&	&0	&36	&6	&	&4	&37	&1	&	&6	&36	&0	&	&8	&34	&0	&	&6	&36	&10	&	&4	&37	&1	\\\cline{2-26}
	&\textit{\textbf{Sum}} 	&	&29	&88	&9	&	&9	&115	&2	&	&36	&90	&0	&	&38	&85	&3	&	&24	&101	&1	&	&8	&117	&1	\\\hline
\multirow{4}*{FSCS + RRT}	&\textit{Block Pattern}	&	&41	&24	&19	&	&11	&64	&9	&	&72	&12	&0	&	&71	&13	&0	&	&14	&59	&11	&	&4	&72	&8	\\
	&\textit{Strip Pattern}	&	&15	&69	&0	&	&1	&82	&1	&	&8	&76	&0	&	&6	&75	&3	&	&7	&77	&0	&	&0	&84	&0	\\
	&\textit{Point Pattern}	&	&0	&57	&27	&	&9	&74	&1	&	&27	&57	&0	&	&27	&57	&0	&	&10	&72	&12	&	&9	&74	&1	\\\cline{2-26}
	&\textit{\textbf{Sum}} 	&	&56	&150	&46	&	&21	&220	&11	&	&107	&145	&0	&	&104	&145	&3	&	&31	&208	&13	&	&13	&230	&9	\\\hline
\end{tabular}}
\end{table*}

\begin{table*}[!t]
\centering
\footnotesize
\caption{\textbf{F-measure Simulation Scenarios} where LSH is Significantly Superior (\ding{52}), Indistinguishable (\ding{109}), or Significantly Inferior (\ding{54}), with respect to \textbf{Dimension}}
\label{TAB:SimulationCollection2}
\resizebox{\textwidth}{!}{
\begin{tabular}{@{}ccccccccccccccccccccccccccc@{}}
\hline
\multirow{2}*{\textbf{ART Version}} &&\multirow{2}*{\textbf{Dimension}} & &\multicolumn{3}{c}{\textit{LSH vs. RT}} & &\multicolumn{3}{c}{\textit{LSH vs. ART}} & &\multicolumn{3}{c}{\textit{LSH vs. RF}} & &\multicolumn{3}{c}{\textit{LSH vs. CR}} & &\multicolumn{3}{c}{\textit{LSH vs. DF}} & &\multicolumn{3}{c}{\textit{LSH vs. KD}}\\\cline{5-7}\cline{9-11}\cline{13-15} \cline{17-19}\cline{21-23} \cline{25-27}
&&&&\ding{52} &\ding{109} &\ding{54} & &\ding{52} &\ding{109} &\ding{54} & &\ding{52} &\ding{109} &\ding{54} & &\ding{52} &\ding{109} &\ding{54} & &\ding{52} &\ding{109} &\ding{54} &&\ding{52} &\ding{109} &\ding{54}
\\\hline
\multirow{7}*{FSCS Version}	&&\textit{$d=1$}	&	&14	&7	&0	&	&0	&21	&0	&	&14	&7	&0	&	&12	&9	&0	&	&0	&21	&0	&	&0	&21	&0	\\
	&&\textit{$d=2$}	&	&7	&14	&0	&	&1	&20	&0	&	&8	&13	&0	&	&7	&14	&0	&	&0	&21	&0	&	&0	&21	&0	\\
	&&\textit{$d=3$}	&	&5	&14	&2	&	&0	&17	&4	&	&8	&13	&0	&	&8	&13	&0	&	&0	&16	&5	&	&0	&17	&4	\\
	&&\textit{$d=4$}	&	&0	&17	&4	&	&0	&18	&3	&	&13	&8	&0	&	&12	&9	&0	&	&0	&18	&3	&	&0	&19	&2	\\
	&&\textit{$d=5$}	&	&0	&7	&14	&	&0	&20	&1	&	&14	&7	&0	&	&13	&8	&0	&	&0	&21	&0	&	&0	&19	&2	\\
	&&\textit{$d=10$}	&	&1	&6	&14	&	&11	&9	&1	&	&14	&7	&0	&	&14	&7	&0	&	&7	&11	&3	&	&5	&16	&0	\\\cline{3-27}
	&&\textit{\textbf{Sum}} 	&	&27	&62	&37	&	&12	&105	&9	&	&71	&55	&0	&	&66	&60	&0	&	&7	&107	&12	&	&5	&113	&8	\\\hline
\multirow{7}*{RRT Version}	&&\textit{$d=1$}	&	&14	&7	&0	&	&0	&21	&0	&	&0	&21	&0	&	&0	&21	&0	&	&14	&7	&0	&	&0	&21	&0	\\
	&&\textit{$d=2$}	&	&7	&14	&0	&	&0	&20	&1	&	&7	&14	&0	&	&7	&14	&0	&	&0	&21	&0	&	&0	&21	&0	\\
	&&\textit{$d=3$}	&	&7	&14	&0	&	&0	&20	&1	&	&6	&15	&0	&	&6	&14	&1	&	&0	&20	&1	&	&0	&20	&1	\\
	&&\textit{$d=4$}	&	&1	&18	&2	&	&0	&21	&0	&	&6	&15	&0	&	&8	&13	&0	&	&0	&21	&0	&	&0	&21	&0	\\
	&&\textit{$d=5$}	&	&0	&20	&1	&	&0	&21	&0	&	&10	&11	&0	&	&10	&10	&1	&	&2	&19	&0	&	&1	&20	&0	\\
	&&\textit{$d=10$}	&	&0	&15	&6	&	&9	&12	&0	&	&7	&14	&0	&	&7	&13	&1	&	&8	&13	&0	&	&7	&14	&0	\\\cline{3-27}
	&&\textit{\textbf{Sum}} 	&	&29	&88	&9	&	&9	&115	&2	&	&36	&90	&0	&	&38	&85	&3	&	&24	&101	&1	&	&8	&117	&1	\\\hline
\multirow{7}*{FSCS + RRT}	&&\textit{$d=1$}	&	&28	&14	&0	&	&0	&42	&0	&	&14	&28	&0	&	&12	&30	&0	&	&14	&28	&0	&	&0	&42	&0	\\
	&&\textit{$d=2$}	&	&14	&28	&0	&	&1	&40	&1	&	&15	&27	&0	&	&14	&28	&0	&	&0	&42	&0	&	&0	&42	&0	\\
	&&\textit{$d=3$}	&	&12	&28	&2	&	&0	&37	&5	&	&14	&28	&0	&	&14	&27	&1	&	&0	&36	&6	&	&0	&37	&5	\\
	&&\textit{$d=4$}	&	&1	&35	&6	&	&0	&39	&3	&	&19	&23	&0	&	&20	&22	&0	&	&0	&39	&3	&	&0	&40	&2	\\
	&&\textit{$d=5$}	&	&0	&27	&15	&	&0	&41	&1	&	&24	&18	&0	&	&23	&18	&1	&	&2	&40	&0	&	&1	&39	&2	\\
	&&\textit{$d=10$}	&	&1	&21	&20	&	&20	&21	&1	&	&21	&21	&0	&	&21	&20	&1	&	&15	&24	&3	&	&12	&30	&0	\\\cline{3-27}
	&&\textit{\textbf{Sum}} 	&	&56	&150	&46	&	&21	&220	&11	&	&107	&145	&0	&	&104	&145	&3	&	&31	&208	&13	&	&13	&230	&9	\\\hline

\end{tabular}}
\end{table*}

\subsubsection{F-measure Simulation Observations: Strip Pattern}
Based on the strip pattern F-measure simulation results (Tables \ref{TAB:FSCS-strip} and \ref{TAB:RRT-strip}), we have the following observations:
\begin{itemize}
    \item[\textbf{\textit{(a)}}]\textbf{\textit{LSH vs. RT:}}
    When $d=1$, LSH (both FSCS and RRT) has much better performance than RT for the strip pattern, regardless of failure rates
    ---
    the strip pattern is equivalent to the block pattern in one-dimensional space.
    When $d \geq 2$, LSH has similar, or slightly better, F-measure performance, for all dimensions and failure rates.

    \item[\textbf{\textit{(b)}}]\textbf{\textit{LSH vs. ART:}}
    The LSH and original ART have very similar F-ratio performance, regardless of failure rates and dimensions.
    All $p$-values (other than two LSH-FSCS cases) indicate a lack of significance; and
    all $\hat{\textrm{A}}_{12}$ are around 0.50, indicating very little difference between LSH and ART performance.

    \item[\textbf{\textit{(c)}}]\textbf{\textit{LSH vs. RF:}}
    The FSCS version of LSH is much better than RF when $d=1$.
    For all other cases (FSCS when $d>1$, all RRT), however, the LSH and RF performances are very similar.

    \item[\textbf{\textit{(d)}}]\textbf{\textit{LSH vs. CR:}}
    The comparison of LSH and CR is very similar to LSH vs. RF.
    When $d=1$, LSH-FSCS is significantly better than CR, but LSH and CR have very similar performance in all other cases.

    \item[\textbf{\textit{(e)}}]\textbf{\textit{LSH vs. DF:}}
    The FSCS version of LSH has very similar F-ratio results to DF, for all failure rates and dimensions.
    The situation is similar for the RRT version, except for when $d=1$ (for which LSH has much better performance than DF).

    \item[\textbf{\textit{(f)}}]\textbf{\textit{LSH vs. KD:}}
    The performance of LSH (both FSCS and RRT) and KD are very similar, regardless of failure rates and dimensions.
    The $p$-values for all the related statistical pairwise comparisons are greater than 0.01, which indicates no significant differences.
\end{itemize}

\subsubsection{F-measure Simulation Observations: Point Pattern}

Based on the point pattern F-measure simulation results (Tables \ref{TAB:FSCS-point} and \ref{TAB:RRT-point}), we have the following observations:
\begin{itemize}
    \item[\textbf{\textit{(a)}}]\textbf{\textit{LSH vs. RT:}}
    The FSCS version of LSH has different performances to the RRT version for the point pattern simulations:
    When $d \leq 3$, LSH-FSCS is similar to, or slightly better than, RT;
    but RT significantly outperforms LSH-FSCS overall when $d \geq 4$.
    LSH-RRT, however, has similar F-measure results to RT, regardless of dimension and failure rates.

    \item[\textbf{\textit{(b)}}]\textbf{\textit{LSH vs. ART:}}
    When $d \leq 5$, LSH is very similar to the original ART (both FSCS and RRT), regardless of failure rates.
    When $d=10$, however, LSH performs slightly better than ART.
    The $p$-values and effect size values, when $d \leq 5$, nearly all indicate no significant difference.
    When $d=10$, however, some $p$-values do indicate significant differences, for both FSCS and RRT, especially when the failure rate is low.

    \item[\textbf{\textit{(c)}}]\textbf{\textit{LSH vs. RF:}}
    LSH-FSCS has similar, or slightly better, F-ratio results than RF when $d \leq 3$;
    and performs significantly better than RF when $d \geq 4$.
    LSH-RRT is different from LSH-FSCS, overall achieving a similar or slightly better performance than RF,
    which is supported by the statistical comparisons.

    \item[\textbf{\textit{(d)}}]\textbf{\textit{LSH vs. CR:}}
    The comparisons between LSH and CR are very similar to those between LSH and RF:
    LSH-FSCS has a similar performance to CR when $d \leq 3$, but is much better than CR when $d \geq 4$, regardless of the failure rates.
    LSH-RRT, however, performs similarly to, or slightly better than, CR for all values of $d$.

    \item[\textbf{\textit{(e)}}]\textbf{\textit{LSH vs. DF:}}
    LSH-FSCS performs similarly to DF when $d \leq 5$;
    significantly better than DF for low failure rates when $d=10$; but
    slightly worse than DF when $d=10$ for high failure rates
    (such as $\theta = 1.0 \times 10^{-2}$ and $\theta = 5.0 \times 10^{-3}$).
    LSH-RRT overall achieves similar, or slightly better, F-ratio results than DF, regardless of failure rates and dimensions.
    For high dimensions ($d=10$), LSH has significantly better performance than DF, especially when the failure rate is low.

    \item[\textbf{\textit{(f)}}]\textbf{\textit{LSH vs. KD:}}
    When $d \leq 5$, LSH has very similar performance to KD, for both FSCS and RRT.
    When $d = 10$, however, LSH performs significantly better, especially for lower failure rates.
\end{itemize}

\subsubsection{Analysis and Summary}

Table \ref{TAB:SimulationCollection1} presents the numbers of simulation scenarios, for each failure pattern, where LSH is significantly superior (\ding{52}), indistinguishable (\ding{109}), or significantly inferior (\ding{54}) to each compared technique, in terms of the F-measure.
Each failure pattern has 42 scenarios for pairwise comparison ---
6 dimensions $\times$ 7 failure rates.
Table \ref{TAB:SimulationCollection2} presents the F-measure comparisons for each dimension:
Each dimension has 21 scenarios for each pairwise comparison (3 failure patterns $\times$ 7 failure rates).

Previous studies have shown that block and strip patterns are more favorable for ART than for RT \cite{Chen2007a}.
However, as the dimensionality increases, ART may perform worse than RT, due to the \textit{curse of dimensionality} \cite{Huang2019}.
Tables \ref{TAB:SimulationCollection1} and \ref{TAB:SimulationCollection2} show that the low-dimension block pattern is most favorable for LSH, followed by the strip pattern, and then the low-dimension point pattern.

Intuitively speaking, LSH may be expected to have poorer fault-detection effectiveness than the original ART (and its variants that do not discard information during test-case generation):
LSH uses an approximate, rather than precise, NN search for each candidate, losing some distance calculation information.
In this study, the original ART (\textit{ART} in the tables) is the only included technique that does not lose information during test case construction.
According to Tables \ref{TAB:SimulationCollection1} and \ref{TAB:SimulationCollection2}, compared with \textit{ART}, LSH performs similarly overall, sometimes achieving slightly worse performance, which is unsurprising and expected.
Surprisingly, however, LSH sometimes performs better than \textit{ART} in high dimensions.
As discussed above, while other ART techniques may suffer from the \textit{curse of dimensionality} \cite{Bellman1957}, LSH may (to some extent) alleviate this problem.

It was expected that LSH would be comparable to some ART variants that discard information when generating test cases,
such as  \textit{RF}, \textit{CR}, \textit{DF}, and \textit{KD}.
Tables \ref{TAB:SimulationCollection1} and \ref{TAB:SimulationCollection2} show that LSH achieves comparable performances to \textit{RF}, \textit{CR}, \textit{DF}, and \textit{KD}, on the whole.
Compared with \textit{RF} and \textit{CR}, LSH has better F-measure performances, especially for the block pattern.
Furthermore, LSH performs slightly better than \textit{DF} and \textit{KD}, especially for the RRT version, and in high dimensions.

\begin{tcolorbox}[colback=white,breakable,colframe=black,arc=0mm,left={0mm},top={0mm},bottom={0mm},right={0mm},boxrule={0.25mm}]
\textit{\textbf{Summary of Answers to RQ1.1 for Simulations:}}
\begin{itemize}
    \item
        \textit{Similar to other ART techniques, compared with RT, LSH also has favorable conditions according to the F-measure:
        Block and strip patterns with low dimensions.}
    \item
        \textit{Compared with the original ART (both FSCS and RRT versions), LSH has similar, or slightly worse, F-measure performance.
        Nevertheless, surprisingly, LSH is better than ART in high dimensions.}
    \item
        \textit{Compared with the ART variants that may discard some information for test-case generation,
        LSH has comparable F-measure results overall,
        and sometimes slightly better performance, especially in high dimensions.}
\end{itemize}
\end{tcolorbox}

\subsection{Answer to RQ1.1: Effectiveness: F-measure Empirical Studies}
\label{SECTION:FmeasureES}

Tables \ref{TAB:FSCS-real} and \ref{TAB:RRT-real} summarize the F-measure results for the empirical studies using the 23 subject programs.
Because the failure rates for the real-life programs were unknown, it was not possible to calculate the theoretical RT F-measure:
Instead, the F-measure data are reported in the tables.

\subsubsection{F-measure Empirical Study Observations}

\begin{table*}[!t]
\centering
\footnotesize
\caption{\textbf{F-measure Empirical Study Scenarios} for which LSH Performance is Significantly Superior (\ding{52}), Indistinguishable (\ding{109}), or Significantly Inferior (\ding{54})}
\label{TAB:realCollectionFmeasure}
\resizebox{\textwidth}{!}{
\begin{tabular}{@{}ccccccccccccccccccccccccc@{}}
\hline
\multirow{2}*{\textbf{ART Version}} & &\multicolumn{3}{c}{\textit{LSH vs. RT}} & &\multicolumn{3}{c}{\textit{LSH vs. ART}} & &\multicolumn{3}{c}{\textit{LSH vs. RF}} & &\multicolumn{3}{c}{\textit{LSH vs. CR}} & &\multicolumn{3}{c}{\textit{LSH vs. DF}} & &\multicolumn{3}{c}{\textit{LSH vs. KD}}\\\cline{3-5}\cline{7-9}\cline{11-13} \cline{15-17}\cline{19-21} \cline{23-25}
&&\ding{52} &\ding{109} &\ding{54} & &\ding{52} &\ding{109} &\ding{54} & &\ding{52} &\ding{109} &\ding{54} & &\ding{52} &\ding{109} &\ding{54} & &\ding{52} &\ding{109} &\ding{54} &&\ding{52} &\ding{109} &\ding{54}
\\\hline
FSCS Version  & &13 &6 &4 & &0 &21 &2 & &8 &12 &3 & &9 &10 &4 & &5 &16 &2 & &0 &22 &1\\\hline
RRT Version   & &9 &14 &0 & &0 &21 &2 & &9 &14 &0 & &8 &15 &0 & &8 &14 &1 & &0 &23 &0\\\hline
\textbf{\textit{Sum}} & &22 &20 &4 & &0 &42 &4 & &17 &26 &3 & &17 &25 &4 & &13 &30 &3 & &0 &45 &1\\\hline
\end{tabular}}
\end{table*}

Based on the results reported in Tables \ref{TAB:FSCS-real} and \ref{TAB:RRT-real}, we have the following observations:
\begin{itemize}
    \item[\textbf{\textit{(a)}}]\textbf{\textit{LSH vs. RT:}}
    The LSH-FSCS F-measure performance is very similar to RT for five programs (P8, P9, P14, P16, P18, and P20);
    better (lower) for 13 programs; and
    worse (higher) for three programs (P15, P17, and P19).
    The statistical analysis fully supports these observations.
    In addition, LSH-RRT performs better than RT for the first seven programs, and for P10 and P12 (the dimensionality, $d$, for all of which is less than 4).
    For the other 13 programs, LSH-RRT and RT have very similar F-measures.
    Overall, LSH-RRT performs similarly to, or better than, RT for all programs.
    In summary, overall, LSH has similar or better performance than RT for most programs, especially when the dimensionality is low.

    \item[\textbf{\textit{(b)}}]\textbf{\textit{LSH vs. ART:}}
    LSH-FSCS has slightly better performance than ART for two programs (P18 and P19);
    worse performance for three  (P12, P21, and P22); and
    very similar performance for the remaining 17 programs.
    LSH-RRT also outperforms ART for P18 and P19;
    is worse for two programs (P11 and P12); and
    has similar performances for the remaining 18.
    However, the statistical analyses show that, overall,
    the F-measure differences between LSH and ART are not significant, for both FSCS and RRT.

    \item[\textbf{\textit{(c)}}]\textbf{\textit{LSH vs. RF:}}
     Overall, LSH has comparable, or significantly better F-measure performance than RF, for all programs apart from P10 and P21 for LSH-FSCS.

    \item[\textbf{\textit{(d)}}]\textbf{\textit{LSH vs. CR:}}
    The comparison between LSH and CR is very similar to that between LSH and RF:
    LSH is much better than CR for eight or nine subject programs, and similar for the rest

    \item[\textbf{\textit{(e)}}]\textbf{\textit{LSH vs. DF:}}
    For both FSCS and RRT versions, the F-measure differences between LSH and DF are very similar for more than half of the programs.
    LSH-FSCS significantly outperforms DF for four programs (P6, P11, P13, and P19);
    and is significantly outperformed by DF for two (P7 and P10).
    LSH-RRT significantly outperforms DF for eight programs (P1--P6, P10 and P19); and
    is outperformed for one (P12).

    \item[\textbf{\textit{(f)}}]\textbf{\textit{LSH vs. KD:}}
    LSH has very similar performances to KD for all programs except P21 for LSH-FSCS.
\end{itemize}

\subsubsection{Analysis and Summary}

Table \ref{TAB:realCollectionFmeasure} presents the numbers of subject programs in the empirical studies for which LSH is significantly superior (\ding{52}), indistinguishable (\ding{109}), or significantly inferior (\ding{54}) to each compared technique, according to the F-measure.

As expected, LSH has similar or better fault-detection performance compared with RT (in $42/46=91.30\%$ of the cases):
This is because LSH-generated test cases may be more diverse \cite{Chen2010,chen2015revisit}.
LSH is also comparable to ART in most scenarios ($42/46=91.30\%$), in spite of the fact that LSH discards some information when generating test cases, but ART does not.

Similar to RF, CR, and DF, LSH discards some information during test-case generation, resulting in similar fault-detection performance.
However, LSH achieves similar, or significantly better results compared with RF, CR, and DF, for most programs
($43/46=93.48\%$, $42/46=91.30\%$, and $43/46=93.48\%$, respectively).
LSH also has a very similar performance to KD (except in one case, the FSCS version for program P21).

\begin{tcolorbox}[colback=white,breakable,colframe=black,arc=0mm,left={0mm},top={0mm},bottom={0mm},right={0mm},boxrule={0.25mm}]
\textit{\textbf{Summary of Answers to RQ1.1 for Empirical Studies:}}
\begin{itemize}
    \item
        \textit{Compared with RT, overall, LSH has similar or better F-measure performance to RT for most programs,
        especially when the dimensionality is low.}
    \item
        \textit{LSH has very similar F-measure performance to the original ART (both FSCS and RRT) for most programs.}
    \item
        \textit{LSH generally has comparable F-measure performances to the ART variants, for most programs,
        and has better F-measure results than RF, CR, and DF for many programs.}
\end{itemize}
\end{tcolorbox}

\subsection{Answer to RQ1.2: Effectiveness: P-measure Simulations}
\label{SECTION:SIM-P}

Tables \ref{TAB:FSCS-block-p} to \ref{TAB:FSCS-point-p} present the FSCS P-measure simulation results for block, strip, and point patterns, respectively.
Tables \ref{TAB:RRT-block-p} to \ref{TAB:RRT-point-p} show the corresponding RRT P-measure simulation results.
Each table presents the mean P-measure results and statistical pairwise comparisons of LSH against other methods.

\subsubsection{General LSH P-measure Simulation Observations}

Based on all the P-measure simulation results (Tables \ref{TAB:FSCS-block-p} to \ref{TAB:RRT-point-p}), we have the following general simulation observations for LSH:
\begin{itemize}
    \item[(a)]
        Similar to the original ART algorithms (both FSCS and RRT), 
        the LSH P-measure depends on many factors, 
        including the failure pattern, and dimension $d$.
    \item[(b)]
        For a fixed failure rate $\theta$, 
        the LSH P-measure generally decreases as $d$ increases, 
        especially for the block and point patterns.
        In other words, LSH generally has poorer P-measure performance in higher dimensions, similar to the F-measure performance.
\end{itemize}

\begin{table*}[!b]	
\centering
\footnotesize
\caption{\textbf{P-measure Simulation Scenarios} where LSH is Significantly Superior (\ding{52}), Indistinguishable (\ding{109}), or Significantly Inferior (\ding{54}), with respect to \textbf{Failure Pattern}}
\label{TAB:SimulationCollection1-p}
\resizebox{\textwidth}{!}{
\begin{tabular}{@{}cccccccccccccccccccccccccc@{}}
\hline
\multirow{2}*{\textbf{ART Version}} &\multirow{2}*{\textbf{Failure Pattern}} & &\multicolumn{3}{c}{\textit{LSH vs. RT}} & &\multicolumn{3}{c}{\textit{LSH vs. ART}} & &\multicolumn{3}{c}{\textit{LSH vs. RF}} & &\multicolumn{3}{c}{\textit{LSH vs. CR}} & &\multicolumn{3}{c}{\textit{LSH vs. DF}} & &\multicolumn{3}{c}{\textit{LSH vs. KD}}\\\cline{4-6}\cline{8-10}\cline{12-14} \cline{16-18}\cline{20-22} \cline{24-26}
&&&\ding{52} &\ding{109} &\ding{54} & &\ding{52} &\ding{109} &\ding{54} & &\ding{52} &\ding{109} &\ding{54} & &\ding{52} &\ding{109} &\ding{54} & &\ding{52} &\ding{109} &\ding{54} &&\ding{52} &\ding{109} &\ding{54}
\\\hline

\multirow{4}*{FSCS Version}	&\textit{Block Pattern}	&	&21	&1	&20	&	&6	&10	&26	&	&39	&2	&1	&	&39	&3	&0	&	&3	&9	&30	&	&4	&9	&29	\\
	&\textit{Strip Pattern}	&	&21	&21	&0	&	&3	&35	&4	&	&12	&24	&6	&	&10	&27	&5	&	&3	&33	&6	&	&2	&36	&4	\\
	&\textit{Point Pattern}	&	&6	&6	&30	&	&19	&23	&0	&	&34	&7	&1	&	&34	&8	&0	&	&14	&22	&6	&	&10	&31	&1	\\\cline{2-26}
	&\textit{\textbf{Sum}}	&	&48	&28	&50	&	&28	&68	&30	&	&85	&33	&8	&	&83	&38	&5	&	&20	&64	&42	&	&16	&76	&34	\\\hline
\multirow{4}*{RRT Version}	&\textit{Block Pattern}	&	&30	&3	&9	&	&13	&12	&17	&	&37	&2	&3	&	&37	&2	&3	&	&22	&6	&14	&	&7	&14	&21	\\
	&\textit{Strip Pattern}	&	&17	&25	&0	&	&1	&41	&0	&	&10	&28	&4	&	&8	&33	&1	&	&8	&32	&2	&	&1	&38	&3	\\
	&\textit{Point Pattern}	&	&10	&8	&24	&	&19	&22	&1	&	&33	&6	&3	&	&32	&8	&2	&	&27	&13	&2	&	&16	&24	&2	\\\cline{2-26}
	&\textit{\textbf{Sum}}	&	&57	&36	&33	&	&33	&75	&18	&	&80	&36	&10	&	&77	&43	&6	&	&57	&51	&18	&	&24	&76	&26	\\\hline
\multirow{4}*{FSCS + RRT}	&\textit{Block Pattern}	&	&51	&4	&29	&	&19	&22	&43	&	&76	&4	&4	&	&76	&5	&3	&	&25	&15	&44	&	&11	&23	&50	\\
	&\textit{Strip Pattern}	&	&38	&46	&0	&	&4	&76	&4	&	&22	&52	&10	&	&18	&60	&6	&	&11	&65	&8	&	&3	&74	&7	\\
	&\textit{Point Pattern}	&	&16	&14	&54	&	&38	&45	&1	&	&67	&13	&4	&	&66	&16	&2	&	&41	&35	&8	&	&26	&55	&3	\\\cline{2-26}
	&\textit{\textbf{Sum}}	&	&105	&64	&83	&	&61	&143	&48	&	&165	&69	&18	&	&160	&81	&11	&	&77	&115	&60	&	&40	&152	&60	\\\hline

\end{tabular}}
\end{table*}

\begin{table*}
\centering
\footnotesize
\caption{\textbf{P-measure Simulation Scenarios} where LSH is Significantly Superior (\ding{52}), Indistinguishable (\ding{109}), or Significantly Inferior (\ding{54}), with respect to \textbf{Dimension}}
\label{TAB:SimulationCollection2-p}
\resizebox{\textwidth}{!}{
\begin{tabular}{@{}ccccccccccccccccccccccccccc@{}}
\hline
\multirow{2}*{\textbf{ART Version}} &&\multirow{2}*{\textbf{Dimension}} & &\multicolumn{3}{c}{\textit{LSH vs. RT}} & &\multicolumn{3}{c}{\textit{LSH vs. ART}} & &\multicolumn{3}{c}{\textit{LSH vs. RF}} & &\multicolumn{3}{c}{\textit{LSH vs. CR}} & &\multicolumn{3}{c}{\textit{LSH vs. DF}} & &\multicolumn{3}{c}{\textit{LSH vs. KD}}\\\cline{5-7}\cline{9-11}\cline{13-15} \cline{17-19}\cline{21-23} \cline{25-27}
&&&&\ding{52} &\ding{109} &\ding{54} & &\ding{52} &\ding{109} &\ding{54} & &\ding{52} &\ding{109} &\ding{54} & &\ding{52} &\ding{109} &\ding{54} & &\ding{52} &\ding{109} &\ding{54} &&\ding{52} &\ding{109} &\ding{54}
\\\hline

\multirow{7}*{FSCS Version}	&&$d=1$	&	&19	&3	&0	&	&0	&17	&4	&	&18	&3	&0	&	&19	&2	&0	&	&0	&16	&5	&	&0	&16	&5	\\
	&&$d=2$	&	&12	&7	&2	&	&0	&15	&6	&	&15	&6	&0	&	&12	&9	&0	&	&0	&15	&6	&	&1	&15	&5	\\
	&&$d=3$	&	&7	&7	&7	&	&0	&13	&8	&	&13	&8	&0	&	&15	&6	&0	&	&2	&11	&8	&	&1	&12	&8	\\
	&&$d=4$	&	&1	&7	&13	&	&7	&8	&6	&	&12	&6	&3	&	&12	&6	&3	&	&4	&10	&7	&	&1	&12	&8	\\
	&&$d=5$	&	&3	&4	&14	&	&7	&8	&6	&	&12	&4	&5	&	&12	&7	&2	&	&5	&6	&10	&	&3	&11	&7	\\
	&&$d=10$	&	&6	&1	&14	&	&14	&7	&0	&	&15	&6	&0	&	&13	&8	&0	&	&9	&6	&6	&	&10	&10	&1	\\\cline{3-27}
	&&\textit{\textbf{Sum}}	&	&48	&29	&50	&	&28	&68	&30	&	&85	&33	&8	&	&83	&38	&5	&	&20	&64	&42	&	&16	&76	&34	\\\hline
\multirow{7}*{RRT Version}	&&$d=1$	&	&21	&0	&0	&	&0	&21	&0	&	&19	&2	&0	&	&18	&3	&0	&	&19	&2	&0	&	&0	&20	&1	\\
	&&$d=2$	&	&15	&6	&0	&	&0	&15	&6	&	&14	&7	&0	&	&12	&9	&0	&	&8	&12	&1	&	&0	&14	&7	\\
	&&$d=3$	&	&8	&9	&4	&	&0	&15	&6	&	&12	&9	&0	&	&12	&8	&1	&	&2	&13	&6	&	&1	&14	&6	\\
	&&$d=4$	&	&9	&6	&6	&	&8	&8	&5	&	&12	&7	&2	&	&12	&7	&2	&	&6	&8	&7	&	&5	&10	&6	\\
	&&$d=5$	&	&4	&8	&9	&	&11	&9	&1	&	&12	&6	&3	&	&12	&8	&1	&	&10	&10	&1	&	&7	&11	&3	\\
	&&$d=10$	&	&0	&7	&14	&	&14	&7	&0	&	&11	&5	&5	&	&11	&8	&2	&	&12	&6	&3	&	&11	&7	&3	\\\cline{3-27}
	&&\textit{\textbf{Sum}}	&	&57	&36	&33	&	&33	&75	&18	&	&80	&36	&10	&	&77	&43	&6	&	&57	&51	&18	&	&24	&76	&26	\\\hline
\multirow{7}*{FSCS + RRT}	&&$d=1$	&	&40	&3	&0	&	&0	&38	&4	&	&37	&5	&0	&	&37	&5	&0	&	&19	&18	&5	&	&0	&36	&6	\\
	&&$d=2$	&	&27	&13	&2	&	&0	&30	&12	&	&29	&13	&0	&	&24	&18	&0	&	&8	&27	&7	&	&1	&29	&12	\\
	&&$d=3$	&	&15	&16	&11	&	&0	&28	&14	&	&25	&17	&0	&	&27	&14	&1	&	&4	&24	&14	&	&2	&26	&14	\\
	&&$d=4$	&	&10	&13	&19	&	&15	&16	&11	&	&24	&13	&5	&	&24	&13	&5	&	&10	&18	&14	&	&6	&22	&14	\\
	&&$d=5$	&	&7	&12	&23	&	&18	&17	&7	&	&24	&10	&8	&	&24	&15	&3	&	&15	&16	&11	&	&10	&22	&10	\\
	&&$d=10$	&	&6	&8	&28	&	&28	&14	&0	&	&26	&11	&5	&	&24	&16	&2	&	&21	&12	&9	&	&21	&17	&4	\\\cline{3-27}
	&&\textit{\textbf{Sum}}	&	&105	&65	&83	&	&61	&143	&48	&	&165	&69	&18	&	&160	&81	&11	&	&77	&115	&60	&	&40	&152	&60	\\\hline	
\end{tabular}}
\end{table*}

\begin{table*}[!t]
\centering
\footnotesize
\caption{\textbf{P-measure Empirical Study Scenarios} for which LSH performance is Significantly Superior (\ding{52}), Indistinguishable (\ding{109}), or Significantly Inferior (\ding{54})}
\label{TAB:realCollectionPmeasure}
\resizebox{\textwidth}{!}{
\begin{tabular}{@{}cccccccccccccccccccccccccc@{}}
\hline
\multirow{2}*{\textbf{ART Version}} && &\multicolumn{3}{c}{\textit{LSH vs. RT}} & &\multicolumn{3}{c}{\textit{LSH vs. ART}} & &\multicolumn{3}{c}{\textit{LSH vs. RF}} & &\multicolumn{3}{c}{\textit{LSH vs. CR}} & &\multicolumn{3}{c}{\textit{LSH vs. DF}} & &\multicolumn{3}{c}{\textit{LSH vs. KD}}\\\cline{4-6}\cline{8-10}\cline{12-14} \cline{16-18}\cline{20-22} \cline{24-26}
&&&\ding{52} &\ding{109} &\ding{54} & &\ding{52} &\ding{109} &\ding{54} & &\ding{52} &\ding{109} &\ding{54} & &\ding{52} &\ding{109} &\ding{54} & &\ding{52} &\ding{109} &\ding{54} &&\ding{52} &\ding{109} &\ding{54}
\\\hline
FSCS Version  &&&23 &0 &0 & &1 &13 &9 & &3 &11 &9 & &1 &15 &7 & &2 &13 &8 & &4 &11 &8\\\hline
RRT Version   && &20 &2 &1 & &19 &4 &0 & &16 &6 &1 & &19 &4 &0 & &11 &9 &3 & &13 &3 &7\\\hline
\textbf{\textit{Sum}} & &&43 &2 &1 & &20 &17 &9 & &19 &17 &10 & &20 &19 &7 & &13 &22 &11 & &17 &14 &15\\\hline
\end{tabular}}
\end{table*}

\subsubsection{P-measure Simulation Observations: Block Pattern}

Based on the block pattern  P-measure simulation results (Tables \ref{TAB:FSCS-block-p} and \ref{TAB:RRT-block-p}), we have the following observations:
\begin{itemize}
    \item[\textit{\textbf{(a)}}]\textit{\textbf{LSH vs. RT:}}
    When $d$ is small, LSH generally achieves better P-measure results than RT, for both FSCS and RRT.
    As $d$ increases, however, LSH generally performs worse than RT.
    Nevertheless, in the high dimension ($d=10$), the LSH-RRT (but not LSH-FSCS) P-measure values  approach those of RT.

    \item[\textit{\textbf{(b)}}]\textit{\textbf{LSH vs. ART:}}
    When $d=1$, LSH and ART have very similar P-measure results;
    when $d=2,3,4$, LSH (both FSCS and RRT) performs significantly worse than ART, especially for low failure rates.
    However, when $d=5$, the performances of LSH-FSCS and LSH-RRT are different:
    LSH-FSCS has similar or significantly worse P-measure performance than ART;
    but LSH-RRT achieves similar or significantly {\em better} performance.
    When $d=10$, both LSH-FSCS and LSH-RRT overall have significantly better P-measure performance than ART.

    \item[\textit{\textbf{(c)}}]\textit{\textbf{LSH vs. RF:}}
    LSH nearly always performs significantly better than RF (except for a few cases with $\theta=1.0\times 10^{-2}$).

    \item[\textit{\textbf{(d)}}]\textit{\textbf{LSH vs. CR:}}
    The case of CR is similar to that of RF:
    Overall, LSH has significantly better P-measure performance than CR (for both FSCS and RRT), irrespective of the failure rate $\theta$ and dimension $d$.

    \item[\textit{\textbf{(e)}}]\textit{\textbf{LSH vs. DF:}}
    For the FSCS version, LSH overall performs worse than DF in most cases (except for some low failure rates when $d=10$).
    For the RRT version, however, LSH generally has {\em better} P-measure results than DF when $d=1, 2,5,10$;
    while for $d=3,4$, DF is better than LSH, overall.

    \item[\textit{\textbf{(f)}}]\textit{\textbf{LSH vs. KD:}}
    The comparisons of LSH against KD are similar to those against DF:
    When $d=2,3,4,5$, LSH overall has similar or significantly worse P-measure performance than KD;
    but when $d=1,10$, LSH generally achieves similar or significantly better performance.
\end{itemize}

\subsubsection{P-measure Simulation Observations: Strip Pattern}

Based on the strip pattern P-measure simulation results  (Tables \ref{TAB:FSCS-strip-p} and \ref{TAB:RRT-strip-p}), we have the following observations:
\begin{itemize}
    \item[\textit{\textbf{(a)}}]\textit{\textbf{LSH vs. RT:}}
    LSH (both FSCS and RRT) has similar or significantly better P-measure performance than RT, regardless of the dimension and failure rates.

    \item[\textit{\textbf{(b)}}]\textit{\textbf{LSH vs. ART:}}
    The LSH P-measure results are very similar to those for ART (both FSCS and RRT), irrespective of the failure rate and dimension.
    The statistical analysis also shows that there is no significant difference between LSH and ART.

    \item[\textit{\textbf{(c)}}]\textit{\textbf{LSH vs. RF:}}
    When $d=1$, LSH performs significantly better than RF for all failure rates.
    For other dimensions, however, LSH is similar to RF, for both FSCS and RRT.
    This is confirmed by the statistical analysis (because most effect size values are around $0.50$).

    \item[\textit{\textbf{(d)}}]\textit{\textbf{LSH vs. CR:}}
    Similar to the case of RF, LSH achieves significantly better P-measure results than CR, when $d=1$.
    LSH and CR have similar P-measure performance in the remaining dimensions.

    \item[\textit{\textbf{(e)}}]\textit{\textbf{LSH vs. DF:}}
    Overall, LSH-FSCS  performs similarly to DF, regardless of the failure rate and dimension.
    LSH-RRT is also similar to DF when $d>1$,
    but is significantly better than DF, for all failure rates, when $d=1$.

    \item[\textit{\textbf{(f)}}]\textit{\textbf{LSH vs. KD:}}
    Both FSCS and RRT versions of LSH have comparable P-measure results to KD, with the P-measure differences not being highly significant.
\end{itemize}

\subsubsection{P-measure Simulation Observations: Point Pattern}

Based on the point pattern P-measure simulation results (Tables \ref{TAB:FSCS-point-p} and \ref{TAB:RRT-point-p}), we have following observations:
\begin{itemize}
    \item[\textit{\textbf{(a)}}]\textit{\textbf{LSH vs. RT:}}
    When $d=1,2$, LSH has similar, or slightly better performance than RT, for both FSCS and RRT versions.
    When $d \geq 3$, LSH is, overall, worse than RT, but the differences between LSH-RRT and RT are much less than those between LSH-FSCS and RT, especially when $d$ is large.
    In other words, LSH-RRT is closer to RT performance than LSH-FSCS is.
    For example, when $d=10$, the LSH-RRT P-measure values range from 0.46 to 0.49, but the LSH-FSCS P-measures range from 0.18 to 0.30.

    \item[\textit{\textbf{(b)}}]\textit{\textbf{LSH vs. ART:}}
    When $d=1,2,3$, LSH has comparable P-measure results to ART.
    When $d=4,5,10$, however, overall, LSH has significantly better P-measure results than ART (for both FSCS and RRT).

    \item[\textit{\textbf{(c)}}]\textit{\textbf{LSH vs. RF:}}
    Overall, LSH has much better P-measure performance than RF, especially when the failure rate is low and the dimensionality is high.

    \item[\textit{\textbf{(d)}}]\textit{\textbf{LSH vs. CR:}}
    The comparison between LSH and CR is very similar to that between LSH and RF:
    LSH generally outperforms CR, especially for low failure rates and high dimensions.

    \item[\textit{\textbf{(e)}}]\textit{\textbf{LSH vs. DF:}}
    When $1\leq d \leq 3$, LSH has a very similar or slightly better performance than DF.
    When $d \geq 4$, LSH has similar or worse P-measure performance than DF for high failure rates;
    however, as the failure rate decreases, LSH performs better than DF.

    \item[\textit{\textbf{(f)}}]\textit{\textbf{LSH vs. KD:}}
    When the dimension $d$ is low ($d=1,2,3$), LSH is very comparable to KD.
    However, when $d$ is high ($d \geq 4$), overall, LSH outperforms KD, especially for lower failure rates.
\end{itemize}

\subsubsection{Analysis and Summary}

Table \ref{TAB:SimulationCollection1-p} presents the numbers of simulation scenarios, for each failure pattern, where LSH is significantly superior (\ding{52}), indistinguishable (\ding{109}), or significantly inferior (\ding{54}) to each compared technique, 
in terms of the P-measure.
Each failure pattern has 42 scenarios for pairwise comparison
---
6 dimensions $\times$ 7 failure rates.
Table \ref{TAB:SimulationCollection2-p} presents the P-measure comparisons for each dimension:
Each dimension has 21 scenarios for each pairwise comparison (3 failure patterns $\times$ 7 failure rates).

Similar to previous studies \cite{Chen2007a}, overall, with respect to the P-measure, LSH outperforms RT for block and strip patterns, but not for the point pattern.
Additionally, similar to the F-measure results, the LSH P-measure also shows evidence of the \textit{curse of dimensionality} \cite{Bellman1957}:
In higher dimensions, LSH generally has worse P-measure performance than RT (especially for the block and point patterns).
Overall, the P-measure observations for LSH compared with RT align with the F-measure observations.

For the same number of test cases,
LSH generally has worse P-measure performance than the original ART for the block pattern;
comparable performance for the strip pattern; and
better performances for the point pattern.
In high dimensions, LSH generally has better P-measure results than ART, even for the block pattern.
When comparing LSH with ART, therefore, the P-measure observations are consistent with those for the F-measure.

Overall, compared with RF and CR, LSH  has better P-measure performance for the block and point patterns, which is unexpected
---
they would be expected to have comparable performance.
All three perform similarly for the strip pattern.
Compared with DF and KD, LSH has comparable P-measure results on the whole, which is expected.
In low dimensions, DF and KD perform slightly better than LSH overall;
while in high dimensions, LSH generally performs slightly better than DF and KD.

In conclusion, in spite of some differences, overall, the P-measure simulation observations are broadly in line with the F-measure simulation observations (Section \ref{SECTION:Fsimulation}).

\begin{tcolorbox}[colback=white,breakable,colframe=black,arc=0mm,left={0mm},top={0mm},bottom={0mm},right={0mm},boxrule={0.25mm}]
\textit{\textbf{Summary of Answers to RQ1.2 for Simulations:}}
\begin{itemize}
    \item
        \textit{Compared with RT,
        LSH has comparable or better P-measure performance for the strip pattern; and for the block and point patterns in low dimensions.}
    \item
        \textit{Compared with the original ART, overall, LSH has similar or better P-measure results for the strip and point patterns, but similar or worse results for the block pattern.
        In high dimensions, however, LSH is similar to or better than ART for all three failure patterns.}
    \item
        \textit{Overall, LSH has better P-measure results than RF and CR; and comparable results to DF and KD.}
\end{itemize}
\end{tcolorbox}

\subsection{Answer to RQ1.2: Effectiveness: P-measure Empirical Studies}
\label{SECTION:ES-P}

Tables \ref{TAB:FSCS-real-p} and \ref{TAB:RRT-real-p} summarize the P-measure results for the empirical study.

\subsubsection{P-measure Empirical Study Observations}

Based on the P-measure empirical study results (Tables \ref{TAB:FSCS-real-p} and \ref{TAB:RRT-real-p}), we have the following observations:
\begin{itemize}
    \item[\textit{\textbf{(a)}}]\textit{\textbf{LSH vs. RT:}} Overall, LSH has significantly better P-measure results than RT, for both FSCS and RRT versions, irrespective of the test set size.

    \item[\textit{\textbf{(b)}}]\textit{\textbf{LSH vs. ART:}} When the test set size is relatively small (for example, $|T|\leq 30$), LSH-FSCS performs significantly worse than ART;
    when the test set size becomes large, however, there is no significant difference compared with ART.
    LSH-RRT has, overall, better P-measure results than ART, especially when the test set size is relatively large.

    \item[\textit{\textbf{(c)}}]\textit{\textbf{LSH vs. RF:}} The comparison between LSH and RF is similar to that between LSH and ART.
    When executing a small test set, LSH-FSCS is worse than RF; and
    when running more test cases, the performance becomes similar to RF.
    LSH-RRT is better than RF, overall, especially when a large number of test cases are executed.

    \item[\textit{\textbf{(d)}}]\textit{\textbf{LSH vs. CR:}} 
    The P-measure comparisons between LSH and CR are very similar to the comparisons with ART and RF, for both FSCS and RRT versions.

    \item[\textit{\textbf{(e)}}]\textit{\textbf{LSH vs. DF:}} When the test set size is relatively small, LSH has similar or slightly worse P-measure results than DF for both FSCS and RRT versions.
    However, when the test set size is large, the LSH performance is different for the different versions:
    when executing a larger number of test cases, LSH-FSCS performs similarly to DF, but LSH-RRT outperforms DF.

    \item[\textit{\textbf{(f)}}]\textit{\textbf{LSH vs. KD:}} The P-measure comparisons between LSH and KD are, overall, very similar to the comparisons with DF:
    When the test set is relatively small, KD outperforms LSH;
    but when the test set is relatively large, LSH overall outperforms KD.
\end{itemize}

\subsubsection{Analysis and Summary}

Table \ref{TAB:realCollectionPmeasure} presents the numbers of testing scenarios in the empirical studies for which LSH, according to the P-measure, is
significantly superior (\ding{52}),
indistinguishable (\ding{109}), or
significantly inferior (\ding{54}) to each compared technique.

According to the P-measure observations, LSH has significantly better performance than RT ($43/46=93.48\%$ of the cases), which is consistent with the observations based on the empirical F-measure study.
LSH has comparable or better P-measure performances than ART in most scenarios ($37/46=80.43\%$), even though LSH may discard some information when generating test cases
---
because LSH adopts the ANN search rather than the NN search. 

When the test set size is less than $\lambda$ (a predefined parameter for RF and CR), both RF and CR are equivalent to ART: 
They both only \emph{forget} previous test cases after already executing $\lambda$ test cases
---
$\lambda$ was set to $60$ in this study.
Even when generating slightly more than $\lambda$ test cases, RF and CR may only be slightly worse than ART.
As shown in Table \ref{TAB:realCollectionPmeasure}, LSH has comparable or significantly better results than RF and CR in most cases ($36/46=78.26\%$ and $39/46=84.78\%$, respectively).

Compared with DF and KD, LSH has different P-measure results for different versions of LSH. More specifically, LSH-FSCS is similar to or worse than DF and KD; while
LSH-RRT performs similarly or better than DF and KD. Overall, LSH achieves comparable P-measure results to DF and KD.

\begin{tcolorbox}[colback=white,breakable,colframe=black,arc=0mm,left={0mm},top={0mm},bottom={0mm},right={0mm},boxrule={0.25mm}]\textit{\textbf{Summary of Answers to RQ1.2 for Empirical Studies:}}
\begin{itemize}
    \item
        \textit{Overall, LSH has significantly better P-measure results than RT in most scenarios.}
    \item
        \textit{LSH has very similar or worse results than the original ART, for the FSCS version;
        and significantly better results for the RRT version.}
    \item
        \textit{LSH generally has better P-measure results than RF and CR, in most cases, and comparable results to DF and KD.}
\end{itemize}
\end{tcolorbox}

\subsection{Answer to RQ2: Efficiency: Test Generation Time\label{SECTION:ExecutionTime}}

We investigated the execution time required by the different techniques to generate $n$ test cases for each $d$-dimensional input domain $[0,1.0)^d$.
The values of $d$ were selected as $1$, $2$, $3$, $4$, $5$, and 10.
The values of $n$ were chosen from $500$ to $10,000$, with an increment of $500$:
$n=500; 1000; \cdots; 10,000$.

\begin{figure*}[!b]
\centering
\graphicspath{{Graphs/time-new/}}
\resizebox{\textwidth}{!}{
    \subfigure[$d=1$]
    {
        \includegraphics[width=0.48\textwidth]{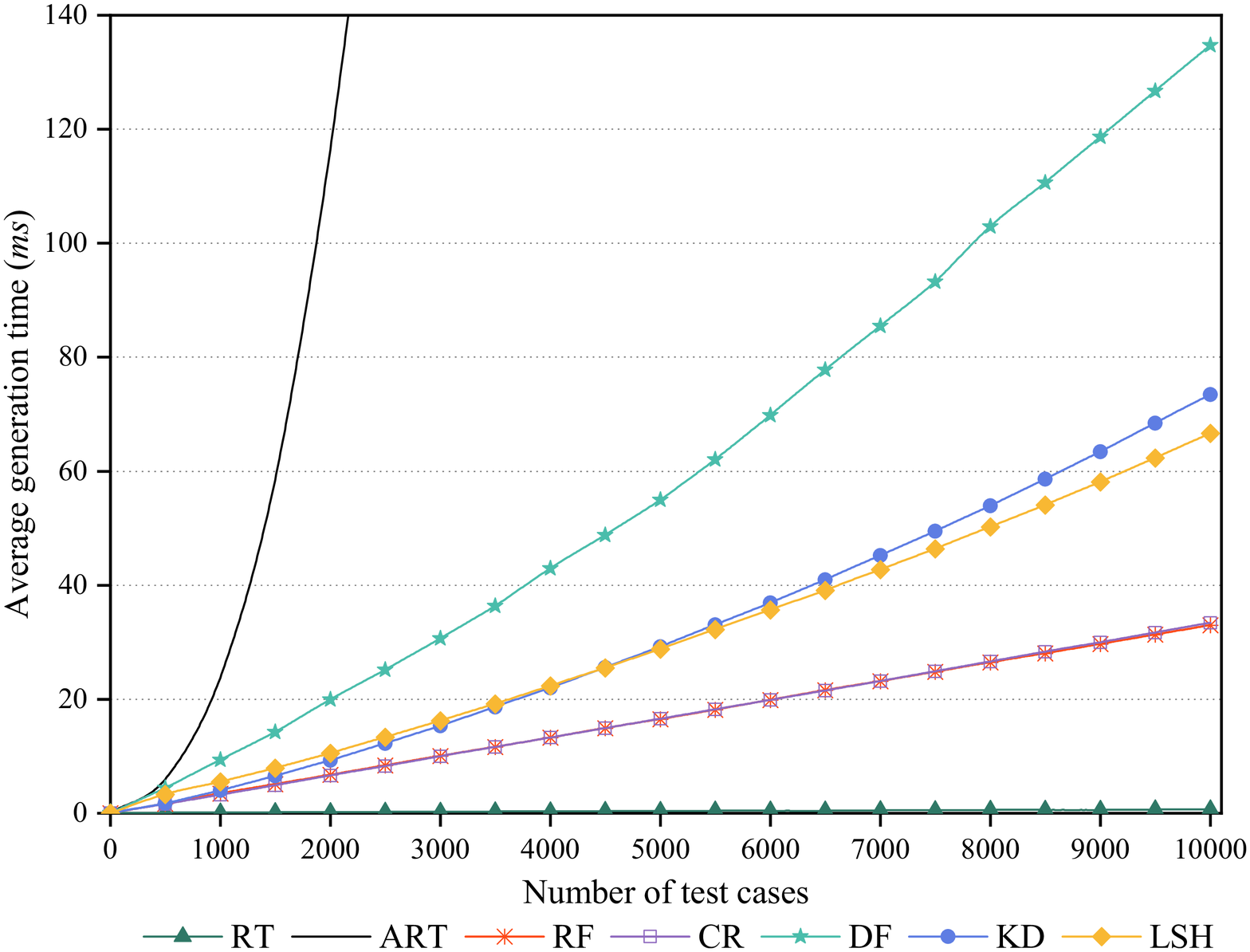}
    }
    \subfigure[$d=2$]
    {
        \includegraphics[width=0.48\textwidth]{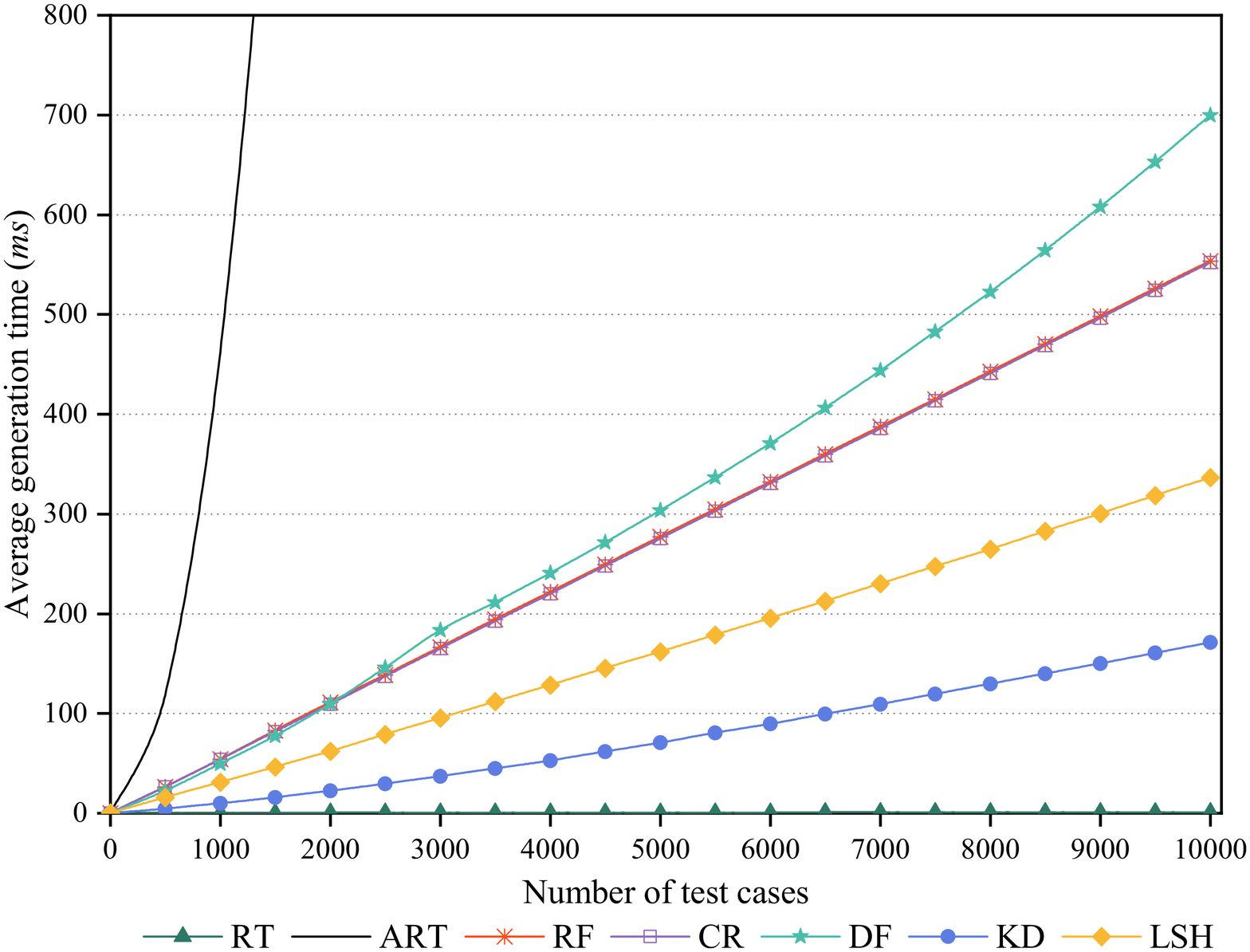}
    }
    \subfigure[$d=3$]
    {
        \includegraphics[width=0.485\textwidth]{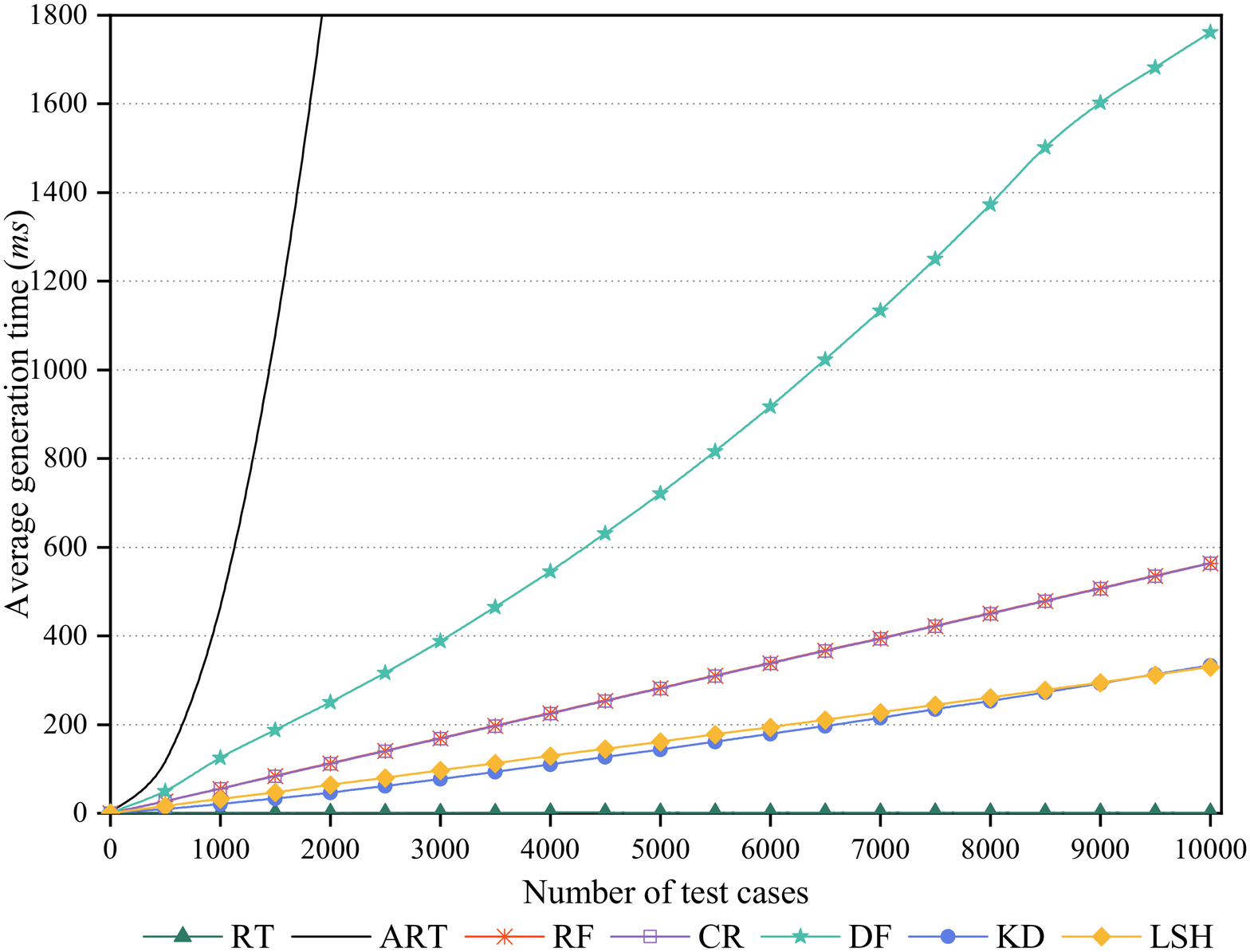}
    }}
    \resizebox{\textwidth}{!}{
        \subfigure[$d=4$]
    {
        \includegraphics[width=0.48\textwidth]{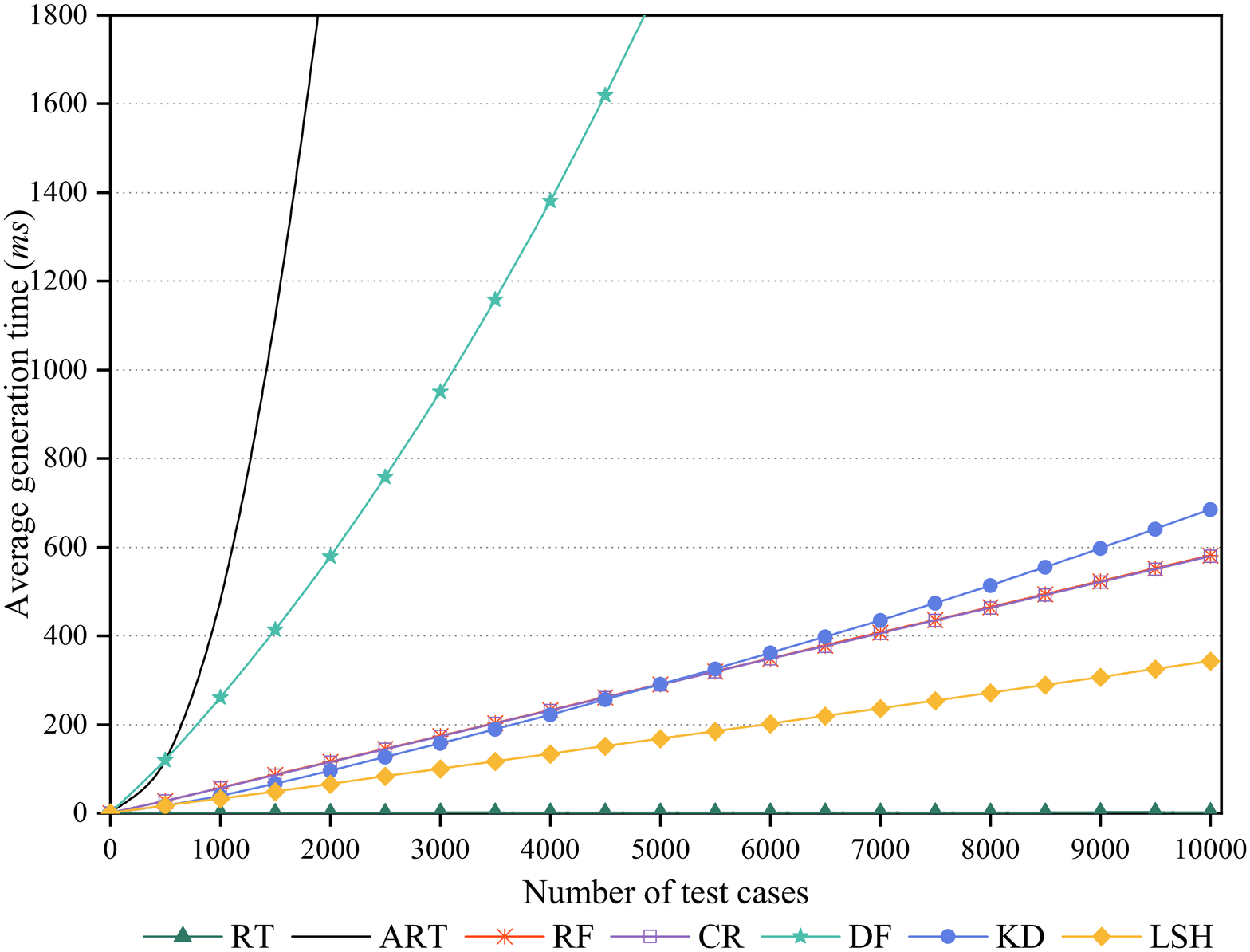}
    }
    \subfigure[$d=5$]
    {
        \includegraphics[width=0.48\textwidth]{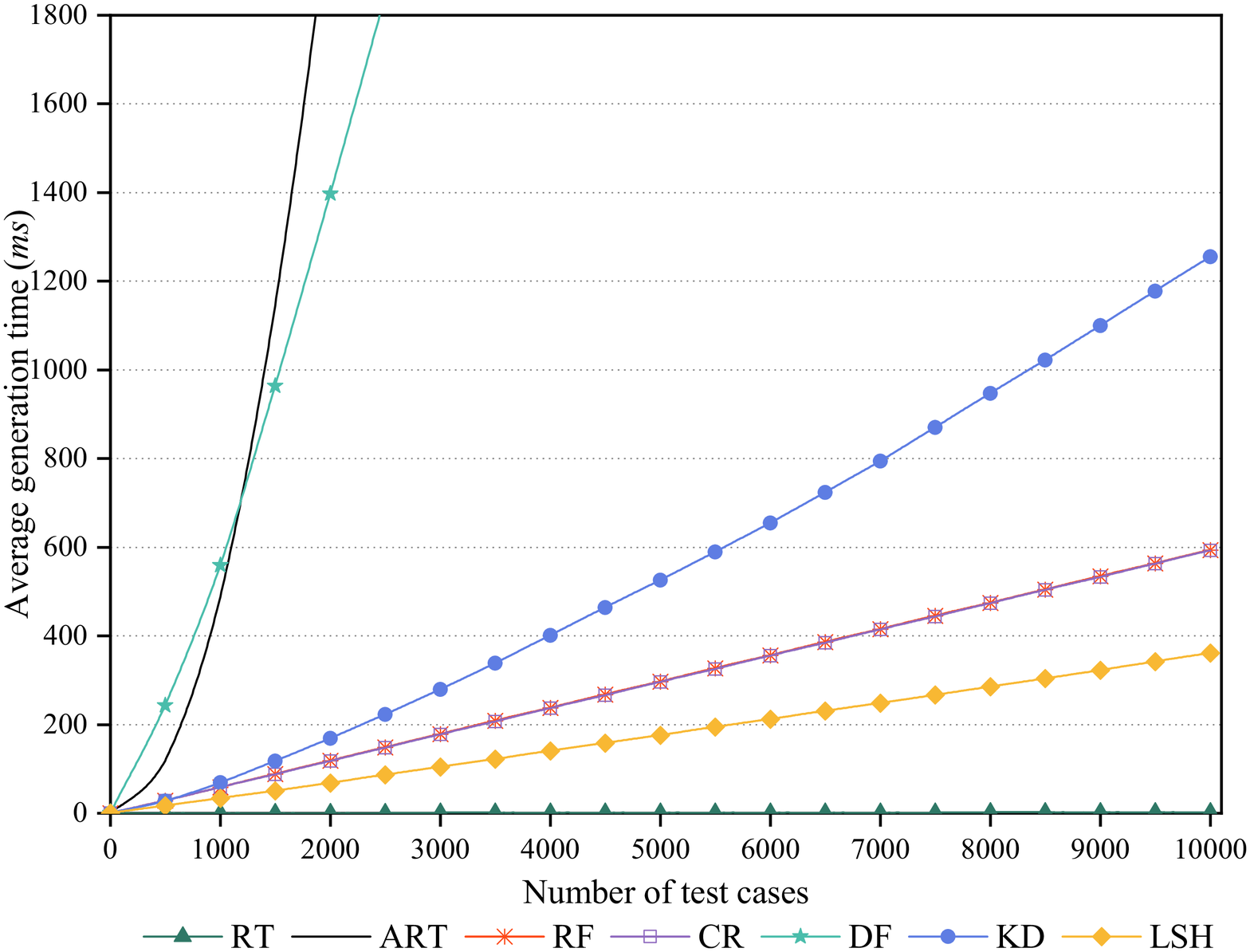}
    }
    \subfigure[$d=10$]
    {
        \includegraphics[width=0.48\textwidth]{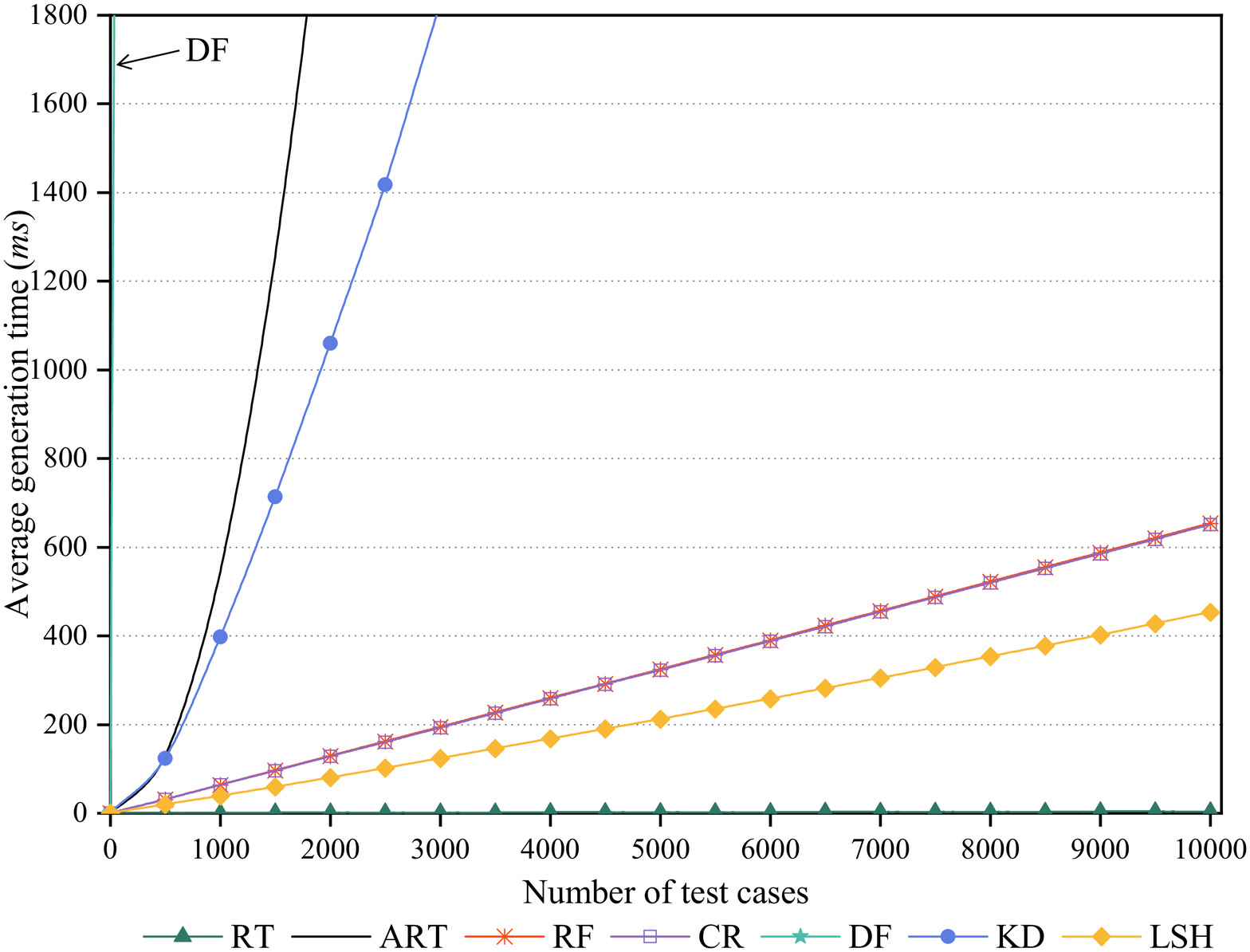}
    }}
    \caption{\textbf{FSCS version}: \textbf{Test-case generation time} for various test set sizes.}    
    \label{FIG:executionTimeFSCS}
\end{figure*}

\begin{figure*}[!t]
\centering
\graphicspath{{Graphs/time-new/}}
\resizebox{\textwidth}{!}{
    \subfigure[$d=1$]
    {
        \includegraphics[width=0.48\textwidth]{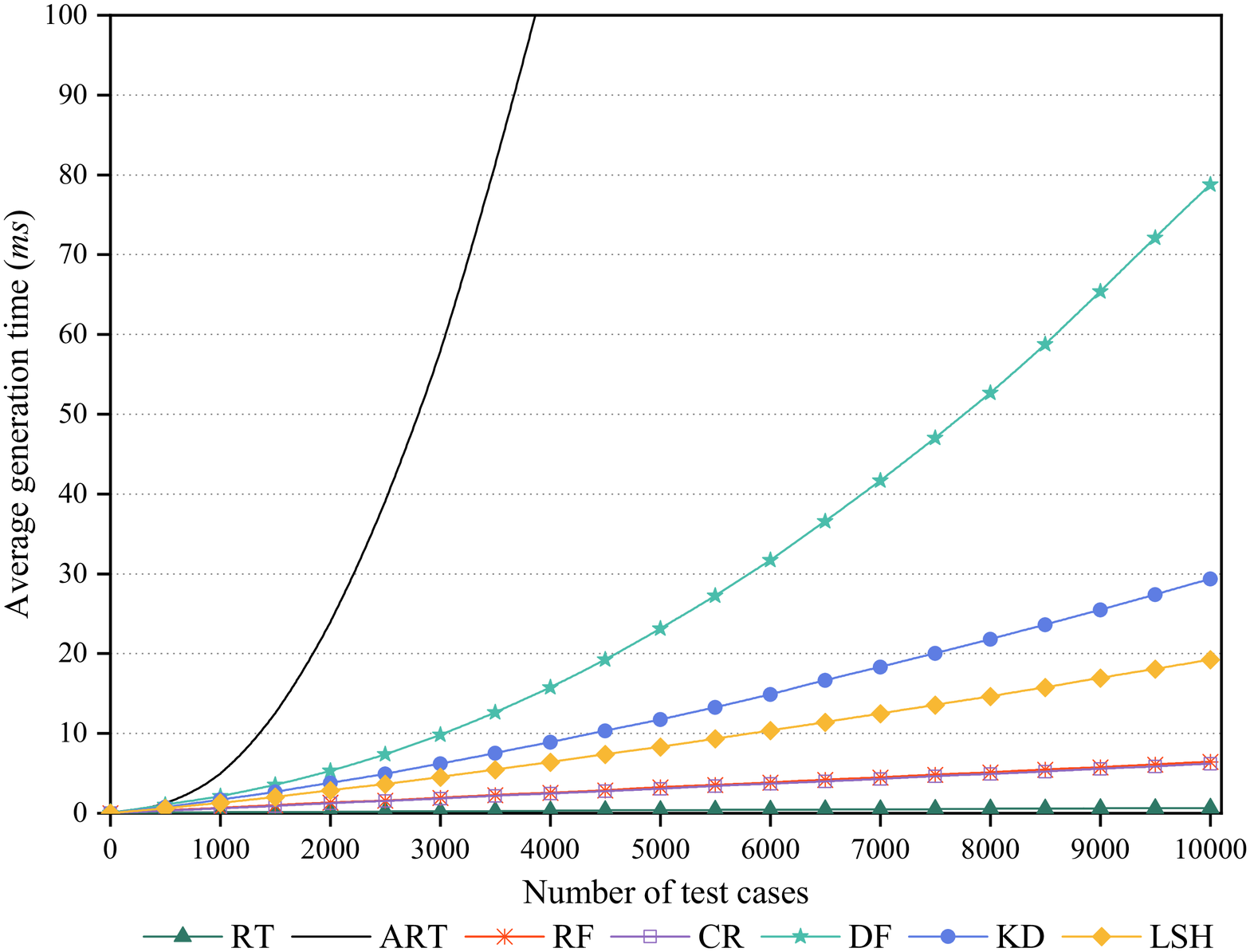}
    }
    \subfigure[$d=2$]
    {
        \includegraphics[width=0.48\textwidth]{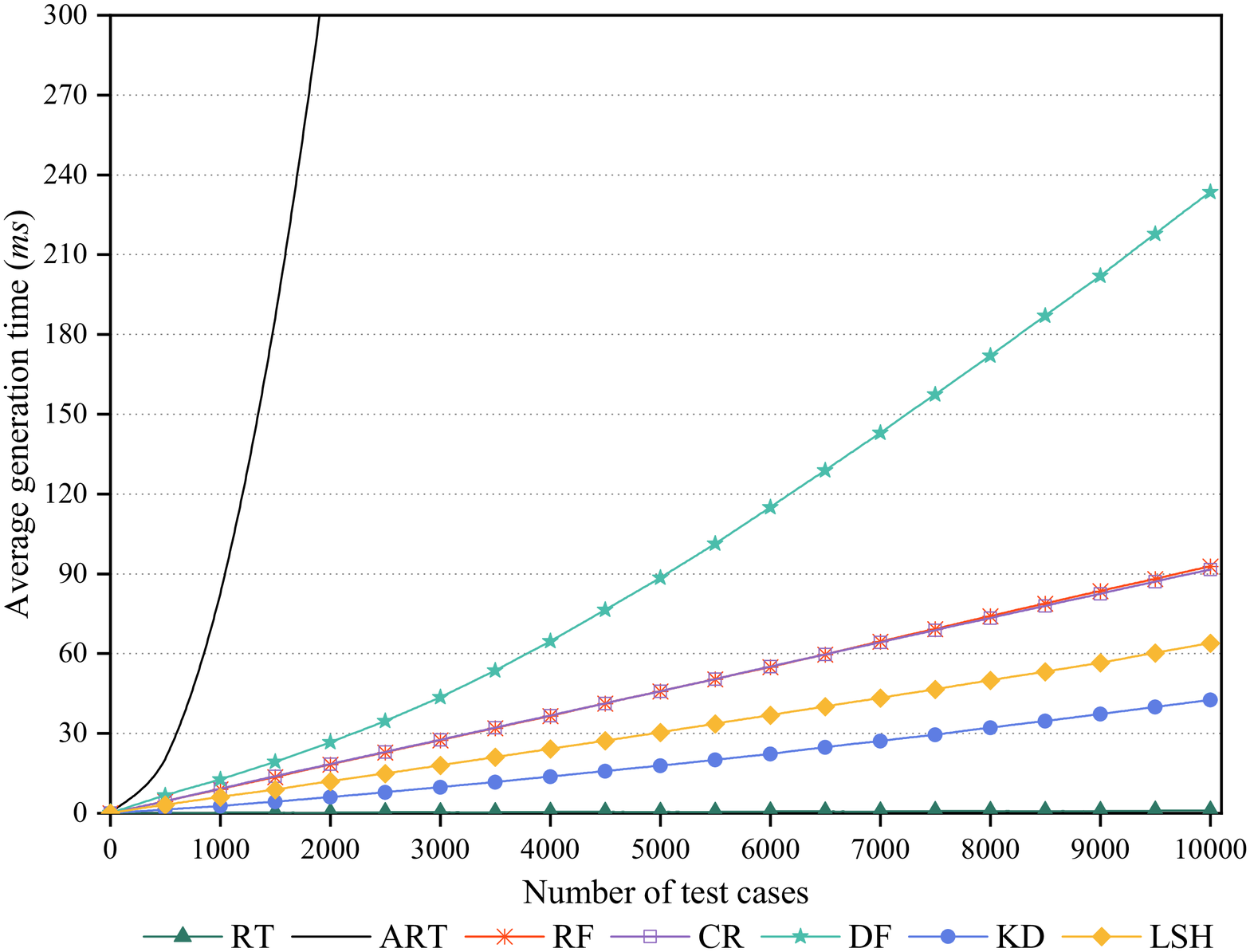}
    }
    \subfigure[$d=3$]
    {
        \includegraphics[width=0.48\textwidth]{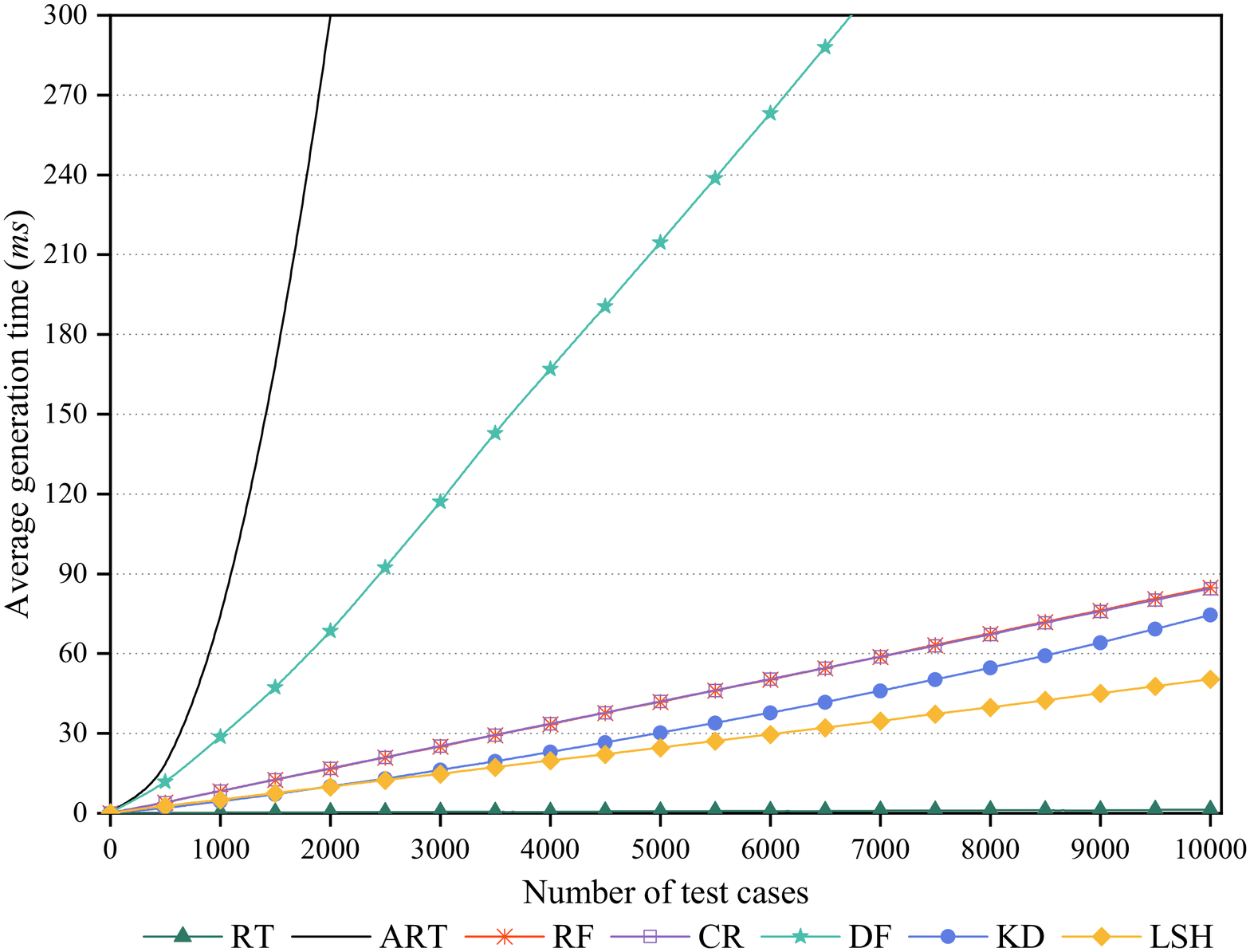}
    }}
    \resizebox{\textwidth}{!}{
        \subfigure[$d=4$]
    {
        \includegraphics[width=0.48\textwidth]{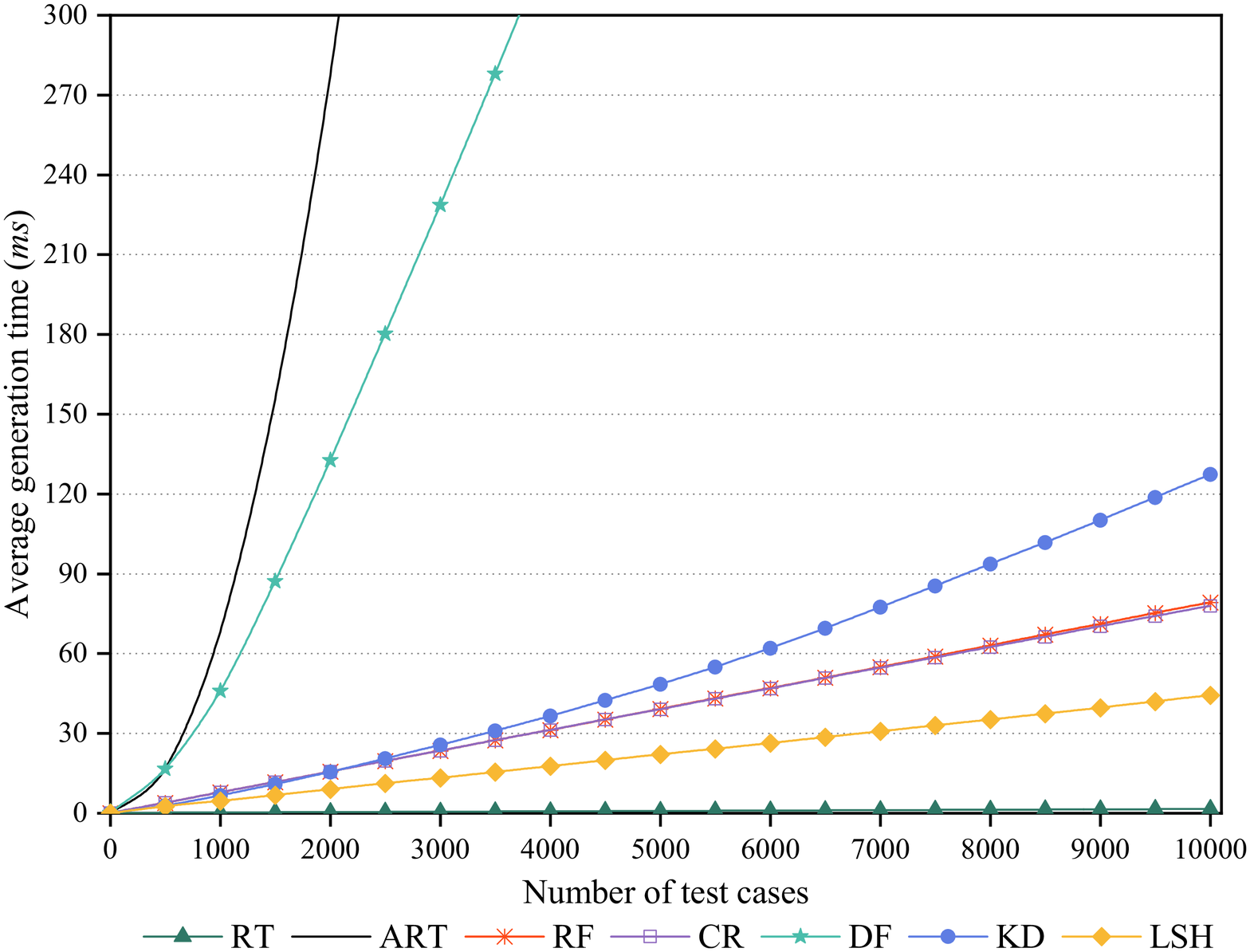}
    }
    \subfigure[$d=5$]
    {
        \includegraphics[width=0.48\textwidth]{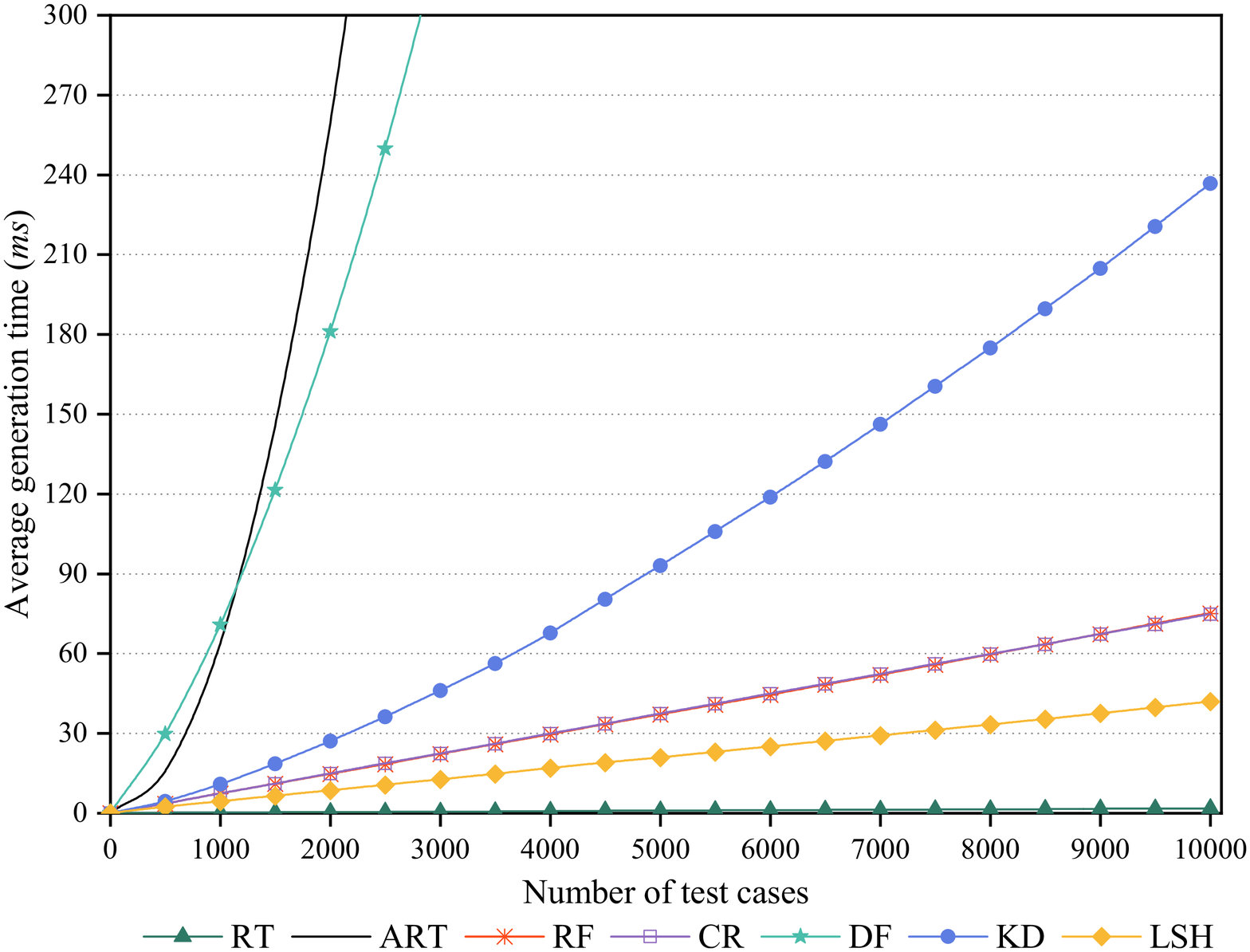}
    }
    \subfigure[$d=10$]
    {
        \includegraphics[width=0.48\textwidth]{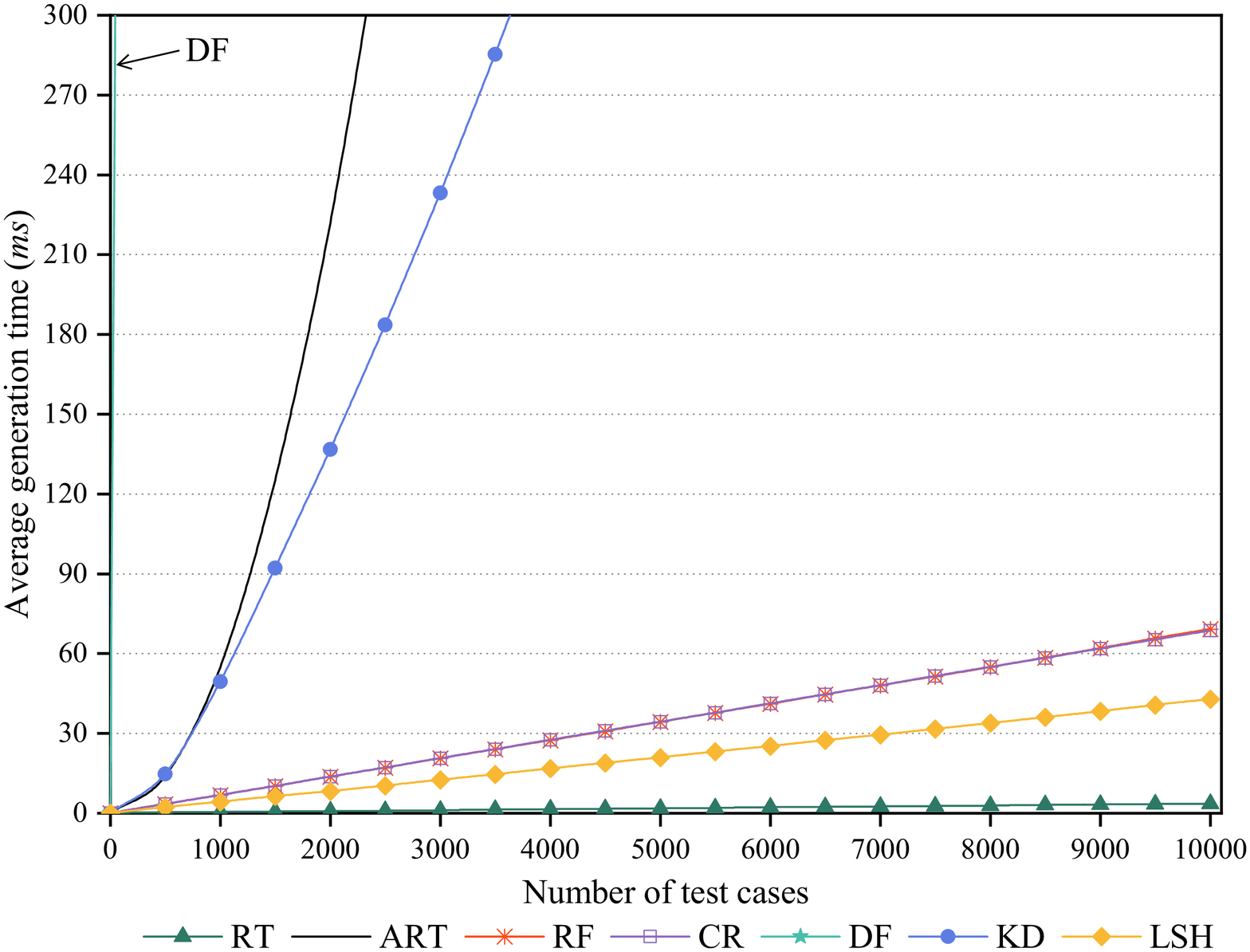}
    }}
    \caption{\textbf{RRT version}: \textbf{Test-case generation time} for various test set sizes.}
    \label{FIG:executionTimeRRT}
\end{figure*}

Figures \ref{FIG:executionTimeFSCS} and \ref{FIG:executionTimeRRT} present the test-case generation time for the various methods under study.
Each subfigure presents data for a specific dimension $d$, with the $x$-axis representing the number of generated test cases $n$, and the $y$-axis showing the time required to generate those $n$ test cases.
Table \ref{TAB:executionTime} presents the test-case generation time for each dimension, when $n$ was fixed at four representative values: 
$500$;
$1000$;
$5000$; and
$10,000$.

\subsubsection{Observations for Fixed Dimensions ($d$)}

Based on the data in Figures \ref{FIG:executionTimeFSCS} and \ref{FIG:executionTimeRRT}, we have the following observations:
\begin{itemize}
    \item[\textit{\textbf{(a)}}]\textit{\textbf{LSH vs. RT:}}
     Compared with RT, LSH always requires more computational time to generate the same number of test cases.
     This is because RT generally uses very little or no information to guide the test-case generation.

    \item[\textit{\textbf{(b)}}]\textit{\textbf{LSH vs. ART:}}
    For all dimensions ($d$) and all numbers of test cases ($n$), LSH has a much lower execution time than the original ART technique when generating the same number of test cases.
    As $n$ increases, the difference in time taken becomes significantly larger.
    LSH is much more efficient than ART when generating the same number of test cases.
    This observation applies to both FSCS and RRT.

    \item[\textit{\textbf{(c)}}]\textit{\textbf{LSH vs. RF:}}
    When $d=1$, LSH takes longer than RF, but when $d>1$, the LSH test-case generation time is always less than that of RF.
    As $n$ increases, the difference between LSH and RF becomes slightly larger.
    These observations are valid for both FSCS and RRT.

    \item[\textit{\textbf{(d)}}]\textit{\textbf{LSH vs. CR:}}
    Because their test-case generation processes are very similar, CR and RF require almost the same amount of time to generate the same number of test cases.
    Accordingly, the observations from comparing LSH with CR are the same as for comparing LSH with RF.

    \item[\textit{\textbf{(e)}}]\textit{\textbf{LSH vs. DF:}}
    For each fixed dimension ($d$), LSH requires much less time than DF to generate the same number of test cases, for all values of $n$.
    As $n$ increases, the difference between LSH and DF also increases.
    These observations apply to both FSCS and RRT.

    \item[\textit{\textbf{(f)}}]\textit{\textbf{LSH vs. KD:}}
    When $d=1$, for $n \leq 5000$, LSH-FSCS needs similar, or slightly more time than KD;
    however, when $n$ increases, LSH-FSCS begins to need slightly less time than KD (to generate the same number of test cases).
    When $d=2$, LSH-FSCS takes more time than KD, regardless of $n$.
    When $d=3$, the LSH-FSCS time is similar, or slightly greater, overall.
    For the remaining dimensions, LSH-FSCS is much more efficient than KD, regardless of $n$.
    In addition, the observations for LSH-RRT are, overall, the same as for LSH-FSCS, when $d=2,4,5,10$.
    However, LSH-RRT is slightly different from LSH-FSCS when $d=1$ or $d=3$:
    LSH-RRT is faster than KD in these dimensions.
    In summary, LSH is more efficient than KD for all values of $n$, except when $d=2$.
\end{itemize}

\subsubsection{Observations for Fixed Number of Test Cases ($n$)}

It can be observed from Table \ref{TAB:executionTime} that, when generating a fixed number $n$ of test cases, as the dimensionality increases, the FSCS techniques generally require an increasing amount of time.
However, this is not the case for the RRT techniques:
For higher dimensions, DF and KD take more time.
However, the time taken by (RRT) ART, RF, CR, and LSH, when $d=2$, is highest among all dimensions, for all values of $n$.
This is mainly due to the characteristics of RRT.

Comparing LSH with the other techniques, we have the following observations:
\begin{itemize}

    \item[\textit{\textbf{(a)}}]{\textit{\textbf{LSH vs. RT:}}}
    LSH is more time-consuming than RT when generating $n$ test cases, regardless of $d$.

    \item[\textit{\textbf{(b)}}]\textit{\textbf{LSH vs. ART:}}
    When $n$ is fixed, LSH is much more efficient than ART, in all dimensions, for all values of $n$.
    As $n$ increases, the differences in time taken also increase, for all dimensions.

    \item[\textit{\textbf{(c)}}]\textit{\textbf{LSH vs. RF:}}
    For all values of $n$, when $d=1$, LSH is less efficient than RF;
    however, when $d \geq 2$, LSH is more efficient.
    Nevertheless, the differences in time taken are relatively small.
    These observations apply to both FSCS and RRT.

    \item[\textit{\textbf{(d)}}]\textit{\textbf{LSH vs. CR:}}
    Since CR has very similar test-case generation time to RF, the comparisons between LSH and CR are very similar to those between LSH and RF:
    Apart from when $d=1$, LSH is more efficient than CR.

    \item[\textit{\textbf{(e)}}]\textit{\textbf{LSH vs. DF:}}
    LSH is much more efficient than DF for all values of $n$ and $d$.
    In addition, as $d$ increases, the difference between LSH and DF also increases.

    \item[\textit{\textbf{(f)}}]\textit{\textbf{LSH vs. KD:}}
    When $n=500$, LSH-FSCS is slightly faster than KD for $d \geq 5$.
    When $n=1000$, KD outperforms LSH-FSCS for $d \leq 3$, but LSH-FSCS is faster when $d >3$.
    When $n=5000$, LSH-FSCS outperforms KD for $d > 3$.
    When $n=10,000$, LSH-FSCS is faster than KD for all dimensions, except $d=2$. 
    LSH-RRT is faster than KD for all values of $n$, and all values of $d$ except $2$. 
    As the dimensionality $d$ increases, for all values of $n$, the differences between LSH and KD also, overall, increase.
    The main reason is that, as $d$ increases, the increment in computation time for KD is much greater than for LSH.
    In other words, for higher dimensions, LSH is much more efficient than KD.    
\end{itemize}

\begin{figure*}[!t]
\centering
\graphicspath{{Graphs/time-new/}}
\resizebox{\textwidth}{!}{
    \subfigure[LSH-FSCS]
    {
        \includegraphics[width=0.48\textwidth]{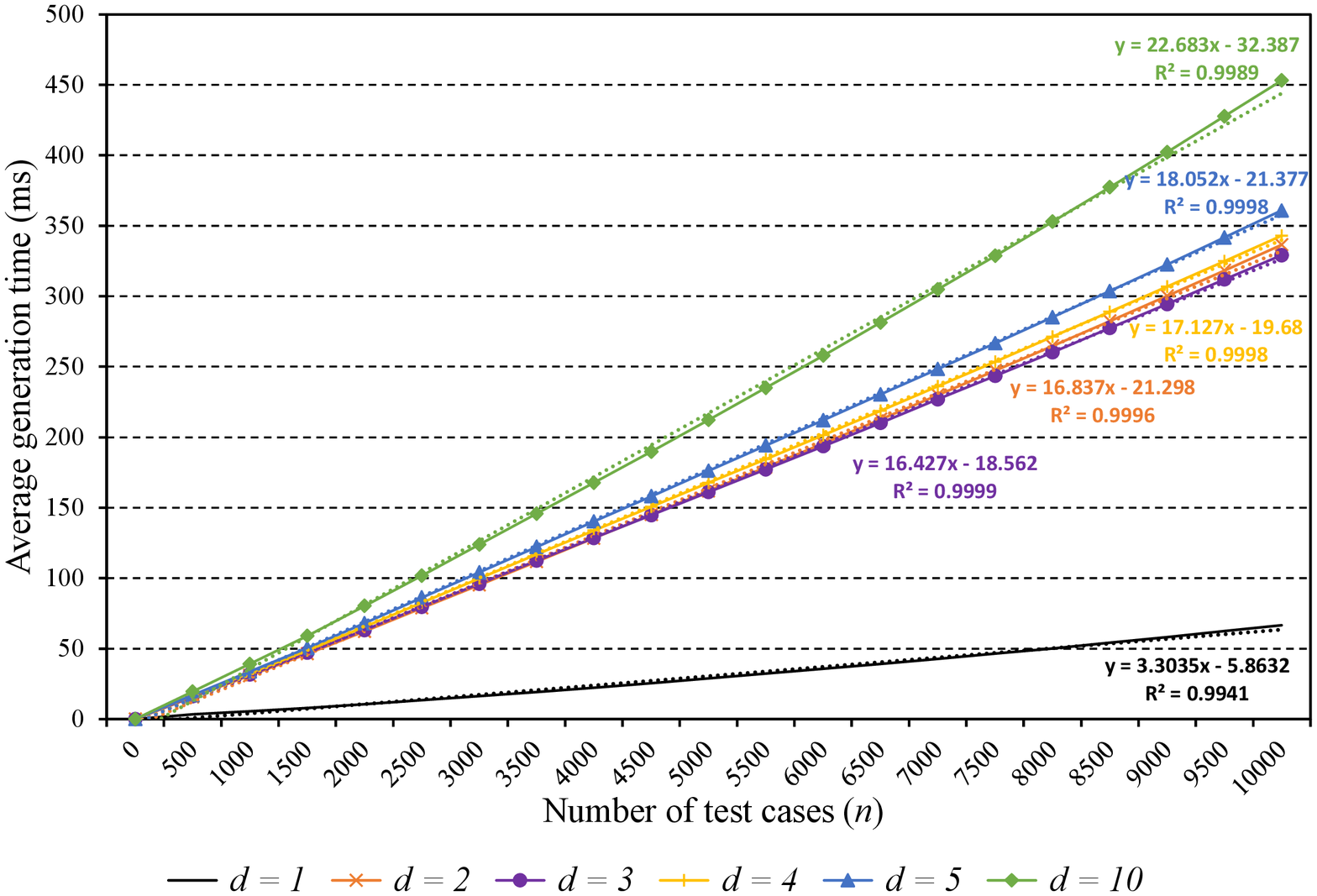}
        \label{FIG:curveTime1}
    }
    \subfigure[LSH-RRT]
    {
        \includegraphics[width=0.48\textwidth]{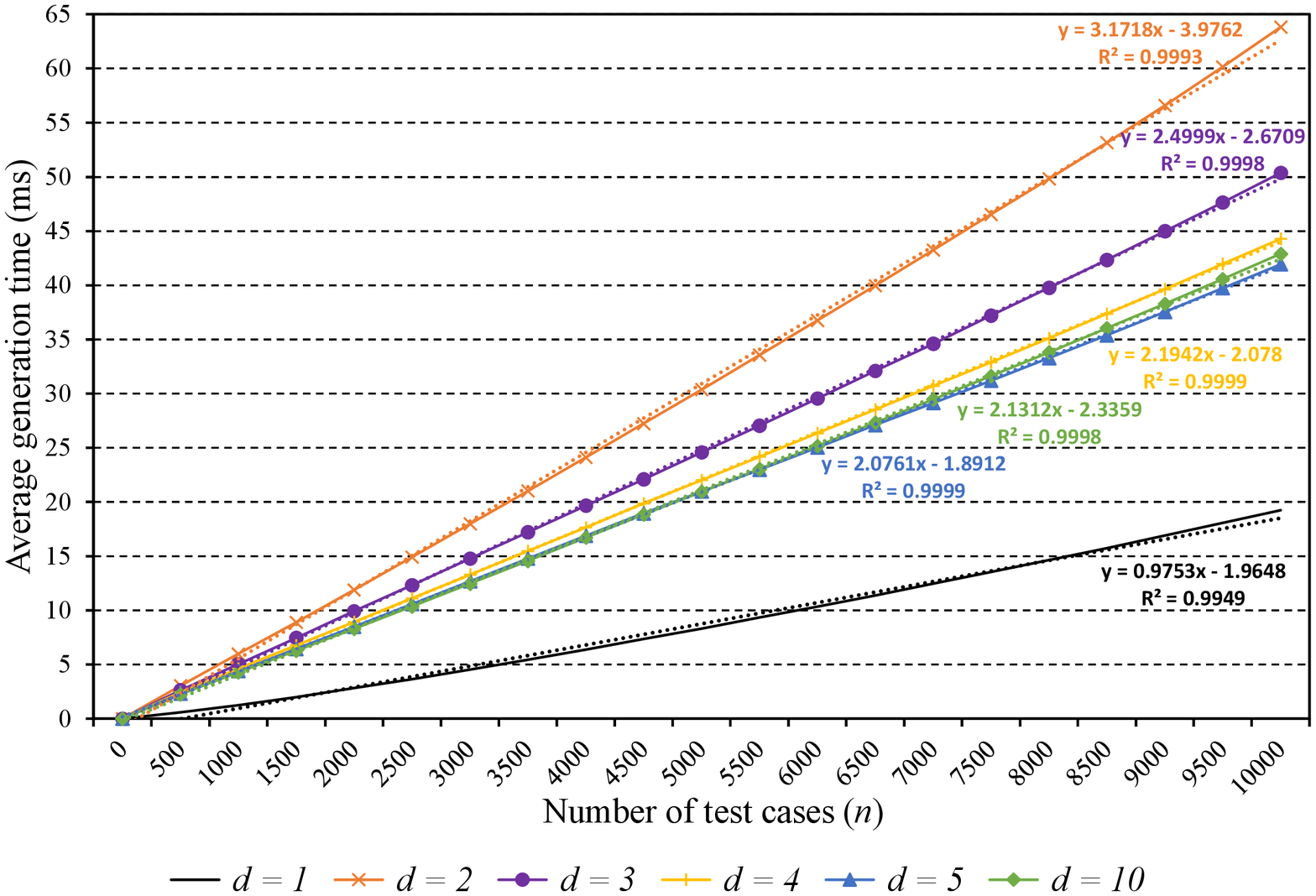}
        \label{FIG:curveTime2}
    }}
    \caption{Curve-fitting for LSH test-case generation times, for various test set sizes and dimensions.}
    \label{FIG:curveTime}
\end{figure*}

\subsubsection{Analysis and Summary}

It has been noted \cite{Chen2004a,Mayer2006} that ART generation of the ``next'' ($(n+1)$-th) test case
---
for both FSCS and RRT
---
incurs $n\times k$ distance calculations between $k$ random candidates and the already-executed $n$ test cases.
In FSCS, $k$ is generally assigned a constant number, typically $k=10$ \cite{Chen2004a};
but in RRT, the average value of $k$ is probably logarithmic to $n$:
$k=\beta_1\log(n)+\beta_2$, where $\beta_1$ and $\beta_2$ are two constants \cite{Mayer2006}.
Application of a \textit{forgetting} strategy involves setting a predefined constant  $\lambda$, and only retaining the information for $\lambda$ executed tests:
The other $n-\lambda$ (where $\lambda < n$) executed test cases are discarded, and distance calculations are only required for $\lambda$ executed test cases, for each candidate.
This means that forgetting strategies need only a constant number of calculations ($\lambda \times k$) to generate a new test case, resulting in a linear time-complexity order, $O(n)$.
As shown in Figures \ref{FIG:executionTimeFSCS} and \ref{FIG:executionTimeRRT}, the LSH trends are very similar to those of the forgetting strategies,  suggesting that the LSH time complexity is also very close to a linear order.

Figure \ref{FIG:curveTime} presents the LSH test-case generation times, for various $n$.
The average times for each dimension $d$ are shown with connected lines, with the associated fitted curves shown as dashed lines.
Figure \ref{FIG:curveTime1} presents the LSH-FSCS data, and Figure \ref{FIG:curveTime2} shows the LSH-RRT data.
An analysis of the cumulative generation time shows that a line function with a high determination coefficient ($R^2 \geq 0.9941$) can be identified for each fitted curve, for all dimensions, for both LSH versions:
The average LSH test-case generation time has a significant linear relationship with the number of test cases, for each dimension.

In conclusion, although the worst-case LSH time complexity order is $O(n \log n)$, as discussed in Section \ref{SECTION:complexity}, the simulation results show the time complexity to be approximately linear.

\begin{tcolorbox}[colback=white,breakable,colframe=black,arc=0mm,left={0mm},top={0mm},bottom={0mm},right={0mm},boxrule={0.25mm}]\textit{\textbf{Summary of Answers to RQ2:}}
\begin{itemize}
    \item
        \textit{LSH is always less efficient than RT when generating the same number of test cases, for all dimensions.}
    \item
        \textit{LSH is much more efficient than the original ART (for both FSCS and RRT) when generating the same number of test cases, in all dimensions.}
    \item
        \textit{LSH is also much more efficient than the DF, RF, and CR variations of ART, when generating the same number of test cases, for all dimensions except $d=1$.}
    \item
        \textit{LSH has similar (sometimes slightly better or worse) efficiency to KD for low dimensions ($d \leq 3$),
        but is much more efficient for high dimensions, especially for larger numbers of test cases.}
\end{itemize}
\end{tcolorbox}

\subsection{Answer to RQ3: Cost-effectiveness: F-time}
\label{SECTION:F-time}

Tables \ref{TAB:FSCS-Ftime} and \ref{TAB:RRT-Ftime} present the F-time results for the 23 subject programs.

\subsubsection{F-time Observations}

\begin{table*}[!t]
\centering
\footnotesize
\caption{Number of \textbf{Real-life Programs} for which LSH performance is significantly superior (\ding{52}), indistinguishable (\ding{109}), or significantly inferior (\ding{54}), with respect to \textbf{F-time}}
\label{TAB:realCollectionFtime}
\resizebox{\textwidth}{!}{
\begin{tabular}{@{}ccccccccccccccccccccccccc@{}}
\hline
\multirow{2}*{\textbf{ART Version}}  & &\multicolumn{3}{c}{\textit{LSH vs. RT}} & &\multicolumn{3}{c}{\textit{LSH vs. ART}} & &\multicolumn{3}{c}{\textit{LSH vs. RF}} & &\multicolumn{3}{c}{\textit{LSH vs. CR}} & &\multicolumn{3}{c}{\textit{LSH vs. DF}} & &\multicolumn{3}{c}{\textit{LSH vs. KD}}\\\cline{3-5}\cline{7-9}\cline{11-13} \cline{15-17}\cline{19-21} \cline{23-25}
& &\ding{52} &\ding{109} &\ding{54} & &\ding{52} &\ding{109} &\ding{54} & &\ding{52} &\ding{109} &\ding{54} & &\ding{52} &\ding{109} &\ding{54} & &\ding{52} &\ding{109} &\ding{54} &&\ding{52} &\ding{109} &\ding{54}
\\\hline
FSCS Version  & &0 &1 &22 & &23 &0 &0 & &18 &2 &3 & &18 &2 &3 & &23 &0 &0 & &16 &1 &6 \\\hline
RRT Version   & &1 &0 &22 & &23 &0 &0 & &20 &3 &0 & &19 &4 &0 & &23 &0 &0 & &18 &1 &4 \\\hline
\textbf{\textit{Sum}} & &1 &1 &44 & &46 &0 &0 & &38 &5 &3 & &37 &6 &3 & &46 &0 &0 & &34 &2 &10 \\\hline
\end{tabular}}
\end{table*}

Based on the results shown in Tables \ref{TAB:FSCS-Ftime} and \ref{TAB:RRT-Ftime}, we have the following observations:
\begin{itemize}
    \item[\textit{\textbf{(a)}}]\textit{\textbf{LSH vs. RT:}}
    LSH (both FSCS and RRT) has better (lower) F-time than RT for the P4 program, indicating that LSH is more cost-effective for this program.
    For the remaining 22 programs, however, LSH is much less cost-effective than RT.

    \item[\textit{\textbf{(b)}}]\textit{\textbf{LSH vs. ART:}}
    LSH requires much less time than ART to detect the first failure for all 23 programs, for both FSCS and RRT:
    LSH is much more cost-effective than the original ART.

    \item[\textit{\textbf{(c)}}]\textit{\textbf{LSH vs. RF:}}
    The F-time differences between LSH-FSCS and RF are very small (around $1$ ms) for the four programs P1, P2, P3, and P5, with the statistical analyses showing no very/significantly different performances.
    LSH-FSCS performs significantly better than RF for the remaining programs, apart from P10.
    In addition,
    LSH-RRT needs less time than RF to identify the first failure for all the programs.
    Overall, the statistical analyses show LSH-RRT to be significantly more cost-effective than RF for all except the first three programs.

    \item[\textit{\textbf{(d)}}]\textit{\textbf{LSH vs. CR:}}
    The comparisons between LSH and CR are very similar to those for RF:
    LSH is more cost-effective than CR for most programs.

    \item[\textit{\textbf{(e)}}]\textit{\textbf{LSH vs. DF:}}
    LSH is much faster than DF at finding the first failure, especially for the subject programs with high dimensions.
    The LSH-FSCS F-time for P22, for example, is only about 111 ms, compared with 423,722 ms for DF (almost four thousand times as long).
    The statistical analyses show LSH to be significantly more cost-effective than DF.

    \item[\textit{\textbf{(f)}}]\textit{\textbf{LSH vs. KD:}}
    LSH is faster than KD for the  1-dimensional programs (P1 to P5), but KD is more cost-effective when $d=2$ (P6 to P8).
    When $d=3$ or $4$ (P9 to P13), LSH is comparable to KD, with LSH performing better than KD for some programs, but also similarly, or worse, for others.
    For the programs with $d \geq 5$ except P23 (i.e., P14 to P22),  LSH is much faster than KD, which is supported by the statistical analyses.
    These observations are valid for both FSCS and RRT.
\end{itemize}

\subsubsection{Analysis and Summary}

Table \ref{TAB:realCollectionFtime} summarizes the number of real-life programs in the empirical studies, for which, according to F-time, the LSH performance is significantly superior (\ding{52}), indistinguishable (\ding{109}), or significantly inferior (\ding{54}) to each compared technique.
Based on the table, we have the following observations:
Compared with RT, LSH is less cost-effective in nearly all scenarios.
Compared with ART and DF, LSH is significantly more cost-effective in all scenarios.
LSH has better cost-effectiveness than RF and CR in most cases:
LSH performs similarly to, or better than, RF and CR in $43/46=93.48\% $ of the scenarios.
In addition, LSH is also usually more cost-effective than KD ($34/46=73.91\%$ of the scenarios).

During the testing, when the first failure was triggered for any program $P$, for any test-case generation approach, the F-measure (denoted $n$) and F-time were both recorded.
The F-time includes two parts:
the time taken to generate the $n$ test cases (denoted $\textit{TG}$); and
the time taken to execute them to test the program $P$ (denoted $\textit{TE}$), i.e.,
$\textrm{F-{time}} = \textit{TE} + \textit{TG}$.

As explained in Section \ref{SECTION:FmeasureES}, LSH generally requires fewer test cases than RT to find the first failure:
This indicates that the LSH $\textit{TE}$ is generally less than the RT $\textit{TE}$ ( $\textit{TE}_\textrm{LSH} < \textit{TE}_\textrm{RT}$).
However, the test-case generation time of LSH is still much higher than that of RT (as shown in Section \ref{SECTION:ExecutionTime}):
$\textit{TG}_\textrm{LSH} \gg \textit{TG}_\textrm{RT}$.
Therefore, in general, $\textit{TE}_\textrm{LSH} + \textit{TG}_\textrm{LSH} \gg \textit{TE}_\textrm{RT}+\textit{TG}_\textrm{RT}$.

Compared with ART and DF, overall, LSH has a similar F-measure performance
---
LSH, ART, and DF need similar numbers of test case executions to identify the first software failure, which means that $\textit{TE}_\textrm{LSH} \approx \textit{TE}_\textrm{ART} \approx \textit{TE}_\textrm{DF}$.
As discussed in Section \ref{SECTION:ExecutionTime}, however, the ART and DF execution times increase rapidly as the number of test cases increases or the dimensionality increases:
To generate the same number of test cases, ART and DF generally require much more computation time than LSH, i.e., $\textit{TG}_\textrm{LSH} \ll \textit{TG}_\textrm{ART}$ and $\textit{TG}_\textrm{LSH} \ll \textit{TG}_\textrm{DF}$.
LSH, therefore, requires less F-time than ART and DF, and is thus more cost-effective.

Although LSH has better F-measure performance than RF and CR for the 1-dimensional programs (Section \ref{SECTION:FmeasureES}), it also has a higher computation time (Section \ref{SECTION:ExecutionTime}), resulting in different F-time performances for different programs.
For the programs with $d \geq 2$, overall, LSH has similar F-measure performances to RF and CR, but requires less execution time to generate the same number of test cases, resulting in better F-time performance.

According to the F-measure observations in Section \ref{SECTION:FmeasureES}, overall, LSH has a similar performance to KD.
According to the computation time observations (Section \ref{SECTION:ExecutionTime}),
LSH takes longer than KD to generate the same number of test cases when $d=2$, but takes a similar, or much less, amount of time in other dimensions.
Consequently, LSH is less cost-effective than KD when $d=2$, but more cost-effective overall, especially when the dimensionality is high.

\begin{tcolorbox}[colback=white,breakable,colframe=black,arc=0mm,left={0mm},top={0mm},bottom={0mm},right={0mm},boxrule={0.25mm}]\textit{\textbf{Summary of Answers to RQ3:}}
\begin{itemize}
    \item
         \textit{LSH is less cost-effective than RT for nearly all programs.}

    \item
        \textit{LSH is much more cost-effective than the original ART (both FSCS and RRT) for all programs.}

    \item
        \textit{LSH is more cost-effective than DF in all the scenarios; and
        has better cost-effectiveness than RF, CR, and KD in most scenarios.}
\end{itemize}
\end{tcolorbox}

\subsection{Threats to Validity}
This section addresses some potential threats to the validity of our study.

\subsubsection{Threats to Experimental Design}

Some potential threats to the study validity relate to the experimental design, and can be examined as follows:

\begin{itemize}
    \item
    Although the simulations only used a limited number of failure patterns, failure rates, and dimensions, this configuration has also been widely used in ART research \cite{Huang2019,Mao2019,Shahbazi2012}.
    Nevertheless, our future work will involve additional simulations with more failure patterns, failure rates, and dimensions:
    This will enable further and deeper evaluation of the proposed approach.

    \item
    In spite of the number and diversity of real-life programs used in the empirical studies, their sizes were all relatively small.
    Nevertheless, these subject programs are very representative of those used in the literature for ART with numeric inputs.
    Our future work, however, will include the exploration of more subject programs, which we anticipate will strengthen the generalizability of the experimental results.
    Section \ref{SECTION:configurable} reports on a preliminary investigation into the performance of our proposed approach in large-scale non-numerical systems.
    
    \item
    Finally, mutants, rather than real faults, were used in the empirical studies.
    A set of real faults is typically restricted \cite{Offutt1992} by the \textit{competent programmer hypothesis} \cite{Morell1990} and the \textit{coupling effect hypothesis} \cite{DeMillo1978}.
    The competent programmer hypothesis states that competent programmers, when producing faulty programs, tend to write programs that are nearly correct:
    Although a program written by a competent programmer may have some faults, it is likely that these will be relatively small faults, and the difference compared with a correct version of the program will be minor \cite{Morell1990}.
    The coupling effect hypothesis states that complex faults are coupled to simple faults, and that a test set that identifies more simple faults in a program will also identify a high percentage of the complex faults \cite{DeMillo1978}.
    Generally speaking, mutation testing introduces simple faults into a program.
    Therefore, if test cases detect more mutants (considered simple faults), they should also be capable of detecting more complex faults \cite{Papadakis2019}.
    Furthermore, previous studies have shown some relatively strong correlations between mutant detection and real fault detection:
    Mutants can thus be used as a substitute for real faults when comparing different test suites \cite{Just2014,Chekam2017,Papadakis2018}.
    The preliminary study reported in Section \ref{SECTION:configurable} uses real-life faults in highly configurable systems as part of the performance assessment of our proposed approach.    

\end{itemize}

\subsubsection{Threats to Evaluation Metric Selection}

A second potential threat to validity relates to the selection of evaluation metrics.
Four evaluation metrics (F-measure, P-measure, computation time, and F-time) were adopted in our study.
These metrics were used to measure testing effectiveness, efficiency, and cost-effectiveness, and have been widely used in other ART studies \cite{Huang2019,Mao2019,Mao2017}.
Nevertheless, more evaluation metrics, related, for example, to test case distribution \cite{Chen2007} or code coverage \cite{Chen2013}, would be useful to further assess our proposed approach.
We look forward to exploring these perspectives in our future work.

\subsubsection{Threats to Parameter Settings}
The third potential threat to validity relates to the parameter settings.
Different parameter settings may produce different results and observations.
However, testing all possible parameter settings was not practical for this study, and so we set the various parameters according to the following principles:
(1) The parameters for some ART approaches were set according to previous ART studies \cite{Mao2019,Chen2004,Mao2017};
(2) some parameters were specifically set to ensure fair comparisons across techniques; and
(3) some parameter settings were determined through preliminary experiments, as discussed in Section \ref{SECTION:Parameter}.

\section{From Numerical to Non-Numerical Domains}
\label{SECTION:configurable}

This section explores the potential of LSH-ART to detect faults in non-numerical input domains.
We discuss the LSH-ART performance, compared with other ART algorithms, and examine the implications of its use in complex (non-numerical) domains.

\subsection{Notation and Concepts}

The behavior of configurable software is affected by multiple \textit{parameters} and their interactions. 
Parameters may represent internal or external triggers, user choices, and other things. 
For an SUT with $n$ parameters, the value for parameter $p_i$ comes from a set $V_i$. 
A test case can be described as $t=(v_1,v_2,\cdots,v_n)$, with $v_i\in V_i$ for $1\leq i\leq n$.

It is also possible that not all combinations of parameters are valid, due to certain  \textit{constraints} among the SUT parameters. 
For example, when setting the font in a text document, the choices \texttt{Superscript} ($sup$) and \texttt{Subscript} ($sub$) cannot be selected at the same time: 
$(sup=``1")$ conflicts with $(sub = ``1")$
---
$``1"$ means that the choice is selected.
This generates a constraint $(sup=``1") \wedge (sub = ``1")$.
The discussions in previous sections assumed no constraints on the parameter combinations.
However, this may be impractical in many real-life situations.
Violation of a constraint may prevent execution of the relevant test cases: 
It is, therefore, essential to apply constraint-handling strategies during test-case generation. 
If the generated test case values do not violate the constraint, then the test case is considered \textit{valid};
otherwise, it is \textit{invalid}.

\begin{figure}[!b]
    \centering
    \fbox{
        \centering
        \parbox{0.95\linewidth}{
            \begin{algorithmic}[1]
                \renewcommand{\algorithmicrequire}{\textbf{Input:}}
                \renewcommand{\algorithmicensure}{\textbf{Output:}}
                \renewcommand{\algorithmicelsif}{\algorithmicelse}
                \renewcommand{\algorithmicthen}{}
            
            \STATE $E\leftarrow \texttt{\textbf{EmptySet()}}$
            \WHILE {A stopping condition is not satisfied}
            \STATE Generate a test case $t$ through an ART strategy
            \WHILE {\texttt{\textbf{isVaild($t$)}} == FALSE}
            \STATE $t \leftarrow $ Generate a new test case with the same strategy
            \ENDWHILE
            
            \STATE $E\leftarrow E \cup t$
            \ENDWHILE
            \STATE Return $E$
            \end{algorithmic}
        }
    }
    \caption{Framework pseudocode for generating valid test cases.}
    \label{FIG:VALID}
\end{figure}

ART algorithms typically generate one test case at a time.
Figure~\ref{FIG:VALID} shows the process for generating a valid test case that does not violate any constraints:
The algorithm first initializes an empty set of executed test cases, $E$. 
It then applies an ART strategy to generate a test case, and checks whether or not it is valid. 
If this test case is not valid, then the same ART strategy repeatedly generates a new test case until a valid one is obtained.
The valid test case can be used to execute the SUT, and is then stored in $E$. 
The algorithm terminates once a specific stopping condition is met.

\begin{table*}[!t]
\centering
\footnotesize
\caption{Subject Programs for the Assessment Framework}
\label{TAB:config-subject-programs}
\resizebox{\textwidth}{!}{
\begin{tabular}{@{}ccccccccccccccc@{}}
\hline
\textbf{ID} & &\textbf{Program} & &\textbf{Description} & &\textbf{Size}(\textit{LOC}) & &\textbf{Input Domain} ($\mathcal{D}$) & &\textbf{Parameters} & &\textbf{Constraints} & &\textbf{Faults}\\\hline
R1 && DRUPAL && web framework && $336,025$ && $2^{47}$ && $47$ && $45$ & &$160$\\\hline
R2 && BUSYBOX && UNIX utilities && $189,722$ && $2^{68}$ && $68$ && $16$ & &$9$\\\hline
R3 && LINUX KERNEL && operation system && $12,594,584$ && $2^{104}$ && $104$ && $83$ && $28$\\\hline
\end{tabular}}
\end{table*}

\subsection{Configurable Systems Selection}

For this preliminary investigation, we selected three real-life, open-source, highly configurable systems as the SUTs:
DRUPAL, BUSYBOX, and LINUX KERNEL,\cite{Medeiros2016,SanchezSPC17}.
All three SUTs are relatively large, both in terms of lines of code (LOC), and the number of parameters. 
DRUPAL \cite{SanchezSPC17} is a modular framework that is commonly used to manage web content;
BUSYBOX \cite{Medeiros2016} contains a range of UNIX utilities within a single executable file; and
LINUX KERNEL \cite{Medeiros2016} is part of a well-known operating system.
Table~\ref{TAB:config-subject-programs} summarizes the SUT details \cite{Huayao2020}.
The three SUTs were used to evaluate the different ART-based testing algorithms, examining how well the methods performed with genuine faults in these large, configurable  (non-numerical) systems.

Each DRUPAL \textit{parameter} corresponds to a module that can be turned on or off to adapt the system's capabilities.
The parameters of BUSYBOX and LINUX KERNEL refer to configuration options that are implemented through conditional compilation by the C preprocessor.
All SUT parameters and constraints were extracted from the relevant documentation and the KConfig files \cite{Medeiros2016,TartlerLSS11}, and are summarized in Table~\ref{TAB:config-subject-programs}.

We used an existing corpus of actual faults that contains details of the specific parameter combinations that trigger each fault \cite{Huayao2020}.
The faults were identified during the SUT development and testing, and most have since been confirmed and corrected.
The number of faults in each SUT is summarized in Table~\ref{TAB:config-subject-programs}.

\subsection{Assessment Setup}

We performed a series of empirical experiments to assess the effectiveness and efficiency of various ART-based approaches in highly configurable (non-numerical) systems.
The experiments examined both FSCS and RRT using five different ART-based strategies:
the original ART, and four enhanced versions (\textit{RF}, \textit{CR}, \textit{KD}, and \textit{LSH}). 
RT was also examined in the experiments.
Because the SUT input domain is binary in each dimension, the DF strategy was not applied in these non-numerical experiments:
Generated test cases could only consist of boundary values within the input domain, such as $(0,0)$ and $(1,0)$
---
if DF were applied to partition such an input domain, no test case could be generated for any sub-domain within the divided input domain.

\subsubsection{Effectiveness Assessment Setup}
The effectiveness of the different approaches was examined from two perspectives: 
the number of test cases needed to identify each fault in each program (\textit{Effectiveness 1}); and
the number of test cases required to detect all faults in each program (\textit{Effectiveness 2}).

\subsubsection{Efficiency Assessment Setup}
We used the time required to detect all faults in each program as the measure of efficiency.

Each fault in the program was represented as ``$\textrm{f}.x$", where $x$ indicates the fault number.
All faults were considered to have equal weight, regardless of the SUT version in which they appeared:
As with previous studies \cite{Huayao2020}, we did not distinguish which version was responsible for the fault.

To measure the effectiveness and efficiency, we generalized the definitions of F-measure and F-time:
$\textrm{F}^{\textrm{f}.x}\textrm{-measure}$ is the number of test cases required to detect fault $\textrm{f}.x$;
$\textrm{F}^\textrm{all}\textrm{-measure}$ is the number of test cases required to detect all faults; and
$\textrm{F}^\textrm{all}\textrm{-time}$ is the time required to detect all faults.

\begin{table*}[!b]
\centering
    \footnotesize
    \caption{\textbf{Mean Number of test cases to detect each fault scenario} for which LSH performance is significantly superior (\ding{52}), indistinguishable (\ding{109}), or significantly inferior (\ding{54})}
    \label{TAB:configurable-sum}
\resizebox{\textwidth}{!}{
\begin{tabular}{@{}ccclccccccccccccccccccc@{}}
\hline
\multirow{2}{*}{\textbf{ART Version}} && \multirow{2}{*}{\textbf{Program}} &  & \multicolumn{3}{c}{\textit{LSH vs. RT}} & \multicolumn{1}{l}{} & \multicolumn{3}{c}{\textit{LSH vs. ART}} & \multicolumn{1}{l}{} & \multicolumn{3}{c}{\textit{LSH vs. RF}} & \multicolumn{1}{l}{} & \multicolumn{3}{c}{\textit{LSH vs. CR}} & \multicolumn{1}{l}{} & \multicolumn{3}{c}{\textit{LSH vs. KD}}\\\cline{5-7}\cline{9-11}\cline{13-15}\cline{17-19}\cline{21-23}
 &  & & & \ding{52} & \ding{109} & \ding{54} & \multicolumn{1}{l}{} & \ding{52} & \ding{109} & \ding{54} & \multicolumn{1}{l}{} & \ding{52} & \ding{109} & \ding{54} & \multicolumn{1}{l}{} & \ding{52} & \ding{109} & \ding{54} & \multicolumn{1}{l}{} & \ding{52} & \ding{109} & \ding{54}\\\hline
\multirow{4}{*}{FSCS Version} && DRUPAL &  & 153 & 7 & 0 &  & 0 & 159 & 1 &  & 0 & 158 & 2 &  & 8 & 152 & 0 &  & 0 & 160 & 0\\
 && BUSYBOX &  & 4 & 5 & 0 &  & 0 & 9 & 0 &  & 0 & 8 & 1 &  & 0 & 9 & 0 &  & 0 & 9 & 0\\
 && LINUX KERNEL &  & 15 & 12 & 1 &  & 1 & 24 & 3 &  & 2 & 24 & 2 &  & 0 & 28 & 0 &  & 1 & 25 & 2\\\cline{3-23}
 && \textit{\textbf{Sum}} &  & 172 & 24 & 1 &  & 1 & 192 & 4 &  & 2 & 190 & 5 &  & 8 & 189 & 0 &  & 1 & 194 & 2\\\hline
\multirow{4}{*}{RRT Version} && DRUPAL &  & 3 & 157 & 0 &  & 0 & 160 & 0 &  & 1 & 159 & 0 &  & 0 & 160 & 0 &  & 0 & 160 & 0\\
 && BUSYBOX &  & 0 & 9 & 0 &  & 0 & 9 & 0 &  & 0 & 9 & 0 &  & 0 & 9 & 0 &  & 0 & 9 & 0\\
 && LINUX KERNEL &  & 1 & 27 & 0 &  & 0 & 28 & 0 &  & 1 & 27 & 0 &  & 1 & 27 & 0 &  & 0 & 28 & 0\\\cline{3-23}
 & &\textit{\textbf{Sum}} &  & 4 & 193 & 0 &  & 0 & 197 & 0 &  & 2 & 195 & 0 &  & 1 & 196 & 0 &  & 0 & 197 & 0\\\hline
\multirow{4}{*}{FSCS + RRT} & &DRUPAL &  & 156 & 164 & 0 &  & 0 & 319 & 1 &  & 1 & 317 & 2 &  & 8 & 312 & 0 &  & 0 & 320 & 0\\
 && BUSYBOX &  & 4 & 14 & 0 &  & 0 & 18 & 0 &  & 0 & 17 & 1 &  & 0 & 18 & 0 &  & 0 & 18 & 0\\
 && LINUX KERNEL &  & 16 & 39 & 1 &  & 1 & 52 & 3 &  & 3 & 51 & 2 &  & 1 & 55 & 0 &  & 1 & 53 & 2\\\cline{3-23}
 && \textit{\textbf{Sum}} &  & {176} & {217} & {1} &  & {1} & {389} & {4} &  & {4} & {385} & {5} &  & {9} & {385} & {0} & & {1} & {391} & {2}\\\hline
\end{tabular}}
\end{table*}

\subsection{Assessment Results}

Here, we present the results of our experiments exploring the effectiveness and efficiency of various ART techniques in non-numerical SUTs. 
We also provide a statistical analysis of the results to allow for a more detailed description of our observations.
All the results are included in the Appendix (see supplementary material); and we have also made them available online \cite{detailLSH}.

\subsubsection{Effectiveness 1 Observations}
The mean number of test cases required to identify each fault in the three configurable SUTs 
---
the $\textrm{F}^{\textrm{f}.x}\textrm{-measure}$
---
is shown in Tables \ref{TAB:fscs-real-drupal} to \ref{TAB:rrt-real-linux}. 
Table \ref{TAB:configurable-sum} presents the number of scenarios for each fault where, compared with each other method, LSH is either 
significantly better (\ding{52}); 
indistinguishable (\ding{109}); or 
significantly worse (\ding{54}).
There were a total of 197 faults across the three SUTs for pairwise comparison:
DRUPAL had 160;
BUSYBOX had nine; and 
LINUX KERNEL had 28. 
Based on the data in these tables, we have the following observations:

\begin{itemize}
    \item[\textit{\textbf{(a)}}]\textit{\textbf{LSH vs. RT:}} 
    In general, LSH-FSCS tends to have a better fault-detection performance, as indicated by the lower F$^{\textrm{f}.x}$-measures, than RT. 
    However, there is an exception (i.e., f.15 in LINUX KERNEL), where RT has the better performance. 
    Almost all $p$-values in the DRUPAL statistical analyses were lower than 0.01 (except for f.36, f.56, f.72, f.74, f.101, f.138, and f.144), indicating significant differences.
    Some $\hat{\textrm{A}}_{12}$ values were above 0.80 (including f.127, f.130, and f.151). 
    Four out of nine BUSYBOX $p$-values indicated significant differences, and the $\hat{\textrm{A}}_{12}$ values were around 0.50. 
    Most LINUX KERNEL $p$-values also show significant differences, with $\hat{\textrm{A}}_{12}$ ranging from 0.46 to 0.61.
    Apart from f.15, the $\hat{\textrm{A}}_{12}$ values are greater than 0.50. 
    LSH-RRT and RT have similar $\textrm{F}^{\textrm{f}.x}\textrm{-measure}$ results, with most $p$-values indicating no significant differences between the two.
    In summary, the LSH-FSCS fault-detection performance is generally as good as, or much better than, RT; and LSH-RRT and RT have comparable performance.

    \item[\textit{\textbf{(b)}}]\textit{\textbf{LSH vs. ART:}} 
    LSH-FSCS and FSCS, and LSH-RRT and RRT, have comparable fault-detection performance.
    Although most $p$-values indicate no significant differences in the compared methods ($389/394=98.73\%$), 
    there are times when LSH-ART had worse performance ($4/394 = 1.02\%$) than ART.
    Overall, LSH-ART has similar fault-detection effectiveness to the original ART, but the original ART can sometimes perform better.

    \item[\textit{\textbf{(c)}}]\textit{\textbf{LSH vs. RF:}} 
    LSH-ART and RF-ART have similar fault-detection performance, with both the $p$-values and the $\hat{\textrm{A}}_{12}$ data indicating no significant difference between the two ($385/394=97.72\%$).
    There are also instances, however, where one strategy performs better than the other
    ---
    LSH performs better in four examples, while RF performs better in five examples.
    Overall, the fault-detection effectiveness of LSH-ART and RF-ART are similar.

    \item[\textit{\textbf{(d)}}]\textit{\textbf{LSH vs. CR:}} 
    The LSH vs. CR comparison is very similar to the LSH vs. RF one, and there are no cases where CR is significantly superior to LSH.
    Compared with CR, LSH has comparable or better fault-detection effectiveness.
    
    \item[\textit{\textbf{(e)}}]\textit{\textbf{LSH vs. KD:}} 
    LSH (both FSCS and RRT) and KD have very similar performance, with most of the $p$-values and $\hat{\textrm{A}}_{12}$ data indicating no significant differences:
    LSH and KD have comparable fault-detection effectiveness. 
\end{itemize}

\subsubsection{Effectiveness 2 Observations}
Table \ref{TAB:fmeasure-configurable} shows the mean number of test cases required to detect all faults (F${}^{\rm all}$-measures) in the three configurable SUTs.
Based on an analysis of Table \ref{TAB:fmeasure-configurable}, we have the following observations:

\begin{itemize}
    \item[\textit{\textbf{(a)}}]\textit{\textbf{LSH vs. RT:}} 
    LSH-FSCS has lower F${}^{\rm all}$-measure scores than RT, indicating better fault-detection effectiveness, with the $p$-values all indicating that the differences are significant, and the $\hat{\textrm{A}}_{12}$ scores are greater than 0.50.
    LSH-RRT also has lower F${}^{\rm all}$-measure data than RT, but this difference is less than LSH-FSCS vs. RT, with the $\hat{\textrm{A}}_{12}$ data at around 0.50. 
    Overall, LSH has similar or better fault-detection effectiveness than RT.

    \item[\textit{\textbf{(b)}}]\textit{\textbf{LSH vs. ART:}} 
    Although LSH has lower F${}^{\rm all}$-measure scores than the original ART in all scenarios, for both FSCS and RRT, there are no significant differences between the two in terms of $p$-values,
    The $\hat{\textrm{A}}_{12}$ scores are also roughly 0.50.
    Overall, LSH has similar fault-detection effectiveness to the original ART.

    \item[\textit{\textbf{(c)}}]\textit{\textbf{LSH vs. RF:}}
    RF-FSCS has lower F${}^{\rm all}$-measures than LSH-FSCS for BUSYBOX and LINUX KERNEL, but LSH has lower F${}^{\rm all}$-measures than RF in the remaining cases.
    The $p$-values show no significant difference between the two. 
    Overall, LSH has similar fault-detection effectiveness to RF.

    \item[\textit{\textbf{(d)}}]\textit{\textbf{LSH vs. CR:}} 
    In all cases, LSH has lower F${}^{\rm all}$-measures than CR, which means that LSH can detect all the faults with fewer test cases. 
    The statistical analyses indicate that LSH outperforms CR in most cases.
    Overall, compared with CR, LSH has similar or better fault-detection effectiveness.

    \item[\textit{\textbf{(e)}}]\textit{\textbf{LSH vs. KD:}} 
    Based on the F${}^{\rm all}$-measure data and statistical analyses, compared with KD, LSH has similar or better fault-detection effectiveness, with the differences between the two not being significant.
\end{itemize}

\begin{table*}[!t]
\centering
\footnotesize
\caption{Mean \textbf{F$^{\textrm{all}}$-measure} Results and Statistical Pairwise Comparisons of LSH for \textbf{Configurable Programs}}
\label{TAB:fmeasure-configurable}
\resizebox{\textwidth}{!}{
\begin{tabular}{@{}cccrrrrrrrrrrrrr@{}}
\hline
\multirow{2}{*}{\textbf{ID}} & \multirow{2}{*}{\textbf{Program}} & \multirow{2}{*}{\textbf{ART Version}} &  & \multicolumn{6}{c}{\textbf{Methods}} &  & \multicolumn{5}{c}{\textit{LSH}}\\\cline{5-10}\cline{12-16}
 &  &  &  & \textit{RT} & \textit{ART} & \textit{RF} & \textit{CR} & \textit{KD} & \textit{LSH} &  & \textit{vs. RT} & \textit{vs. ART} & \textit{vs. RF} & \textit{vs. CR} & \textit{vs. KD}\\\hline
\multirow{2}{*}{R1} & \multirow{2}{*}{DRUPAL} & FSCS Version&  & 173.43  & 54.08  & 54.07  & 62.63  & 54.04  & 53.61  &   & \ding{52} (0.96)  & \ding{109} (0.51)  & \ding{109} (0.50)  & \ding{52} (0.58)   & \ding{109} (0.51)\\
 &  & RRT Version&  & 173.43 & 169.94 & 173.01 & 175.26  & 172.30 & 164.84 &   & \ding{52} (0.52)  & \ding{109} (0.51)  & \ding{52} (0.52)  & \ding{52} (0.52)  & \ding{109} (0.52) \\\hline
\multirow{2}{*}{R2} & \multirow{2}{*}{BUSYBOX} & FSCS Version&  & 34.04  & 29.63  & 28.78  & 30.62   & 29.34  & 29.50  &   & \ding{52} (0.53)  & \ding{109} (0.50)  & \ding{109} (0.49)  & \ding{109} (0.50)   & \ding{109} (0.49)\\
 &  & RRT Version&  & 34.04 & 33.54 & 34.61 & 33.51  & 33.21 & 32.90 &   & \ding{109} (0.51)  & \ding{109} (0.51)  & \ding{109} (0.52)  & \ding{109} (0.51)  & \ding{109} (0.50)\\\hline
\multirow{2}{*}{R3} & \multirow{2}{*}{LINUX KERNEL} & FSCS Version&  & 87.95  & 71.59  & 69.05  & 71.82   & 72.70  & 70.58  &   & \ding{52} (0.59)  & \ding{109} (0.49)  & \ding{54} (0.49)  & \ding{109} (0.50)   & \ding{109} (0.50)\\
 &  & RRT Version&  & 87.95 & 87.52 & 88.08  & 90.52  & 89.48 & 82.87 &   & \ding{52} (0.52)  & \ding{52} (0.52)  & \ding{52} (0.52)  & \ding{52} (0.53)  & \ding{52} (0.53)\\\hline
\end{tabular}}
\end{table*}

\begin{table*}[!t]
\centering
\footnotesize
\caption{Mean \textbf{F$^{\textrm{all}}$-time} Results (ms) and Statistical Pairwise Comparisons of LSH for \textbf{{Configurable Programs}}}
\label{TAB:ftime-configurable}
\resizebox{\textwidth}{!}{
\begin{tabular}{@{}cccrrrrrrrrrrrrr@{}}
\hline
\multirow{2}{*}{\textbf{ID}} & \multirow{2}{*}{\textbf{Program}} & \multirow{2}{*}{\textbf{ART Version}} &  & \multicolumn{6}{c}{\textbf{Methods}} &  & \multicolumn{5}{c}{\textit{LSH}}\\\cline{5-10}\cline{12-16}
 &  &  &  & \textit{RT} & \textit{ART} & \textit{RF} & \textit{CR} & \textit{KD} & \textit{LSH} &  & \textit{vs. RT} & \textit{vs. ART} & \textit{vs. RF} & \textit{vs. CR} & \textit{vs. KD}\\\hline
\multirow{2}{*}{R1} & \multirow{2}{*}{DRUPAL} & FSCS Version&  & 415.21 & 1087.04 & 1097.49 & 1113.11 & 1103.50 & 1112.04 &  & \ding{54} (0.06) & \ding{109} (0.49) & \ding{109} (0.49) & \ding{52} (0.54) & \ding{109} (0.50)\\
 &  & RRT Version&  & 415.21 & 411.14 & 421.81 & 427.75 & 416.83 & 398.63 &  & \ding{109} (0.51) & \ding{109} (0.51) & \ding{52} (0.52) & \ding{52} (0.53) & \ding{109} (0.51)\\\hline
\multirow{2}{*}{R2} & \multirow{2}{*}{BUSYBOX} & FSCS Version&  & 51.94 & 416.62 & 409.70 & 444.65 & 414.69 & 403.33 &  & \ding{54} (0.04) & \ding{109} (0.50) & \ding{109} (0.50) & \ding{109} (0.51) & \ding{109} (0.49)\\
 &  & RRT Version&  & 51.94 & 51.44 & 53.45 & 52.02 & 50.43 & 50.46 &  & \ding{109} (0.51) & \ding{109} (0.51) & \ding{109} (0.52) & \ding{109} (0.51) & \ding{109} (0.50)\\\hline
\multirow{2}{*}{R3} & \multirow{2}{*}{LINUX KERNEL} & FSCS Version&  & 408.04 & 3169.89 & 3127.62 & 3188.52 & 3294.56 & 3097.64 &  & \ding{54} (0.01) & \ding{109} (0.49) & \ding{109} (0.49) & \ding{109} (0.50) & \ding{109} (0.51)\\
 &  & RRT Version&  & 408.04 & 409.33 & 426.52 & 439.51 & 427.40 & 398.83 &  & \ding{109} (0.50) & \ding{109} (0.51) & \ding{52} (0.53) & \ding{52} (0.54) & \ding{52} (0.52)\\\hline
\end{tabular}}
\end{table*}

\begin{table*}[!b]
\caption{Summary of Statistical Analysis from Numerical and Non-Numerical Experiments.}
\label{TAB:StatisticalAnalysissSummary}
    \resizebox{\textwidth}{!}{
        \renewcommand{\arraystretch}{1}
        \begin{tabular}{@{}cllccccccccccccccccccccccc@{}}
        \hline
        \multirow{2}{*}{\textbf{Domain type}} & \multirow{2}{*}{\textbf{Experiment type}} & \multirow{2}{*}{\textbf{Source}} & \multicolumn{3}{c}{\textit{LSH vs. RT}} && \multicolumn{3}{c}{\textit{LSH vs. ART}}&& \multicolumn{3}{c}{\textit{LSH vs. RF}} && \multicolumn{3}{c}{\textit{LSH vs. CR}} && \multicolumn{3}{c}{\textit{LSH vs. DF}} & \multicolumn{1}{l}{} & \multicolumn{3}{c}{\textit{LSH vs. KD}} \\ \cline{4-6} \cline{8-10} \cline{12-14} \cline{16-18} \cline{20-22} \cline{24-26} 
         & & & \ding{52} & \ding{109} & \ding{54} && \ding{52} & \ding{109} & \ding{54} && \ding{52} & \ding{109} & \ding{54} && \ding{52} & \ding{109} & \ding{54} && \ding{52} & \ding{109} & \ding{54} && \ding{52} & \ding{109} & \ding{54} \\ \hline
        \multirow{6}{*}{Numerical} & F-measure (simulations)& Table \ref{TAB:SimulationCollection1} & 56& 150& 46&& 21& 220& 11&& 107 & 145& 0 && 104 & 145& 3 && 31& 208& 13&& 13& 230& 9 \\
         & F-measure (real programs)& Table \ref{TAB:realCollectionFmeasure} & 22& 20 & 4 && 0 & 42 & 4 && 17& 26 & 3 && 17& 25 & 4 && 13& 30 & 3 && 0 & 45 & 1 \\
         & P-measure (simulations)& Table \ref{TAB:SimulationCollection1-p} & 105 & 64 & 83&& 61& 143& 48&& 165 & 69 & 18&& 160 & 81 & 11&& 77& 115& 60&& 40& 152& 60\\
         & P-measure (real programs)& Table \ref{TAB:realCollectionPmeasure} & 43& 2& 1 && 20& 17 & 9 && 19& 17 & 10&& 20& 19 & 7 && 13& 22 & 11&& 17& 14 & 15\\
         & F-time (real programs) & Table \ref{TAB:realCollectionFtime} & 1 & 1& 44&& 46& 0& 0 && 38& 5& 3 && 37& 6& 3 && 46& 0& 0 && 34& 2& 10\\ \cline{2-26} 
         & \multicolumn{2}{c}{\textit{\textbf{Sum}}}& 227 & 237& 178 && 148 & 422& 72&& 346 & 262& 34&& 338 & 276& 28&& 180 & 375& 87&& 104 & 443& 95\\ \hline

        \multirow{4}{*}{Non-Numerical} & $\textrm{F}^{\textrm{f}.x}\textrm{-measure}$ & Table \ref{TAB:configurable-sum} & 176 & 217& 1 && 1 & 389& 4&& 4& 385& 5&& 9& 385& 0 && --& -- & --&& 1 & 391& 2 \\
         & $\textrm{F}^{\textrm{all}}\textrm{-measure}$ & Table \ref{TAB:fmeasure-configurable} & 5 & 1& 0 && 1 & 5& 0 && 2 & 3& 1 && 3 & 3& 0 && --& -- & --&& 1 & 5& 0 \\
         & $\textrm{F}^{\textrm{all}}\textrm{-time}$& Table \ref{TAB:ftime-configurable} & 0 & 3& 3 && 0 & 6& 0 && 2 & 4& 0 && 3 & 3& 0 && --& -- & --&& 1 & 5& 0 \\ \cline{2-26} 
         & \multicolumn{2}{c}{\textit{\textbf{Sum}}}& 181 & 221& 4 && 2 & 400& 4&& 8& 392& 6&& 15& 391&0 && --& -- & --&& 3 & 401& 2 \\ \hline

        \end{tabular}
    }
\end{table*}

\subsubsection{Efficiency Observations}

Based on an analysis of the time required to detect all faults in the three configurable SUTs (Table \ref{TAB:ftime-configurable}), we have the following observations:
\begin{itemize}
    \item[\textit{\textbf{(a)}}]\textit{\textbf{LSH vs. RT:}} 
    LSH-FSCS takes longer than RT to detect the faults, with the statistical analyses showing significant differences between the two methods:
    RT significantly outperforms LSH-FSCS, as indicated by the $\hat{\textrm{A}}_{12}$ scores being less than 0.06. 
    LSH-RRT has lower F${}^{\rm all}$-times than RT, but the difference between the two is insignificant: The $p$-values were 0.01, and the $\hat{\textrm{A}}_{12}$ scores were around 0.50. 
    Overall, RT is more efficient at detecting the faults.
    
    \item[\textit{\textbf{(b)}}]\textit{\textbf{LSH vs. ART:}} 
    For both FSCS and RRT, LSH and the original ART have similar F${}^{\rm all}$-times, with no significant differences, as indicated by the $\hat{\textrm{A}}_{12}$ value of 0.50: 
    LSH and the original ART have similar fault-detection efficiency.

    \item[\textit{\textbf{(c)}}]\textit{\textbf{LSH vs. RF:}} 
    LSH-FSCS has a larger F${}^{\rm all}$-time than RF-FSCS in DRUPAL, but in other cases, LSH has lower F${}^{\rm all}$-time scores than RF.
    Although the $p$-values do not show a significant difference between LSH-FSCS and RF-FSCS, they do indicate one between LSH-RRT and RF-RRT (except for BUSYBOX). 
    Overall, compared with RF, LSH has similar or better fault-detection efficiency.

    \item[\textit{\textbf{(d)}}]\textit{\textbf{LSH vs. CR:}} 
    LSH takes less time to detect all faults than CR. 
    Overall, LSH has better fault-detection efficiency than CR.

    \item[\textit{\textbf{(e)}}]\textit{\textbf{LSH vs. KD:}} 
    The $p$-values for comparisons between LSH and KD indicated no significant difference (except for LSH-RRT on LINUX KERNEL):
    LSH and KD have comparable failure-detection efficiency.
\end{itemize}

\subsubsection{Analysis and Summary}

From the perspectives of testing effectiveness and efficiency, this investigation has shown that both LSH and other ART-based strategies can be applied to fault detection in highly configurable (non-numerical) software.
Based on the analyses of the results and observations, we have the following conclusions:

\textbf{\textit{(a) Effectiveness:}}
    LSH-FSCS can detect faults in these programs with fewer test cases than RT, and its effectiveness is not significantly different from other ART variants: 
    All ART variants were able to detect all faults with a relatively small number of test cases.
    This can be explained according to two reasons.
    First, the failure-causing inputs in these programs were located at the boundaries of the input domain.
    Previous studies~\cite{Huang2019} have shown that ART tends to generate test cases around the input domain boundary. 
    Accordingly, all ART variants could detect the faults with fewer test cases.
    Second, as the input domain dimensionality increases, the Euclidean distance may cease to measure similarity/dissimilarity effectively \cite{HuangCST20}, which reduces the ART testing effectiveness.

    Due to the high dimensionality of the programs (\texttt{LINUX KERNEL} has $d=104$), the performance of all RRT variants is similar to that of RT.
    The reason for this relates to the initial exclusion radius being about $2.53$ for \texttt{LINUX KERNEL} (according to Eq. (\ref{EQ:RRT})), and gradually decreases as testing progresses. 
    The distance between the two most-distant test cases ($(0,0,\cdots,0)$ and $(1,1,\cdots,1)$), however, is about $10.20$. 
    This leads to a large number of test cases being generated outside the exclusion regions, which is similar to the RT generation method \cite{chan2007controlling}, thus making their performance very similar.

    \textbf{\textit{(b) Efficiency:}}
    Due to the parameter constraints in these highly configurable programs, only valid test cases were considered during the test case generation process. 
    However, the determination of valid test cases can be very time-consuming. 
    We used \texttt{LINUX KERNEL} (which has the highest number of parameters and constraints) to examine this, by generating 500 test cases using the FSCS versions of ART, KD, and LSH.
    Based on 3000 trials, we recorded the average time taken to generate $k=10$ valid candidate test cases ($T_{\textit{\textrm{GEN}}}$), and the average time taken to select the next test case from these candidates ($T_{\textit{\textrm{NNS}}}$).
    The generation times were:
    $T_{\textit{\textrm{GEN}}}(\textit{\textrm{ART}}) = 35,631.08$ ms; $T_{\textit{\textrm{GEN}}}(\textit{\textrm{KD}})= 35,817.95$ ms; and $T_{\textit{\textrm{GEN}}}(\textit{\textrm{LSH}}) = 35,948.77$ ms.
    The selection times were:
    $T_{\textit{\textrm{NNS}}}(\textit{\textrm{ART}}) = 468.31$ ms; $T_{\textit{\textrm{NNS}}}(\textit{\textrm{KD}}) = 443.10$ ms; and $T_{\textit{\textrm{NNS}}}(\textit{\textrm{LSH}}) = 22.59$ ms.
    These results clearly show that $T_{\textit{\textrm{GEN}}}$ is much greater than $T_{\textit{\textrm{NNS}}}$ for all ART variants. 
    Consequently, even though FSCS-LSH takes the least amount of time for the NNS process, overall, the total time taken is similar to the other ART approaches. 

    Because the generation process for RRT-LSH is similar to RT, their generation times are also similar. 
    However, unlike FSCS, RRT does not need to generate $k$ candidate test cases for each new test case, and thus the overall time required is less.
  
\textbf{\textit{(c) Comparison with numerical experiments:}}
    Based on the results of this study (i.e., non-numerical experiments), the LSH testing effectiveness is found to be superior to RT.
    It is also better than the other ART algorithms, in some cases.
    This finding is consistent with the results of the earlier numerical experiments  (Tables \ref{TAB:realCollectionFmeasure} to \ref{TAB:realCollectionPmeasure}).
    Furthermore, due to the characteristics of the programs studied in this section, the performance difference between LSH and the other ART algorithms has become less.
    The LSH testing efficiency was better than the other ART algorithms in the earlier experiments.
    In this section, however, although the LSH nearest-neighbor search time is consistent with the numerical experiments (Table \ref{TAB:realCollectionFtime}), when considering the time required to determine valid test cases, the LSH performance becomes similar to the other ART variants.

This section has reported on an investigation into the potential for our proposed LSH strategy to effectively and efficiently detect faults in non-numerical programs.
As analyzed and discussed above, our method can detect faults in these programs, and, compared with other ART variants, demonstrates equal or better performance.
In particular, LSH maintains its cost-effectiveness when searching for (approximate) nearest neighbors, which is consistent with its performance in numerical input domains. 
Table \ref{TAB:StatisticalAnalysissSummary} summarizes the statistical results for both the numerical and the non-numerical input domains. 
Compared with RT, LSH has comparable or even better performance in both numerical and non-numerical programs, in most cases. 
Furthermore, LSH has superior performance to the other ART variants. 
In conclusion, the performance of LSH-ART in non-numerical input domains aligns with its performance in numerical input domains, surpassing both RT and other ART variants.

\section{Differences between Failures and Faults}
\label{SECTION:discussions}

In this section, we discuss some differences between software failures and faults.

\begin{figure*}[!b]
\centering
\graphicspath{{Graphs/}}
\resizebox{\textwidth}{!}{
    \subfigure[FSCS version]
    {
        \includegraphics[width=0.48\textwidth]{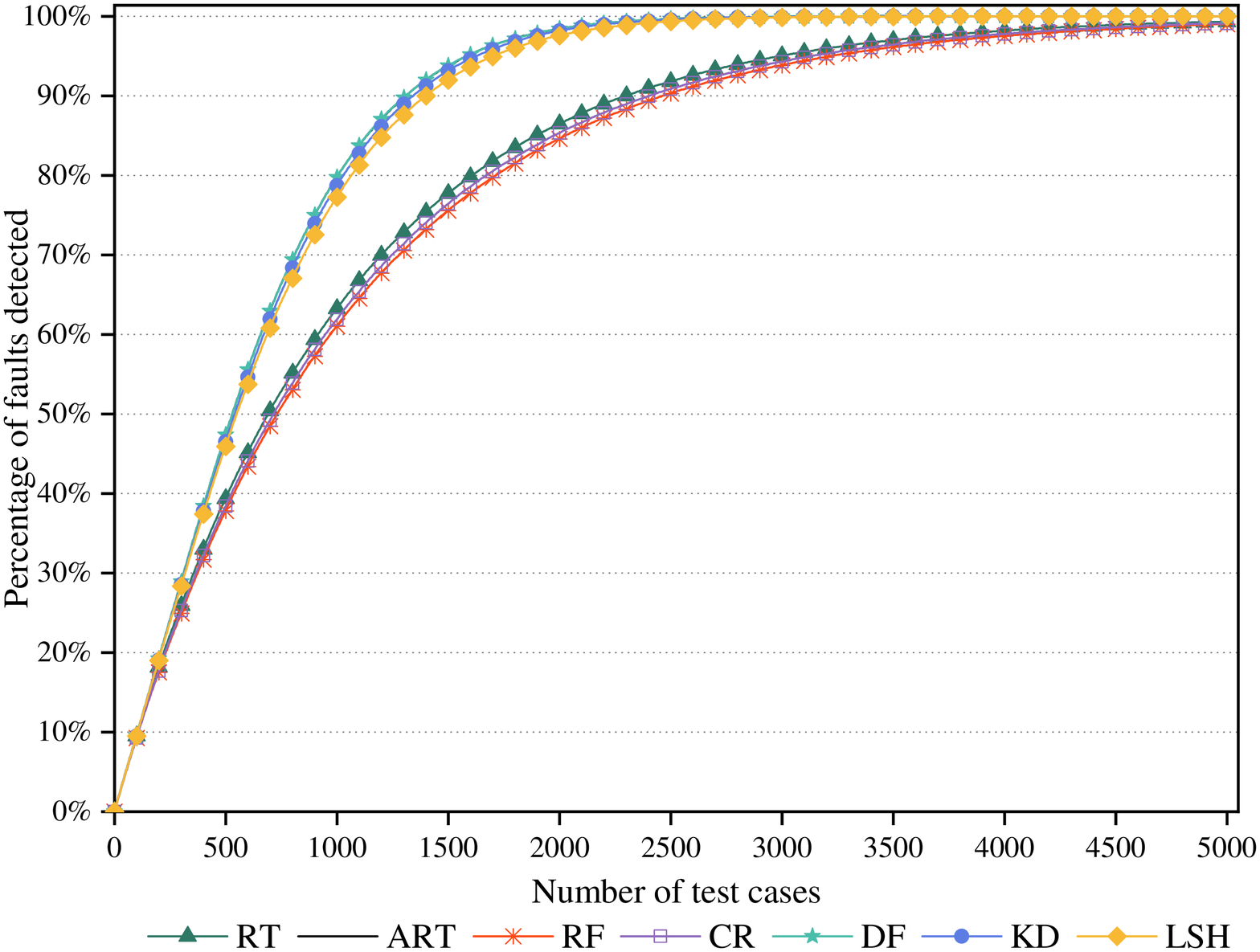}
        \label{FIG:FDE1}
    }
    \subfigure[RRT version]
    {
        \includegraphics[width=0.48\textwidth]{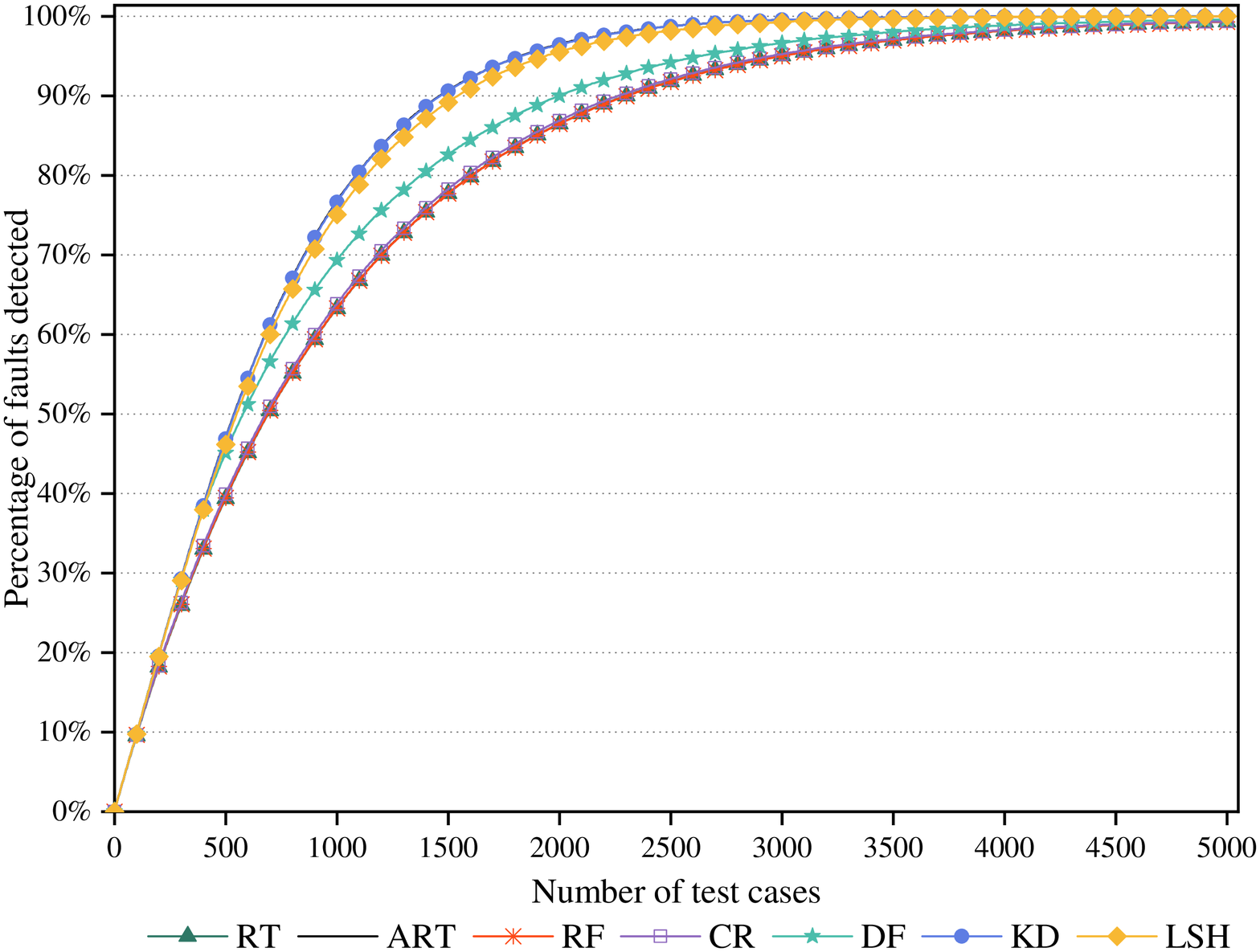}
        \label{FIG:FDE2}
    }}
    \caption{Percentage of faults detected by different test-case generation approaches, for various test set sizes.}
    \label{FIG:FDE}
\end{figure*}

\begin{figure*}[!t]
\centering
\graphicspath{{Graphs/}}
\resizebox{\textwidth}{!}{
    \subfigure[FSCS version]
    {
        \includegraphics[width=0.48\textwidth]{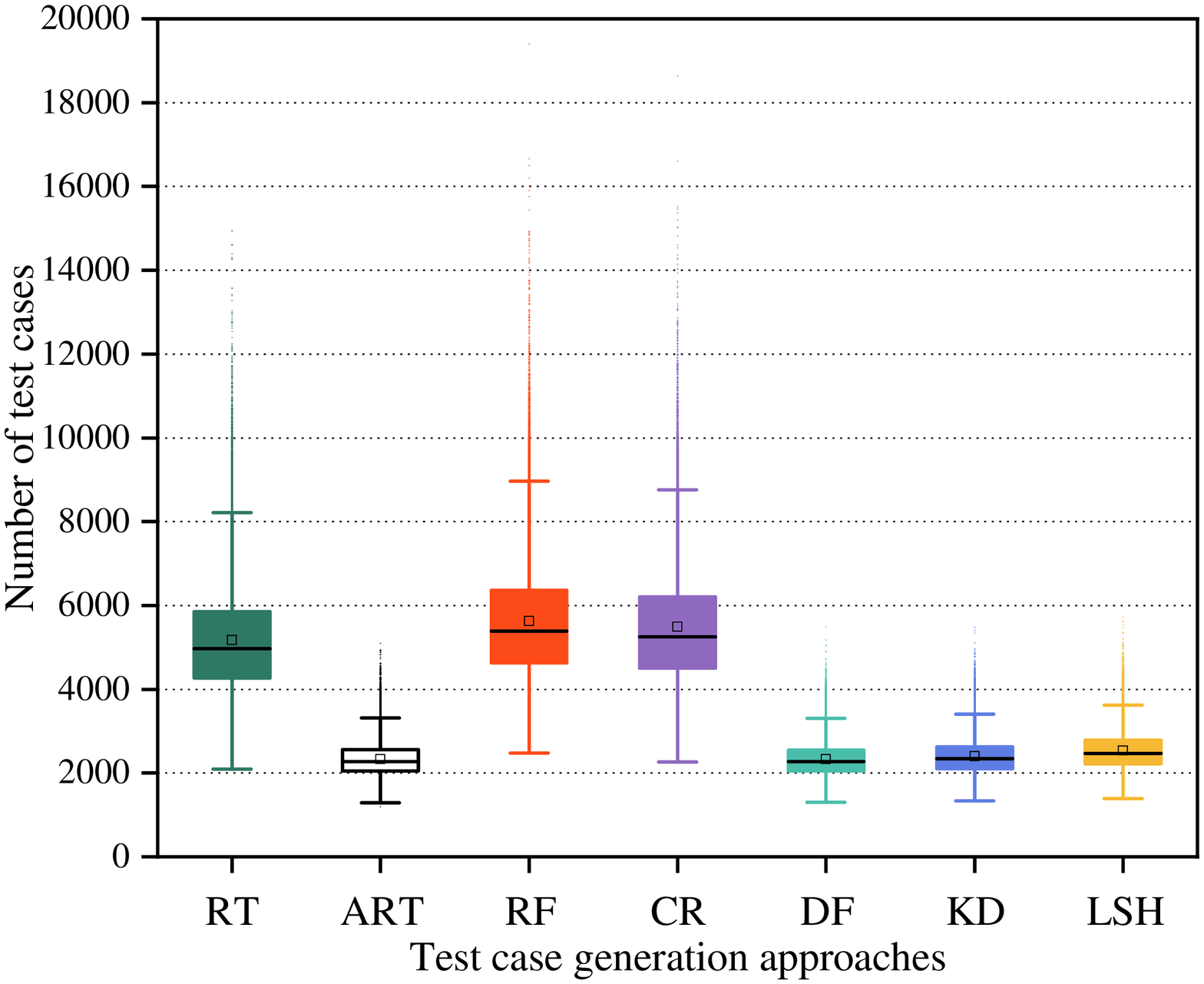}
        \label{FIG:cover1}
    }
    \subfigure[RRT version]
    {
        \includegraphics[width=0.48\textwidth]{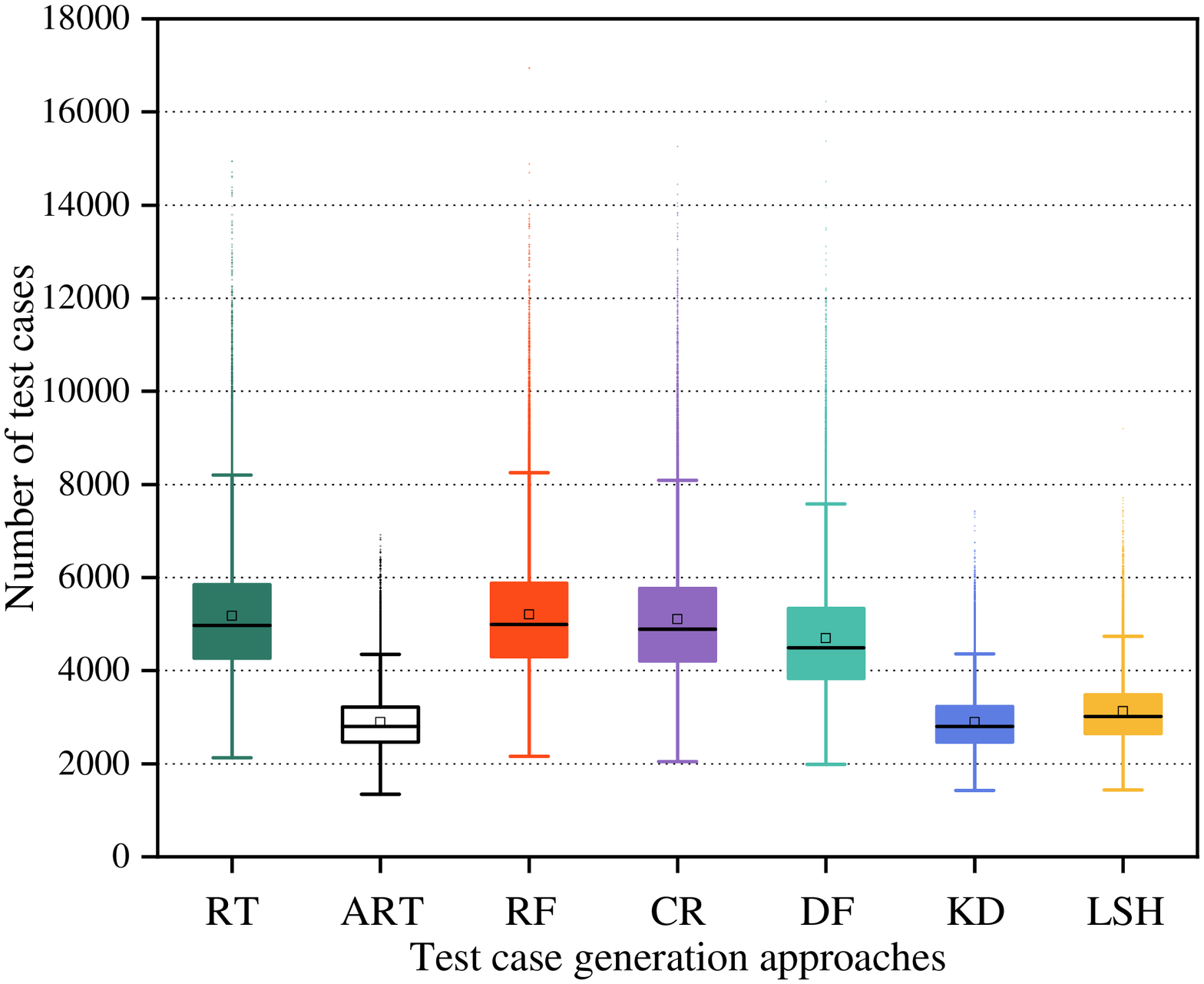}
        \label{FIG:cover2}
    }}
    \caption{The number of test cases required to find all faults, for different test-case generation approaches.}
    \label{FIG:covered}
\end{figure*}

\subsection{Software Failures versus Software Faults}
According to the IEEE~\cite{STD2010}, a software developer may introduce a \emph{fault} (\emph{defect} or \emph{bug}) in the software, through some mistake that they make.
When executing this software (SUT) with a test case $t$, a software \emph{failure} may be produced
---
for example, the SUT may behave unexpectedly:
The output/behavior with $t$ is different from that of the test oracle \cite{Weyuker1982,barr2014oracle,zhou2018introduction}.
Generally speaking, the presence of a failure means that the SUT has a fault(s).
However, the presence of faults does not mean that the software failures can be revealed:
The triggering conditions may not be satisfied~\cite{STD2010}.

As discussed in Section \ref{SECTION:background}, test cases that can trigger software failures are called \emph{failure-causing inputs}.
The set of all failure-causing inputs is called the \emph{failure region}(s).
As noted by Chan et al. \cite{Chan1996},
a software fault may result in a single failure region (such as the block pattern or a strip pattern);
or many failure regions (for example, the point failure pattern).
Similarly, multiple software faults may also result in a single failure region (depending on the inputs that trigger them) or multiple failure regions (if the inputs that trigger the faults are independent).
In other words, there is no obvious relationship between the number of faults and the number of failure regions.

We adopted three representative failure patterns (block, strip, and point) in the framework for our simulation-based study.
Each failure pattern may be caused by a single software fault or multiple faults in actual programs:
Each simulated failure pattern may represent diverse software faults.
We also considered two evaluation frameworks in the empirical study, one for the F-measure, and one for the P-measure.
The F-measure framework seeded one to nine faults into the program source code, constructing one or more failure regions with relatively small failure rates for each program.
These empirical study failure patterns were not limited to the block, strip, and point failure patterns.
The P-measure framework used a mutation tool to construct many mutants, producing many different failure patterns and failure rates.
In summary, both simulation and empirical study frameworks provided diverse faults (or simulated faults) with diverse failure rates.

\subsection{Failure Detection versus Fault Detection}
Failure detection is said to take place when the actual output/behavior of the SUT, by executing test case $t$, is different from expectation
---
the output/behavior is different from that of the oracle.
In this case, $t$ is also called a failure-causing input.
In contrast, {\em fault} detection refers to detection/localization of the specific fault in the SUT:
This focuses on identifying the reason for the observed failure(s), and thus relates to software debugging (i.e., \emph{software fault localization}) \cite{Wong2016}.

In our simulation framework, when a test case was generated inside a constructed failure region, a failure was said to be detected.
As discussed above, a failure region may represent multiple faults, but the concept of fault detection plays no role in the simulations.
In our empirical study framework, when the SUT output for a test case was different from its expected output, a failure was considered to have been found.
At this point, the SUT was shown to have some faults, however, the reason(s) or location(s) leading to the fault was not yet known
---
processes such as fault localization or debugging would need to be carried out before the answers to this could be determined.

Adaptive Random Testing (ART) was motivated by the goal of more quickly finding evidence of a problem in the SUT.
This means that the ART algorithms focus on finding the first failure, often with an expectation that testing will stop once this has been found \cite{Chen2010,Huang2019}.
This also explains the popularity of the F-measure
---
the expected number of test case executions to detect the first failure
---
in ART studies.
Expanding the current study to examine ART performance for more failures or faults, beyond the first found, was therefore not considered:
Once ART finds the first failure of an SUT, a different strategy should then be used to find other failures or faults in that SUT.

Sections \ref{SECTION:SIM-P} and \ref{SECTION:ES-P} provided the results of our simulations and empirical studies involving multiple failures.
We also conducted a preliminary, simulation-based investigation into the number of faults (failures, in fact) that our proposed approach could identify.
This involved a new simulation, with a two-dimensional input domain, in which we randomly constructed 100 equally-sized block-pattern failure regions $\theta_i=1.0\times 10^{-3}~(1\leq i \leq 100)$.
Each failure region corresponded to a single fault
---
once the failure region was detected, the fault was considered to be identified.
All failure regions were independent of each other, thus overlapping regions were possible. To overcome randomness, we repeated the process of random failure region construction 100 times, and independently ran 1000 sets of test cases for each technique, resulting in 100,000 data points.
We adopted two evaluation metrics related to the number of faults:
(1) the number of faults identified by a fixed number of test cases; and
(2) the number of test cases required to detect all faults.

Figure \ref{FIG:FDE} shows the percentage of faults detected by the different test-case generation approaches, for various test set sizes.
According to the results, it can be observed that:
\begin{itemize}
    \item
    When the number of executed test cases was small (for example, $n \leq 200$) or when the number of test cases was large (for example, $n \geq 4000$), LSH had very similar performance to all other approaches (RT, ART, CR, RF, DF, and KD), for both FSCS and RRT versions.

    \item
    For medium amounts of test cases ($200 < n < 4000$), LSH was better than RT, RF, and CR;
    and had a very similar performance to ART and KD, overall, for both FSCS and RRT versions.

    \item
    LSH-FSCS had a similar performance to DF; but LSH-RRT had better performance than DF.
\end{itemize}

Figure \ref{FIG:covered} presents the number of test cases required to detect all faults for the different test-case generation approaches.
The figure shows the distribution of the 100,000 data points (100 rounds of failure-region construction, each with 1000 generated test cases):
Each box plot shows the mean (square in the box);
median (line in the box);
upper and lower quartiles (lines above and below the box);
and minimum and maximum values, for each approach.
From the data, it can be seen that:
\begin{itemize}
    \item
    LSH-FSCS performs similarly to ART, DF, and KD;
    but requires fewer test cases to identify all faults than RT, RF, and CR.

    \item
    LSH-RRT has a similar performance to LSH-FSCS, and also outperforms DF.
\end{itemize}
Overall, the observations from Figure \ref{FIG:covered} are consistent with those from Figure \ref{FIG:FDE}.

\subsection{Failure Pattern versus Fault Pattern}

As discussed in Section \ref{SECTION:background}, a failure pattern refers to the distributions of failure-causing inputs over the input domain, including both their locations and geometric shapes.
Failure patterns have been categorized into three types:
block;
strip; and
point \cite{Chan1996}.
Although not labeled patterns, different types of faults have been identified and categorized
\cite{DiGiuseppe2015,Smith1992,Hayes1994,Kung1998,Hayes2011}.
Among the various taxonomies,  Hayes et al. \cite{Hayes2011} provide a detailed hierarchy of faults based on their location and usage.

As discussed, an aim of ART is to (quickly) find the first failure.
The focus of our investigation has been to balance the tradeoff between ART effectiveness and efficiency by replacing the NN search with an ANN search.
Although investigation of the relationships between failure patterns and fault patterns is beyond the scope of this study, it would be very interesting to examine such things, which we look forward to pursuing in our future work.

\section{Related Work}
\label{SECTION:relatedWork}

This section briefly summarizes some related work about ART.
As explained by Huang et al.~\cite{Huang2019},
in addition to the \textit{Select-Test-From-Candidates Strategy} (STFCS),
there are five other ART implementation strategies:
\textit{Partitioning-Based Strategy} (PBS);
\textit{Test-Profile-Based Strategy} (TPBS);
\textit{Quasi-Random Strategy} (QRS);
\textit{Search-Based Strategy} (SBS); and
\textit{Hybrid Strategies} (HSs).

\subsection{Partitioning-Based Strategy (PBS)}

A \textit{Partitioning-Based Strategy} (PBS) has two components:
The \textit{partitioning schema}; and
the \textit{subdomain selection}.
The partitioning schema defines how to partition the input domain $\mathcal{D}$ into $m$ disjoint subdomains
$\mathcal{D}_1,\mathcal{D}_2,\cdots, \mathcal{D}_m~(m > 1)$,
such that:
$\mathcal{D}_i \bigcap \mathcal{D}_j = \emptyset~(1\leq i \neq j \leq m)$, and
$\mathcal{D}=\mathcal{D}_1\bigcup \mathcal{D}_2 \bigcup \cdots \bigcup \mathcal{D}_m$.
The subdomain-selection component defines how to choose the subdomain in which the next test case will be generated.

The partitioning schema can be \textit{static} \cite{Chen2004a,Sabor2015}, or \textit{dynamic} \cite{Chen2004b,Chen2006,Chow2013}.
A static schema means that the input domain is divided before test-case generation begins, and then no further partitioning takes place once testing begins.
A dynamic schema, in contrast, divides the input domain dynamically, often as each new test case is generated.

Many criteria exist to support the subdomain selection, including choosing the largest \cite{Chen2004a}, or the one with the least number of already-generated test cases \cite{Chow2013}.

\subsection{Test-Profile-Based Strategy (TPBS)}
A \textit{Test-Profile-Based Strategy} (TPBS)~\cite{Liu2011} generates test cases based on a well-designed test profile, not uniform distribution, dynamically updating the profile after each test-case generation.
A test profile can be considered the distribution of selection probability for all test inputs in the input domain $\mathcal{D}$, with inputs in different locations  (potentially) having different selection probabilities.
When a test case is executed without revealing failure, its selection probability is then assigned a value of $0$ \cite{Liu2011}. 

According to the principles of ART, a test profile should be adjusted to satisfy the following criteria \cite{Liu2011}:
(1) The farther away a test input is from previously-executed test cases, the {\em higher} the selection probability that it is assigned should be;
(2) The closer a test input is to the previously-executed test cases, the {\em lower} the selection probability that it is assigned should be;
and (3) The probability distribution should be dynamically adjusted to maintain these two features.

\subsection{Quasi-Random Strategy (QRS)}

The \textit{Quasi-Random Strategy} (QRS) \cite{Chen2007b}
makes use of
\textit{quasi-random sequences}
---
point sequences with low discrepancy and low dispersion
---
to implement ART.
QRS generally has two main components:
The \textit{quasi-random-sequence-selection} component, which constructs each point; and
the \textit{randomization} component, which randomizes each constructed point, making the next test case.

There are many different quasi-random sequences, including:
\textit{Halton}~\cite{Halton1964};
\textit{Sobol}~\cite{Sobol1976}; and
\textit{Niederreiter}~\cite{Niederreiter1988}.
There are also a number of different randomization methods, such as:
\textit{Cranley-Patterson Rotation}~\cite{Cranley1976,Kollig2002};
\textit{Owen Scrambling}~\cite{Owen1995}; and
\textit{Random Shaking and Rotation}~\cite{Liu2016}.

\subsection{Search-Based Strategy (SBS)}

\textit{Search-Based Strategies} (SBSs), which come from \textit{Search-Based Software Testing}~\cite{McMinn2004,Harman2010}, make use of search-based algorithms to evenly spread the test cases over the input domain.
Because ART needs to retain some randomness in the generated test cases, 
an SBS creates an initial test set population $PT$ of randomly generated test sets.
A search-based algorithm is then adopted to iteratively evolve $PT$ into its next generation, according to the predefined criteria.
Once a stopping condition has been satisfied, the best solution from $PT$ is selected as the final test set.
Two core elements of SBS, therefore, are the choice of search-based algorithm for evolving $PT$, and the evaluation (\textit{fitness}) function for each solution.

A number of search-based algorithms have been used to evolve ART test sets, including:
\textit{Hill Climbing}~\cite{Schneckenburger2008};
\textit{Simulated Annealing}~\cite{Bueno2014};
\textit{Genetic Algorithm}~\cite{Bueno2014};
\textit{Simulated Repulsion}~\cite{Bueno2007};
\textit{Local Spreading}~\cite{Huang2017}; and
\textit{Random Border Centroidal Voronoi Tessellations}~\cite{Shahbazi2012}.

\subsection{Hybrid Strategies (HSs)}
\textit{Hybrid Strategies} (HSs) are combinations of multiple ART strategies, usually designed to enhance the testing effectiveness (e.g., fault-detection effectiveness) or efficiency (e.g., test-generation cost).

Chow et al.~\cite{Chow2013}, for example, combined STFCS and PBS, producing an efficient and effective method called \textit{ART with divide-and-conquer} that independently applies STFCS to each bisection-partitioned subdomain.
Mayer~\cite{Mayer2006} used bisection partitioning to control the STFCS test-case-identification component, only calculating distances between the candidate $c$ and already-executed test cases in its neighboring regions, not those in other regions.
Liu et al.~\cite{Liu2011} augmented TPBS with PBS and STFCS:
When the PBS chooses a subdomain within which to generate test cases \cite{Chow2013}, all inputs in this subdomain are given a probability of being selected, and those outside are given none.

\section{Conclusions and Future Work
\label{SECTION:conclusions}}

\textit{Adaptive Random Testing} (ART) \cite{Huang2019,Chen2004} is a family of testing approaches that enhance the fault-detection effectiveness of \textit{Random Testing} (RT).
Previous studies have demonstrated that, compared with RT, ART generates more diverse test cases, spread more evenly over the input domain, and that this can deliver better testing effectiveness \cite{Huang2019}.
There are many ART strategies and implementations, among which the most well-known is the \textit{Select-Test-From-Candidates Strategy} (STFCS) \cite{Huang2019}.
STFCS makes use of the concept of dissimilarity among test cases, and selects an element from random candidates as the next test case such that it has the largest dissimilarity to the previously-generated test cases.
Although popular, STFCS suffers from a problem of high computational costs.
Many enhanced STFCS algorithms exist that aim to reduce computation time, but they also face challenges in balancing the testing effectiveness and efficiency.
In this paper, based on the concept of \textit{Approximate Nearest Neighbor} (ANN), we have proposed a new ART approach that uses \textit{Locality-Sensitive Hashing} (LSH) to support and implement STFCS:
\textit{LSH-based ART} (LSH-ART).
LSH-ART makes use of all previously generated test cases to maintain the fault-detection ability, but uses ANNs for each candidate to reduce computational overheads.
The results of our simulations and empirical studies show that, overall, LSH-ART achieves comparable and even better fault detection than the original ART and its variants.
LSH-ART incurs lower computational costs when generating the same number of test cases, especially when the input domain dimensionality is high, resulting in better cost-effectiveness.

LSH-ART uses an SLSH table that can re-hash elements in a hash bucket, to store
previously generated test cases.
The traditional E$^2$LSH makes use of multiple hash tables to improve the search accuracy of ANNs \cite{Datar2004}. Therefore, an important research direction for future work will be about how multiple SLSH tables could be used to improve LSH-ART.

Another important research direction will be the application of other ANN approaches to ART in the future.
In addition to LSH, there are many other approaches to support the ANN process, including:
\textit{vector quantization} \cite{Gray1998};
\textit{semantic hashing} \cite{Salakhutdinov2009}; and
\textit{production quantization} \cite{Jegou2011}.
It will be interesting to investigate how these other ANN approaches can guide the ART process, and we look forward to comparing and analyzing their favorable (and unfavorable) conditions.

As discussed by Huang et al.~\cite{Huang2019}, a misconception related to ART is that it should always replace RT.
Although ART may have better testing effectiveness than RT, it also incurs more computational overheads. 
In practical situations, the application scenarios are different for ART and RT:
It is necessary to balance the trade-off between testing effectiveness and efficiency, and then choose between RT and ART. 
In particular, if the LSH-ART test-case generation time is less than the test setup and execution time, then it is appropriate to use LSH-ART instead of RT to generate fewer test cases to execute.
When this is not the case, however, RT may be the more suitable choice \cite{Anand2013, Huang2019}.
From the perspective of testing-time costs, RT remains an effective option. 
Although our proposed approach significantly reduces the computational costs related to traditional ART approaches, a significant gap remains compared with RT. 
Our future research, therefore, will also include further exploration of more approaches to reduce time costs:
Efficient test-case generation will continue to be a key goal.

We have also reported on a preliminary investigation into expanding the scope of LSH-ART from numerical to non-numerical domains. 
The study involved three widely used, highly configurable systems, for which we successfully modified and adapted the usage scenarios of LSH-ART and other ART variants (except DF). 
The results demonstrate the versatility of LSH-ART, showing how it can be successfully applied to non-numerical domain programs with similar (or even better) effectiveness and efficiency than other ART variants. 
The study showed LSH efficiency to be superior to other ART variants, with significant advantages over the others when searching for the (approximate) nearest neighbors in these non-numerical domains.
These findings highlight the potential for LSH-ART to be extended to other types of programs, and we look forward to exploring such a possibility in our future research.

A final direction for related future work will include the investigation of more strategies to help reduce the validation time required during the test-case generation process of these constrained programs.
This is also something that we look forward to sharing our findings about.

\ifCLASSOPTIONcompsoc
  \section*{Acknowledgments}
\else
  \section*{Acknowledgment}
\fi
We would like to thank the anonymous reviewers for their many constructive comments. We would also like to thank Andrea Arcuri, Chengying Mao, and Huayao Wu for providing us with the source code and fault data for the subject programs used in their papers \cite{Mao2019,Arcuri2011,Huayao2020}. Finally, we would also like to thank T. Y. Chen for his many helpful suggestions for this paper.

This work is supported by the Science and Technology Development Fund of Macau, Macau SAR, under Grant Nos. 0021/2023/RIA1 and 0046/2021/A, and a Faculty Research Grant of Macau University of Science and Technology under Grant No. FRG-22-103-FIE. This work is also partially supported by the National Natural Science Foundation of China, under Grant Nos. 61872167 and 61502205.

\ifCLASSOPTIONcaptionsoff
  \newpage
\fi
{
\bibliographystyle{IEEEtran}
\bibliography{IEEEabrv,lsh}
}

\clearpage

\newpage
\setcounter{page}{1} 
\pagenumbering{arabic} 

\begin{figure*}[!t]
\parbox{\textwidth}{

\begin{center}
    \Huge{Appendix: Detailed Results for LSH-ART}
    \vspace{10pt}
    
    \normalsize{Rubing Huang, Chenhui Cui, Junlong Lian, Dave Towey, Weifeng Sun, Haibo Chen}
    \vspace{10pt}
\end{center}
}

\parbox{\textwidth}{
\appendices
This appendix contains the detailed experimental data and corresponding statistical analyses for the paper ``Toward Cost-effective Adaptive Random Testing: An Approximate Nearest Neighbor Approach''. 
The data are intended to supplement the discussion and conclusions in the main text, providing a deeper understanding and validation. 
The data are presented in tables, which were not shown in the main text, due to space limitations.
}

\parbox{\textwidth}{
\section{Detailed Experiment Results for Numerical Input Domains}
\vspace{5pt}
}

\parbox{\textwidth}{This section contains the following experimental results for numerical input domains: 
F-measure results, P-measure results, test case generation time results, and F-time results.}

\parbox{\textwidth}{
\subsection{F-measure Results}
\vspace{5pt}

}

\parbox{\textwidth}{

Tables \ref{TAB:FSCS-block}, \ref{TAB:FSCS-strip}, and  \ref{TAB:FSCS-point} present the FSCS F-measure simulation results for block, strip, and point patterns, respectively. 
Tables \ref{TAB:RRT-block}, \ref{TAB:RRT-strip}, and \ref{TAB:RRT-point} show the corresponding RRT F-measure simulation results.
Tables \ref{TAB:FSCS-real} and \ref{TAB:RRT-real} summarize the F-measure results for the empirical studies using the 23 subject programs.
In these figures, when comparing two methods $\mathcal{M}_1$ and $\mathcal{M}_2$, we used
the \ding{109} symbol to indicate that there was no statistical difference between them (their $p$-value was greater than 0.01);
the \ding{52} symbol to indicate that $\mathcal{M}_1$ was significantly better ($p$-value was less than 0.01, and the effect size was greater than 0.50); and
the \ding{54} symbol to indicate that $\mathcal{M}_2$ was significantly better ($p$-value was less than 0.01, and the effect size was less than 0.50).
Each effect size value
---
$\hat{\textrm{A}}_{12}(\mathcal{M}_1,\mathcal{M}_2)$
---
is listed in the parenthesis immediately following the comparison symbol.
}
\end{figure*}

\renewcommand{\thetable}{\thesection.\arabic{table}}
\addtocounter{table}{-15}

\begin{table*}[!ht]
\centering
\footnotesize
\caption{\textbf{FSCS Version:} Mean \textbf{F-ratio} Results and Statistical Pairwise Comparisons of LSH against other Methods for \textbf{{Block Pattern}} Simulations}
\label{TAB:FSCS-block}
\resizebox{\textwidth}{!}{
 
\begin{tabular}{@{}ccrrrrrrrrrrrrrr@{}}
\hline
\textbf{Dimension} &\textbf{Failure Rate}  &\multicolumn{7}{c}{\textbf{Methods}} & &\multicolumn{6}{c}{\textit{LSH}} \\\cline{3-9}\cline{11-16}
($d$)  &($\theta$)  &\textit{RT} &\textit{ART} &\textit{RF} &\textit{CR} &\textit{DF} &\textit{KD} &\textit{LSH} & &\textit{vs. RT} &\textit{vs. ART} &\textit{vs. RF} &\textit{vs. CR} &\textit{vs. DF} &\textit{vs. KD}  \\\hline
\multirow{7}*{$d=1$}	&$1.0\times 10^{-2}$	&0.9866	&0.5669	&0.7012	&0.6439	&0.5706	&0.5682	&0.5564	&	&\ding{52} (0.61)	&\ding{109} (0.51)	&\ding{52} (0.53)	&\ding{109} (0.52)	&\ding{109} (0.51)	&\ding{109} (0.51)	\\
	&$5.0\times 10^{-3}$	&0.9875	&0.5712	&0.8729	&0.8312	&0.5552	&0.5586	&0.5605	&	&\ding{52} (0.61)	&\ding{109} (0.51)	&\ding{52} (0.57)	&\ding{52} (0.57)	&\ding{109} (0.50)	&\ding{109} (0.50)	\\
	&$2.0\times 10^{-3}$	&0.9574	&0.5651	&0.9718	&0.9310	&0.5687	&0.5659	&0.5533	&	&\ding{52} (0.61)	&\ding{109} (0.51)	&\ding{52} (0.61)	&\ding{52} (0.60)	&\ding{109} (0.51)	&\ding{109} (0.51)	\\
	&$1.0\times 10^{-3}$	&1.0053	&0.5605	&0.9830	&0.9956	&0.5694	&0.5570	&0.5643	&	&\ding{52} (0.62)	&\ding{109} (0.50)	&\ding{52} (0.60)	&\ding{52} (0.61)	&\ding{109} (0.50)	&\ding{109} (0.49)	\\
	&$5.0\times 10^{-4}$	&1.0160	&0.5679	&0.9913	&0.9957	&0.5532	&0.5557	&0.5546	&	&\ding{52} (0.62)	&\ding{109} (0.51)	&\ding{52} (0.62)	&\ding{52} (0.61)	&\ding{109} (0.50)	&\ding{109} (0.50)	\\
	&$2.0\times 10^{-4}$	&0.9955	&0.5527	&0.9820	&0.9775	&0.5702	&0.5530	&0.5697	&	&\ding{52} (0.60)	&\ding{109} (0.49)	&\ding{52} (0.60)	&\ding{52} (0.60)	&\ding{109} (0.50)	&\ding{109} (0.49)	\\
\hline
\end{tabular}}
\end{table*}

\begin{table*}[!ht]
\centering
\footnotesize
\caption{\textbf{FSCS Version:} Mean \textbf{F-ratio} Results and Statistical Pairwise Comparisons of LSH against other Methods for \textbf{{Strip Pattern}} Simulations}
\label{TAB:FSCS-strip}

\resizebox{\textwidth}{!}{
\begin{tabular}{@{}ccrrrrrrrrrrrrrr@{}}
\hline
\textbf{Dimension} &\textbf{Failure Rate}  &\multicolumn{7}{c}{\textbf{Methods}} & &\multicolumn{6}{c}{\textit{LSH}} \\\cline{3-9}\cline{11-16}
($d$)  &($\theta$)  &\textit{RT} &\textit{ART} &\textit{RF} &\textit{CR} &\textit{DF} &\textit{KD} &\textit{LSH} & &\textit{vs. RT} &\textit{vs. ART} &\textit{vs. RF} &\textit{vs. CR} &\textit{vs. DF} &\textit{vs. KD}  \\\hline
\multirow{7}*{$d=1$}	&$1.0\times 10^{-2}$	&0.9866	&0.5669	&0.7012	&0.6439	&0.5706	&0.5682	&0.5564	&	&\ding{52} (0.61)	&\ding{109} (0.51)	&\ding{52} (0.53)	&\ding{109} (0.52)	&\ding{109} (0.51)	&\ding{109} (0.51)	\\
	&$5.0\times 10^{-3}$	&0.9875	&0.5712	&0.8729	&0.8312	&0.5552	&0.5586	&0.5605	&	&\ding{52} (0.61)	&\ding{109} (0.51)	&\ding{52} (0.57)	&\ding{52} (0.57)	&\ding{109} (0.50)	&\ding{109} (0.50)	\\
	&$2.0\times 10^{-3}$	&0.9574	&0.5651	&0.9718	&0.9310	&0.5687	&0.5659	&0.5533	&	&\ding{52} (0.61)	&\ding{109} (0.51)	&\ding{52} (0.61)	&\ding{52} (0.60)	&\ding{109} (0.51)	&\ding{109} (0.51)	\\
	&$1.0\times 10^{-3}$	&1.0053	&0.5605	&0.9830	&0.9956	&0.5694	&0.5570	&0.5643	&	&\ding{52} (0.62)	&\ding{109} (0.50)	&\ding{52} (0.60)	&\ding{52} (0.61)	&\ding{109} (0.50)	&\ding{109} (0.49)	\\
	&$5.0\times 10^{-4}$	&1.0160	&0.5679	&0.9913	&0.9957	&0.5532	&0.5557	&0.5546	&	&\ding{52} (0.62)	&\ding{109} (0.51)	&\ding{52} (0.62)	&\ding{52} (0.61)	&\ding{109} (0.50)	&\ding{109} (0.50)	\\
	&$2.0\times 10^{-4}$	&0.9955	&0.5527	&0.9820	&0.9775	&0.5702	&0.5530	&0.5697	&	&\ding{52} (0.60)	&\ding{109} (0.49)	&\ding{52} (0.60)	&\ding{52} (0.60)	&\ding{109} (0.50)	&\ding{109} (0.49)	\\
\hline
\end{tabular}}
\end{table*}

\begin{table*}[!ht]
\centering
\footnotesize
\caption{\textbf{FSCS Version:} Mean \textbf{F-ratio} Results and Statistical Pairwise Comparisons of LSH against other Methods for \textbf{{Point Pattern}} Simulations}
\label{TAB:FSCS-point}

\resizebox{\textwidth}{!}{
\begin{tabular}{@{}ccrrrrrrrrrrrrrr@{}}
\hline
\textbf{Dimension} &\textbf{Failure Rate}  &\multicolumn{7}{c}{\textbf{Methods}} & &\multicolumn{6}{c}{\textit{LSH}} \\\cline{3-9}\cline{11-16}
($d$)  &($\theta$)  &\textit{RT} &\textit{ART} &\textit{RF} &\textit{CR} &\textit{DF} &\textit{KD} &\textit{LSH} & &\textit{vs. RT} &\textit{vs. ART} &\textit{vs. RF} &\textit{vs. CR} &\textit{vs. DF} &\textit{vs. KD}  \\\hline
\multirow{7}*{$d=1$}	&$1.0\times 10^{-2}$	&1.0273	&0.9467	&1.0015	&1.0352	&0.9655	&0.9861	&0.9540	&	&\ding{109} (0.52)	&\ding{109} (0.49)	&\ding{109} (0.51)	&\ding{109} (0.52)	&\ding{109} (0.50)	&\ding{109} (0.51)	\\
	&$5.0\times 10^{-3}$	&1.0057	&0.9719	&0.9949	&0.9899	&0.9797	&0.9435	&0.9583	&	&\ding{109} (0.51)	&\ding{109} (0.50)	&\ding{109} (0.50)	&\ding{109} (0.50)	&\ding{109} (0.50)	&\ding{109} (0.50)	\\
	&$2.0\times 10^{-3}$	&0.9869	&0.9658	&1.0017	&0.9985	&0.9690	&0.9322	&0.9510	&	&\ding{109} (0.51)	&\ding{109} (0.51)	&\ding{109} (0.51)	&\ding{109} (0.50)	&\ding{109} (0.51)	&\ding{109} (0.49)	\\
	&$1.0\times 10^{-3}$	&0.9626	&0.9598	&0.9640	&0.9839	&0.9643	&0.9548	&0.9664	&	&\ding{109} (0.49)	&\ding{109} (0.50)	&\ding{109} (0.49)	&\ding{109} (0.50)	&\ding{109} (0.50)	&\ding{109} (0.50)	\\
	&$5.0\times 10^{-4}$	&1.0055	&0.9628	&0.9950	&0.9913	&0.9793	&0.9728	&0.9511	&	&\ding{109} (0.51)	&\ding{109} (0.50)	&\ding{109} (0.50)	&\ding{109} (0.50)	&\ding{109} (0.50)	&\ding{109} (0.51)	\\
	&$2.0\times 10^{-4}$	&1.0111	&0.9589	&0.9974	&0.9961	&1.0095	&0.9822	&0.9637	&	&\ding{109} (0.51)	&\ding{109} (0.49)	&\ding{109} (0.51)	&\ding{109} (0.50)	&\ding{109} (0.50)	&\ding{109} (0.50)	\\
\hline
\end{tabular}}
\end{table*}

\begin{table*}
\centering
\footnotesize
\caption{\textbf{RRT Version:} Mean \textbf{F-ratio} Results and Statistical Pairwise Comparisons of LSH against other Methods for  \textbf{{Block Pattern}} Simulations}
\label{TAB:RRT-block}

\resizebox{\textwidth}{!}{
\begin{tabular}{@{}ccrrrrrrrrrrrrrr@{}}
\hline
\textbf{Dimension} &\textbf{Failure Rate}  &\multicolumn{7}{c}{\textbf{Methods}} & &\multicolumn{6}{c}{\textit{LSH}} \\\cline{3-9}\cline{11-16}
($d$)  &($\theta$)  &\textit{RT} &\textit{ART} &\textit{RF} &\textit{CR} &\textit{DF} &\textit{KD} &\textit{LSH} & &\textit{vs. RT} &\textit{vs. ART} &\textit{vs. RF} &\textit{vs. CR} &\textit{vs. DF} &\textit{vs. KD}  \\\hline
\multirow{7}*{$d=1$}	&$1.0\times 10^{-2}$	&0.9866	&0.5980 	&0.6166 	&0.6004 	&0.7816 	&0.5945 	&0.6039 	&	&\ding{52} (0.59)	&\ding{109} (0.50)	&\ding{109} (0.51)	&\ding{109} (0.50)	&\ding{52} (0.53)	&\ding{109} (0.49)	\\
	&$5.0\times 10^{-3}$	&0.9875	&0.5883 	&0.6048 	&0.5987 	&0.8748 	&0.6025 	&0.5975 	&	&\ding{52} (0.60)	&\ding{109} (0.49)	&\ding{109} (0.51)	&\ding{109} (0.50)	&\ding{52} (0.56)	&\ding{109} (0.51)	\\
	&$2.0\times 10^{-3}$	&0.9574	&0.6074 	&0.5949 	&0.5949 	&0.9450 	&0.5887 	&0.6056 	&	&\ding{52} (0.59)	&\ding{109} (0.50)	&\ding{109} (0.50)	&\ding{109} (0.50)	&\ding{52} (0.58)	&\ding{109} (0.50)	\\
	&$1.0\times 10^{-3}$	&1.0053	&0.5899 	&0.5938 	&0.5910 	&0.9820 	&0.5954 	&0.6020 	&	&\ding{52} (0.60)	&\ding{109} (0.49)	&\ding{109} (0.50)	&\ding{109} (0.49)	&\ding{52} (0.59)	&\ding{109} (0.50)	\\
	&$5.0\times 10^{-4}$	&1.0160	&0.5851 	&0.6122 	&0.6097 	&0.9904 	&0.6028 	&0.6067 	&	&\ding{52} (0.60)	&\ding{109} (0.49)	&\ding{109} (0.50)	&\ding{109} (0.50)	&\ding{52} (0.59)	&\ding{109} (0.50)	\\
	&$2.0\times 10^{-4}$	&0.9955	&0.5929 	&0.5821 	&0.6055 	&1.0392 	&0.5923 	&0.5991 	&	&\ding{52} (0.59)	&\ding{109} (0.49)	&\ding{109} (0.49)	&\ding{109} (0.50)	&\ding{52} (0.61)	&\ding{109} (0.49)	\\
\hline
\end{tabular}}
\end{table*}

\begin{table*}
\centering
\footnotesize
\caption{\textbf{RRT Version:} Mean \textbf{F-ratio} Results and Statistical Pairwise Comparisons of LSH against other Methods for  \textbf{{Strip Pattern}} Simulations}
\label{TAB:RRT-strip}

\resizebox{\textwidth}{!}{
\begin{tabular}{@{}ccrrrrrrrrrrrrrr@{}}
\hline
\textbf{Dimension} &\textbf{Failure Rate}  &\multicolumn{7}{c}{\textbf{Methods}} & &\multicolumn{6}{c}{\textit{LSH}} \\\cline{3-9}\cline{11-16}
($d$)  &($\theta$)  &\textit{RT} &\textit{ART} &\textit{RF} &\textit{CR} &\textit{DF} &\textit{KD} &\textit{LSH} & &\textit{vs. RT} &\textit{vs. ART} &\textit{vs. RF} &\textit{vs. CR} &\textit{vs. DF} &\textit{vs. KD}  \\\hline
\multirow{7}*{$d=1$}	&$1.0\times 10^{-2}$	&0.9866	&0.5980 	&0.6166 	&0.6004 	&0.7816 	&0.5945 	&0.6039 	&	&\ding{52} (0.59)	&\ding{109} (0.50)	&\ding{109} (0.51)	&\ding{109} (0.50)	&\ding{52} (0.53)	&\ding{109} (0.49)	\\
	&$5.0\times 10^{-3}$	&0.9875	&0.5883 	&0.6048 	&0.5987 	&0.8748 	&0.6025 	&0.5975 	&	&\ding{52} (0.60)	&\ding{109} (0.49)	&\ding{109} (0.51)	&\ding{109} (0.50)	&\ding{52} (0.56)	&\ding{109} (0.51)	\\
	&$2.0\times 10^{-3}$	&0.9574	&0.6074 	&0.5949 	&0.5949 	&0.9450 	&0.5887 	&0.6056 	&	&\ding{52} (0.59)	&\ding{109} (0.50)	&\ding{109} (0.50)	&\ding{109} (0.50)	&\ding{52} (0.58)	&\ding{109} (0.50)	\\
	&$1.0\times 10^{-3}$	&1.0053	&0.5899 	&0.5938 	&0.5910 	&0.9820 	&0.5954 	&0.6020 	&	&\ding{52} (0.60)	&\ding{109} (0.49)	&\ding{109} (0.50)	&\ding{109} (0.49)	&\ding{52} (0.59)	&\ding{109} (0.50)	\\
	&$5.0\times 10^{-4}$	&1.0160	&0.5851 	&0.6122 	&0.6097 	&0.9904 	&0.6028 	&0.6067 	&	&\ding{52} (0.60)	&\ding{109} (0.49)	&\ding{109} (0.50)	&\ding{109} (0.50)	&\ding{52} (0.59)	&\ding{109} (0.50)	\\
	&$2.0\times 10^{-4}$	&0.9955	&0.5929 	&0.5821 	&0.6055 	&1.0392 	&0.5923 	&0.5991 	&	&\ding{52} (0.59)	&\ding{109} (0.49)	&\ding{109} (0.49)	&\ding{109} (0.50)	&\ding{52} (0.61)	&\ding{109} (0.49)	\\
\hline
\end{tabular}}
\end{table*}

\begin{table*}
\centering
\footnotesize
\caption{\textbf{RRT Version:} Mean \textbf{F-ratio} Results and Statistical Pairwise Comparisons of LSH against other Methods for \textbf{{Point Pattern}} Simulations}
\label{TAB:RRT-point}

\resizebox{\textwidth}{!}{
\begin{tabular}{@{}ccrrrrrrrrrrrrrr@{}}
\hline
\textbf{Dimension} &\textbf{Failure Rate}  &\multicolumn{7}{c}{\textbf{Methods}} & &\multicolumn{6}{c}{\textit{LSH}} \\\cline{3-9}\cline{11-16}
($d$)  &($\theta$)  &\textit{RT} &\textit{ART} &\textit{RF} &\textit{CR} &\textit{DF} &\textit{KD} &\textit{LSH} & &\textit{vs. RT} &\textit{vs. ART} &\textit{vs. RF} &\textit{vs. CR} &\textit{vs. DF} &\textit{vs. KD}  \\\hline
\multirow{7}*{$d=1$}	&$1.0\times 10^{-2}$	&1.0273	&0.9752 	&0.9478 	&0.9431 	&0.9779 	&0.9551 	&0.9780 	&	&\ding{109} (0.51)	&\ding{109} (0.50)	&\ding{109} (0.50)	&\ding{109} (0.49)	&\ding{109} (0.49)	&\ding{109} (0.50)	\\
	&$5.0\times 10^{-3}$	&1.0057	&0.9291 	&0.9541 	&0.9514 	&0.9966 	&0.9353 	&0.9646 	&	&\ding{109} (0.50)	&\ding{109} (0.49)	&\ding{109} (0.50)	&\ding{109} (0.50)	&\ding{109} (0.51)	&\ding{109} (0.49)	\\
	&$2.0\times 10^{-3}$	&0.9869	&0.9325 	&0.9359 	&0.9655 	&0.9724 	&0.9699 	&0.9639 	&	&\ding{109} (0.50)	&\ding{109} (0.49)	&\ding{109} (0.49)	&\ding{109} (0.50)	&\ding{109} (0.49)	&\ding{109} (0.50)	\\
	&$1.0\times 10^{-3}$	&0.9626	&0.9805 	&0.9443 	&0.9666 	&0.9616 	&0.9611 	&0.9388 	&	&\ding{109} (0.50)	&\ding{109} (0.51)	&\ding{109} (0.51)	&\ding{109} (0.52)	&\ding{109} (0.51)	&\ding{109} (0.51)	\\
	&$5.0\times 10^{-4}$	&1.0055	&0.9608 	&0.9877 	&0.9583 	&1.0172 	&0.9623 	&0.9444 	&	&\ding{109} (0.51)	&\ding{109} (0.51)	&\ding{109} (0.52)	&\ding{109} (0.51)	&\ding{109} (0.52)	&\ding{109} (0.51)	\\
	&$2.0\times 10^{-4}$	&1.0111	&0.9541 	&0.9361 	&0.9716 	&1.0041 	&0.9504 	&0.9536 	&	&\ding{109} (0.51)	&\ding{109} (0.49)	&\ding{109} (0.49)	&\ding{109} (0.50)	&\ding{109} (0.50)	&\ding{109} (0.49)	\\
\hline
\end{tabular}}
\end{table*}

\begin{table*}[!ht]
\centering
\footnotesize
\caption{\textbf{FSCS Version:} Mean \textbf{F-measure} Results and Statistical Pairwise Comparisons of LSH for  \textbf{{Real-life Programs}}}
\label{TAB:FSCS-real}
\resizebox{\textwidth}{!}{
\begin{tabular}{@{}ccrrrrrrrrrrrrrr@{}}
\hline
\multirow{2}*{\textbf{ID}} &\multirow{2}*{\textbf{Program}}  &\multicolumn{7}{c}{\textbf{Methods}} & &\multicolumn{6}{c}{\textit{LSH}} \\\cline{3-9}\cline{11-16}
  & &\textit{RT} &\textit{ART} &\textit{RF} &\textit{CR} &\textit{DF} &\textit{KD} &\textit{LSH} & &\textit{vs. RT} &\textit{vs. ART} &\textit{vs. RF} &\textit{vs. CR} &\textit{vs. DF} &\textit{vs. KD}  \\\hline
P1	&airy	&1444.64 	&794.02 	&1400.15 	&1440.76 	&788.61 	&803.70 	&788.23 	&	&\ding{52} (0.62)	&\ding{109} (0.50)	&\ding{52} (0.61)	&\ding{52} (0.62)	&\ding{109} (0.50)	&\ding{109} (0.51) 	\\
P2	&bessj0	&760.24 	&437.02 	&750.17 	&712.10 	&443.36 	&445.48 	&443.03 	&	&\ding{52} (0.61)	&\ding{109} (0.50)	&\ding{52} (0.60)	&\ding{52} (0.59)	&\ding{109} (0.50)	&\ding{109} (0.50) 	\\
P3	&erfcc	&1867.60 	&1032.28 	&1897.29 	&1839.75 	&1033.04 	&1045.93 	&1018.20 	&	&\ding{52} (0.62)	&\ding{109} (0.51)	&\ding{52} (0.62)	&\ding{52} (0.62)	&\ding{109} (0.51)	&\ding{109} (0.51)	\\
P4	&probks	&2590.02 	&1468.08 	&2527.86 	&2681.32 	&1428.13 	&1440.23 	&1473.70 	&	&\ding{52} (0.60)	&\ding{109} (0.50)	&\ding{52} (0.60)	&\ding{52} (0.62)	&\ding{109} (0.49)	&\ding{109} (0.49) 	\\
P5	&tanh	&556.28 	&303.59 	&551.62 	&528.89 	&303.43 	&308.54 	&313.58 	&	&\ding{52} (0.61)	&\ding{109} (0.49)	&\ding{52} (0.61)	&\ding{52} (0.60)	&\ding{109} (0.49)	&\ding{109} (0.49) 	\\
P6	&bessj	&778.47 	&433.47 	&516.64 	&498.96 	&544.76 	&457.08 	&462.24 	&	&\ding{52} (0.62)	&\ding{109} (0.49)	&\ding{109} (0.52)	&\ding{109} (0.51)	&\ding{52} (0.54)	&\ding{109} (0.50) 	\\
\hline
\end{tabular}}
\end{table*}

\begin{table*}[!ht]
\centering
\footnotesize
\caption{\textbf{RRT Version:} Mean \textbf{F-measure} Results and Statistical Pairwise Comparisons of LSH for \textbf{{Real-life Programs}}}
\label{TAB:RRT-real}
\resizebox{\textwidth}{!}{
\begin{tabular}{@{}ccrrrrrrrrrrrrrr@{}}
\hline
\multirow{2}*{\textbf{ID}} &\multirow{2}*{\textbf{Program}}  &\multicolumn{7}{c}{\textbf{Methods}} & &\multicolumn{6}{c}{\textit{LSH}} \\\cline{3-9}\cline{11-16}
  & &\textit{RT} &\textit{ART} &\textit{RF} &\textit{CR} &\textit{DF} &\textit{KD} &\textit{LSH} & &\textit{vs. RT} &\textit{vs. ART} &\textit{vs. RF} &\textit{vs. CR} &\textit{vs. DF} &\textit{vs. KD}  \\\hline
P1	&airy	&1444.64 	&853.28 	&1434.04 	&1397.80 	&1363.03 	&864.91 	&859.32 	&	&\ding{52} (0.61)	&\ding{109} (0.50) 	&\ding{52} (0.59) 	&\ding{52} (0.59) 	&\ding{52} (0.58) 	&\ding{109} (0.50) 	\\
P2	&bessj0	&760.24 	&453.21 	&734.54 	&721.98 	&722.39 	&464.37 	&478.19 	&	&\ding{52} (0.60)	&\ding{54} (0.48) 	&\ding{52} (0.58) 	&\ding{52} (0.57) 	&\ding{52} (0.58) 	&\ding{109} (0.49) 	\\
P3	&erfcc	&1867.60 	&1087.85 	&1857.94 	&1869.51 	&1797.57 	&1092.82 	&1119.16 	&	&\ding{52} (0.60)	&\ding{109} (0.49) 	&\ding{52} (0.60) 	&\ding{52} (0.60) 	&\ding{52} (0.59) 	&\ding{109} (0.49)	\\
P4	&probks	&2590.02 	&1552.39 	&2597.79 	&2570.65 	&2572.19 	&1554.80 	&1533.70 	&	&\ding{52} (0.60)	&\ding{109} (0.51) 	&\ding{52} (0.60) 	&\ding{52} (0.60) 	&\ding{52} (0.60) 	&\ding{109} (0.51) 	\\
P5	&tanh	&556.28 	&335.62 	&532.49 	&534.69 	&527.56 	&328.49 	&331.06 	&	&\ding{52} (0.60)	&\ding{109} (0.51) 	&\ding{52} (0.59) 	&\ding{52} (0.60) 	&\ding{52} (0.58) 	&\ding{109} (0.50) 	\\
P6	&bessj	&778.47 	&540.29 	&630.89 	&626.24 	&669.32 	&534.82 	&532.36 	&	&\ding{52} (0.58)	&\ding{109} (0.51) 	&\ding{52} (0.54) 	&\ding{52} (0.53) 	&\ding{52} (0.54) 	&\ding{109} (0.50) 	\\
\hline
\end{tabular}}
\end{table*}
\clearpage
\newpage
\begin{table*}
\parbox{\textwidth}{
\subsection{P-measure Results}
\vspace{10pt}
}
\parbox{\textwidth}{
Tables \ref{TAB:FSCS-block-p}, \ref{TAB:FSCS-strip-p}, and \ref{TAB:FSCS-point-p} present the FSCS P-measure simulation results for block, strip, and point patterns, respectively.
Tables \ref{TAB:RRT-block-p}, \ref{TAB:RRT-strip-p}, and \ref{TAB:RRT-point-p} show the corresponding RRT P-measure simulation results.
Tables \ref{TAB:FSCS-real-p} and \ref{TAB:RRT-real-p} summarize the P-measure results for the empirical study.
In these figures, when comparing two methods $\mathcal{M}_1$ and $\mathcal{M}_2$, we used
the \ding{109} symbol to indicate that there was no statistical difference between them (their $p$-value was greater than 0.01);
the \ding{52} symbol to indicate that $\mathcal{M}_1$ was significantly better ($p$-value was less than 0.01, and the effect size was greater than 1.0); and
the \ding{54} symbol to indicate that $\mathcal{M}_2$ was significantly better ($p$-value was less than 0.01, and the effect size was less than 1.0).
Each effect size value
---
$\psi(\mathcal{M}_1,\mathcal{M}_2)$ 
---
is listed in the parenthesis immediately following the comparison symbol.
}
\end{table*}

\begin{table*}[!ht]									
\centering		
\footnotesize	
\caption{\textbf{FSCS Version:} Mean \textbf{P-measure} Results and Statistical Pairwise Comparisons of LSH against other Methods for \textbf{{Block Pattern}} Simulations}		
\label{TAB:FSCS-block-p}

\resizebox{\textwidth}{!}{

\begin{tabular}{@{}ccrrrrrrrrrrrrrr@{}}		
\hline
\textbf{Dimension} &\textbf{Failure Rate}  &\multicolumn{7}{c}{\textbf{Methods}} & &\multicolumn{6}{c}{\textit{LSH}} \\\cline{3-9}\cline{11-16}																
($d$)  &($\theta$)  &\textit{RT} &\textit{ART} &\textit{RF} &\textit{CR} &\textit{DF} &\textit{KD} &\textit{LSH} & &\textit{vs. RT} &\textit{vs. ART} &\textit{vs. RF} &\textit{vs. CR} &\textit{vs. DF} &\textit{vs. KD}  \\\hline																
\multirow{7}*{$d=1$}	&$1.0\times 10^{-2}$	&0.5016 	&0.6646 	&0.6546 	&0.6525 	&0.6672 	&0.6670 	&0.6653 	&	&\ding{52} (1.9751)	&\ding{109} (1.0033)	&\ding{52} (1.0489)	&\ding{52} (1.0587)	&\ding{109} (0.9918)	&\ding{109} (0.9924)	\\
	&$5.0\times 10^{-3}$	&0.5022 	&0.6664 	&0.5513 	&0.5641 	&0.6683 	&0.6673 	&0.6672 	&	&\ding{52} (1.9877)	&\ding{109} (1.0038)	&\ding{52} (1.6316)	&\ding{52} (1.5492)	&\ding{109} (0.9954)	&\ding{109} (0.9995)	\\
	&$2.0\times 10^{-3}$	&0.4991 	&0.6700 	&0.5095 	&0.5202 	&0.6700 	&0.6691 	&0.6677 	&	&\ding{52} (2.0169)	&\ding{109} (0.9898)	&\ding{52} (1.9347)	&\ding{52} (1.8533)	&\ding{109} (0.9898)	&\ding{109} (0.9936)	\\
	&$1.0\times 10^{-3}$	&0.4997 	&0.6715 	&0.5036 	&0.5119 	&0.6704 	&0.6712 	&0.6678 	&	&\ding{52} (2.0130)	&\ding{54} (0.9836)	&\ding{52} (1.9819)	&\ding{52} (1.9169)	&\ding{109} (0.9885)	&\ding{54} (0.9850)	\\
	&$5.0\times 10^{-4}$	&0.5001 	&0.6701 	&0.5002 	&0.5035 	&0.6710 	&0.6709 	&0.6679 	&	&\ding{52} (2.0102)	&\ding{109} (0.9901)	&\ding{52} (2.0093)	&\ding{52} (1.9828)	&\ding{109} (0.9861)	&\ding{109} (0.9864)	\\
	&$2.0\times 10^{-4}$	&0.5000 	&0.6705 	&0.4994 	&0.5017 	&0.6724 	&0.6702 	&0.6680 	&	&\ding{52} (2.0118)	&\ding{109} (0.9889)	&\ding{52} (2.0169)	&\ding{52} (1.9979)	&\ding{54} (0.9803)	&\ding{109} (0.9898)	\\
\hline

\end{tabular}}																
\end{table*}

\begin{table*}[!ht]																
\centering																
\footnotesize																
\caption{\textbf{FSCS Version:} Mean \textbf{P-measure} Results and Statistical Pairwise Comparisons of LSH against other Methods for \textbf{{Strip Pattern}} Simulations}	
\label{TAB:FSCS-strip-p}										 
\resizebox{\textwidth}{!}{
\begin{tabular}{@{}ccrrrrrrrrrrrrrr@{}}											\hline																
\textbf{Dimension} &\textbf{Failure Rate}  &\multicolumn{7}{c}{\textbf{Methods}} & &\multicolumn{6}{c}{\textit{LSH}} \\\cline{3-9}\cline{11-16}																
($d$)  &($\theta$)  &\textit{RT} &\textit{ART} &\textit{RF} &\textit{CR} &\textit{DF} &\textit{KD} &\textit{LSH} & &\textit{vs. RT} &\textit{vs. ART} &\textit{vs. RF} &\textit{vs. CR} &\textit{vs. DF} &\textit{vs. KD}  \\\hline																
\multirow{7}*{$d=1$}	&$1.0\times 10^{-2}$	&0.4999 	&0.6657 	&0.6536 	&0.6535 	&0.6674 	&0.6644 	&0.6643 	&	&\ding{52} (1.9796)	&\ding{109} (0.9935)	&\ding{52} (1.0486)	&\ding{52} (1.0492)	&\ding{54} (0.9859)	&\ding{109} (0.9993)	\\
	&$5.0\times 10^{-3}$	&0.5033 	&0.6676 	&0.5519 	&0.5652 	&0.6672 	&0.6685 	&0.6644 	&	&\ding{52} (1.9538)	&\ding{54} (0.9857)	&\ding{52} (1.6075)	&\ding{52} (1.5230)	&\ding{109} (0.9877)	&\ding{54} (0.9818)	\\
	&$2.0\times 10^{-3}$	&0.4986 	&0.6703 	&0.5081 	&0.5209 	&0.6706 	&0.6700 	&0.6656 	&	&\ding{52} (2.0014)	&\ding{54} (0.9789)	&\ding{52} (1.9265)	&\ding{52} (1.8307)	&\ding{54} (0.9775)	&\ding{54} (0.9801)	\\
	&$1.0\times 10^{-3}$	&0.5001 	&0.6707 	&0.5038 	&0.5102 	&0.6702 	&0.6705 	&0.6680 	&	&\ding{52} (2.0108)	&\ding{109} (0.9879)	&\ding{52} (1.9816)	&\ding{52} (1.9314)	&\ding{109} (0.9900)	&\ding{109} (0.9888)	\\
	&$5.0\times 10^{-4}$	&0.4994 	&0.6715 	&0.5016 	&0.5051 	&0.6691 	&0.6710 	&0.6698 	&	&\ding{52} (2.0333)	&\ding{109} (0.9925)	&\ding{52} (2.0157)	&\ding{52} (1.9878)	&\ding{109} (1.0031)	&\ding{109} (0.9946)	\\
	&$2.0\times 10^{-4}$	&0.4977 	&0.6709 	&0.4995 	&0.5025 	&0.6701 	&0.6702 	&0.6685 	&	&\ding{52} (2.0358)	&\ding{109} (0.9891)	&\ding{52} (2.0209)	&\ding{52} (1.9967)	&\ding{109} (0.9929)	&\ding{109} (0.9925)	\\
\hline

\end{tabular}}																
\end{table*}			

\begin{table*}[!ht]																	
\centering																
\footnotesize																
\caption{\textbf{FSCS Version:} Mean \textbf{P-measure} Results and Statistical Pairwise Comparisons of LSH against other Methods for \textbf{{Point Pattern}} Simulations}		
\label{TAB:FSCS-point-p}										 
				\resizebox{\textwidth}{!}{
\begin{tabular}{@{}ccrrrrrrrrrrrrrr@{}}											\hline																
\textbf{Dimension} &\textbf{Failure Rate}  &\multicolumn{7}{c}{\textbf{Methods}} & &\multicolumn{6}{c}{\textit{LSH}} \\\cline{3-9}\cline{11-16}																
($d$)  &($\theta$)  &\textit{RT} &\textit{ART} &\textit{RF} &\textit{CR} &\textit{DF} &\textit{KD} &\textit{LSH} & &\textit{vs. RT} &\textit{vs. ART} &\textit{vs. RF} &\textit{vs. CR} &\textit{vs. DF} &\textit{vs. KD}  \\\hline																
\multirow{7}*{$d=1$}	&$1.0\times 10^{-2}$	&0.4999 	&0.5047 	&0.5032 	&0.5035 	&0.5068 	&0.5050 	&0.5033 	&	&\ding{52} (1.0134)	&\ding{109} (0.9941)	&\ding{109} (1.0003)	&\ding{109} (0.9991)	&\ding{54} (0.9858)	&\ding{109} (0.9929)	\\
	&$5.0\times 10^{-3}$	&0.4993 	&0.5047 	&0.5018 	&0.5002 	&0.5048 	&0.5021 	&0.5054 	&	&\ding{52} (1.0247)	&\ding{109} (1.0028)	&\ding{52} (1.0144)	&\ding{52} (1.0209)	&\ding{109} (1.0026)	&\ding{109} (1.0133)	\\
	&$2.0\times 10^{-3}$	&0.5007 	&0.5047 	&0.5012 	&0.5000 	&0.5061 	&0.5048 	&0.5039 	&	&\ding{109} (1.0125)	&\ding{109} (0.9965)	&\ding{109} (1.0109)	&\ding{52} (1.0155)	&\ding{109} (0.9911)	&\ding{109} (0.9962)	\\
	&$1.0\times 10^{-3}$	&0.5010 	&0.5052 	&0.5021 	&0.5010 	&0.5044 	&0.5048 	&0.5047 	&	&\ding{52} (1.0147)	&\ding{109} (0.9981)	&\ding{109} (1.0102)	&\ding{52} (1.0148)	&\ding{109} (1.0012)	&\ding{109} (0.9993)	\\
	&$5.0\times 10^{-4}$	&0.4992 	&0.5049 	&0.4996 	&0.5019 	&0.5051 	&0.5059 	&0.5049 	&	&\ding{52} (1.0233)	&\ding{109} (1.0001)	&\ding{52} (1.0216)	&\ding{109} (1.0121)	&\ding{109} (0.9992)	&\ding{109} (0.9962)	\\
	&$2.0\times 10^{-4}$	&0.5021 	&0.5061 	&0.4991 	&0.4996 	&0.5054 	&0.5037 	&0.5041 	&	&\ding{109} (1.0080)	&\ding{109} (0.9921)	&\ding{52} (1.0204)	&\ding{52} (1.0182)	&\ding{109} (0.9947)	&\ding{109} (1.0015)	\\
\hline																

\end{tabular}}																
\end{table*}			

\begin{table*}[!ht]																	
\centering																
\footnotesize																
\caption{\textbf{RRT Version:} Mean \textbf{P-measure} Results and Statistical Pairwise Comparisons of LSH against other Methods for \textbf{{Block Pattern}} Simulations}	
\label{TAB:RRT-block-p}															\resizebox{\textwidth}{!}{														
\begin{tabular}{@{}ccrrrrrrrrrrrrrr@{}}																
\hline																
\textbf{Dimension} &\textbf{Failure Rate}  &\multicolumn{7}{c}{\textbf{Methods}} & &\multicolumn{6}{c}{\textit{LSH}} \\\cline{3-9}\cline{11-16}																
($d$)  &($\theta$)  &\textit{RT} &\textit{ART} &\textit{RF} &\textit{CR} &\textit{DF} &\textit{KD} &\textit{LSH} & &\textit{vs. RT} &\textit{vs. ART} &\textit{vs. RF} &\textit{vs. CR} &\textit{vs. DF} &\textit{vs. KD}  \\\hline																
\multirow{7}*{$d=1$}	&$1.0\times 10^{-2}$	&0.5016 	&0.6558 	&0.6518 	&0.6526 	&0.5912 	&0.6570 	&0.6573 	&	&\ding{52} (1.9051)	&\ding{109} (1.0065)	&\ding{52} (1.0246)	&\ding{52} (1.0207)	&\ding{52} (1.3260)	&\ding{109} (1.0012)	\\
	&$5.0\times 10^{-3}$	&0.5022 	&0.6554 	&0.5711 	&0.5760 	&0.5462 	&0.6565 	&0.6540 	&	&\ding{52} (1.8742)	&\ding{109} (0.9939)	&\ding{52} (1.4197)	&\ding{52} (1.3916)	&\ding{52} (1.5708)	&\ding{109} (0.9890)	\\
	&$2.0\times 10^{-3}$	&0.4991 	&0.6573 	&0.5136 	&0.5229 	&0.5221 	&0.6567 	&0.6562 	&	&\ding{52} (1.9158)	&\ding{109} (0.9953)	&\ding{52} (1.8078)	&\ding{52} (1.7414)	&\ding{52} (1.7476)	&\ding{109} (0.9977)	\\
	&$1.0\times 10^{-3}$	&0.4997 	&0.6589 	&0.5050 	&0.5098 	&0.5147 	&0.6591 	&0.6573 	&	&\ding{52} (1.9208)	&\ding{109} (0.9931)	&\ding{52} (1.8806)	&\ding{52} (1.8448)	&\ding{52} (1.8088)	&\ding{109} (0.9922)	\\
	&$5.0\times 10^{-4}$	&0.5001 	&0.6587 	&0.4999 	&0.5073 	&0.5065 	&0.6586 	&0.6588 	&	&\ding{52} (1.9307)	&\ding{109} (1.0008)	&\ding{52} (1.9319)	&\ding{52} (1.8755)	&\ding{52} (1.8819)	&\ding{109} (1.0011)	\\
	&$2.0\times 10^{-4}$	&0.5000 	&0.6587 	&0.4996 	&0.5039 	&0.5024 	&0.6576 	&0.6578 	&	&\ding{52} (1.9224)	&\ding{109} (0.9959)	&\ding{52} (1.9252)	&\ding{52} (1.8924)	&\ding{52} (1.9037)	&\ding{109} (1.0008)	\\
\hline

\end{tabular}}																
\end{table*}

\begin{table*}[!ht]																	
\centering																
\footnotesize																
\caption{\textbf{RRT Version:} Mean \textbf{P-measure} Results and Statistical Pairwise Comparisons of LSH against other Methods for \textbf{{Strip Pattern}} Simulations}
\label{TAB:RRT-strip-p}															\resizebox{\textwidth}{!}{
\begin{tabular}{@{}ccrrrrrrrrrrrrrr@{}}																
\hline																
\textbf{Dimension} &\textbf{Failure Rate}  &\multicolumn{7}{c}{\textbf{Methods}} & &\multicolumn{6}{c}{\textit{LSH}} \\\cline{3-9}\cline{11-16}																
($d$)  &($\theta$)  &\textit{RT} &\textit{ART} &\textit{RF} &\textit{CR} &\textit{DF} &\textit{KD} &\textit{LSH} & &\textit{vs. RT} &\textit{vs. ART} &\textit{vs. RF} &\textit{vs. CR} &\textit{vs. DF} &\textit{vs. KD}  \\\hline																
\multirow{7}*{$d=1$}	&$1.0\times 10^{-2}$	&0.4999 	&0.6543 	&0.6498 	&0.6521 	&0.5901 	&0.6564 	&0.6566 	&	&\ding{52} (1.9131)	&\ding{109} (1.0105)	&\ding{52} (1.0305)	&\ding{52} (1.0200)	&\ding{52} (1.3282)	&\ding{109} (1.0010)	\\
	&$5.0\times 10^{-3}$	&0.5033 	&0.6559 	&0.5723 	&0.5779 	&0.5489 	&0.6574 	&0.6559 	&	&\ding{52} (1.8813)	&\ding{109} (0.9999)	&\ding{52} (1.4249)	&\ding{52} (1.3923)	&\ding{52} (1.5667)	&\ding{109} (0.9936)	\\
	&$2.0\times 10^{-3}$	&0.4986 	&0.6562 	&0.5136 	&0.5242 	&0.5232 	&0.6560 	&0.6548 	&	&\ding{52} (1.9075)	&\ding{109} (0.9937)	&\ding{52} (1.7966)	&\ding{52} (1.7217)	&\ding{52} (1.7284)	&\ding{109} (0.9945)	\\
	&$1.0\times 10^{-3}$	&0.5001 	&0.6579 	&0.5040 	&0.5113 	&0.5122 	&0.6619 	&0.6566 	&	&\ding{52} (1.9112)	&\ding{109} (0.9945)	&\ding{52} (1.8823)	&\ding{52} (1.8279)	&\ding{52} (1.8209)	&\ding{54} (0.9768)	\\
	&$5.0\times 10^{-4}$	&0.4994 	&0.6584 	&0.5022 	&0.5061 	&0.5064 	&0.6577 	&0.6561 	&	&\ding{52} (1.9122)	&\ding{109} (0.9897)	&\ding{52} (1.8908)	&\ding{52} (1.8617)	&\ding{52} (1.8593)	&\ding{109} (0.9930)	\\
	&$2.0\times 10^{-4}$	&0.4977 	&0.6597 	&0.4992 	&0.5031 	&0.5031 	&0.6586 	&0.6568 	&	&\ding{52} (1.9314)	&\ding{109} (0.9868)	&\ding{52} (1.9196)	&\ding{52} (1.8895)	&\ding{52} (1.8897)	&\ding{109} (0.9918)	\\
\hline

\end{tabular}}																
\end{table*}			

\begin{table*}[!ht]																	
\centering																
\footnotesize																
\caption{\textbf{RRT Version:} Mean \textbf{P-measure} Results and Statistical Pairwise Comparisons of LSH against other Methods for \textbf{{Point Pattern}} Simulations}		
\label{TAB:RRT-point-p}															\resizebox{\textwidth}{!}{
\begin{tabular}{@{}ccrrrrrrrrrrrrrr@{}}																
\hline																
\textbf{Dimension} &\textbf{Failure Rate}  &\multicolumn{7}{c}{\textbf{Methods}} & &\multicolumn{6}{c}{\textit{LSH}} \\\cline{3-9}\cline{11-16}																
($d$)  &($\theta$)  &\textit{RT} &\textit{ART} &\textit{RF} &\textit{CR} &\textit{DF} &\textit{KD} &\textit{LSH} & &\textit{vs. RT} &\textit{vs. ART} &\textit{vs. RF} &\textit{vs. CR} &\textit{vs. DF} &\textit{vs. KD}  \\\hline																
\multirow{7}*{$d=1$}	&$1.0\times 10^{-2}$	&0.4999 	&0.5055 	&0.5040 	&0.5036 	&0.5046 	&0.5040 	&0.5049 	&	&\ding{52} (1.0201)	&\ding{109} (0.9974)	&\ding{109} (1.0035)	&\ding{109} (1.0051)	&\ding{109} (1.0009)	&\ding{109} (1.0034)	\\
	&$5.0\times 10^{-3}$	&0.4993 	&0.5036 	&0.5023 	&0.5021 	&0.5021 	&0.5050 	&0.5041 	&	&\ding{52} (1.0193)	&\ding{109} (1.0020)	&\ding{109} (1.0071)	&\ding{109} (1.0081)	&\ding{109} (1.0080)	&\ding{109} (0.9963)	\\
	&$2.0\times 10^{-3}$	&0.5007 	&0.5045 	&0.4992 	&0.4994 	&0.5009 	&0.5053 	&0.5045 	&	&\ding{52} (1.0149)	&\ding{109} (0.9999)	&\ding{52} (1.0211)	&\ding{52} (1.0202)	&\ding{52} (1.0143)	&\ding{109} (0.9967)	\\
	&$1.0\times 10^{-3}$	&0.5010 	&0.5042 	&0.4987 	&0.5006 	&0.5017 	&0.5055 	&0.5065 	&	&\ding{52} (1.0222)	&\ding{109} (1.0093)	&\ding{52} (1.0317)	&\ding{52} (1.0240)	&\ding{52} (1.0193)	&\ding{109} (1.0041)	\\
	&$5.0\times 10^{-4}$	&0.4992 	&0.5042 	&0.4993 	&0.5013 	&0.5006 	&0.5048 	&0.5041 	&	&\ding{52} (1.0199)	&\ding{109} (0.9996)	&\ding{52} (1.0192)	&\ding{109} (1.0112)	&\ding{52} (1.0141)	&\ding{109} (0.9970)	\\
	&$2.0\times 10^{-4}$	&0.5021 	&0.5048 	&0.4996 	&0.4993 	&0.4995 	&0.5064 	&0.5057 	&	&\ding{52} (1.0147)	&\ding{109} (1.0036)	&\ding{52} (1.0248)	&\ding{52} (1.0262)	&\ding{52} (1.0254)	&\ding{109} (0.9975)	\\
\hline

\end{tabular}}																
\end{table*}	

\begin{table*}[!ht]															
\centering																
\footnotesize																
\caption{\textbf{FSCS Version:} Mean \textbf{P-measure} Results and Statistical Pairwise Comparisons of LSH against other Methods for \textbf{Real-life Programs}}																\label{TAB:FSCS-real-p}																
\resizebox{\textwidth}{!}{
\begin{tabular}{@{}crrrrrrrrrrrrrr@{}}																
\hline																
\textbf{Test Set Size}  &\multicolumn{7}{c}{\textbf{Methods}} & &\multicolumn{6}{c}{\textit{LSH}} \\\cline{2-8}\cline{10-15}																
($|T|$)  &\textit{RT} &\textit{ART} &\textit{RF} &\textit{CR} &\textit{DF} &\textit{KD} &\textit{LSH} & &\textit{vs. RT} &\textit{vs. ART} &\textit{vs. RF} &\textit{vs. CR} &\textit{vs. DF} &\textit{vs. KD}  \\\hline	
2	&0.0022 	&0.0028 	&0.0027 	&0.0026 	&0.0025 	&0.0026 	&0.0026 	&	&\ding{52} (1.1661)	&\ding{54} (0.9323)	&\ding{54} (0.9477)	&\ding{109} (0.9914 )	&\ding{52} (1.0453)	&\ding{109} (0.9779)	\\
4	&0.0048 	&0.0058 	&0.0055 	&0.0053 	&0.0053 	&0.0055 	&0.0054 	&	&\ding{52} (1.1196)	&\ding{54} (0.9288)	&\ding{54} (0.9746)	&\ding{52} (1.0216 )	&\ding{52} (1.0242)	&\ding{54} (0.9790)	\\
6	&0.0072 	&0.0082 	&0.0083 	&0.0082 	&0.0081 	&0.0081 	&0.0079 	&	&\ding{52} (1.1081)	&\ding{54} (0.9615)	&\ding{54} (0.9563)	&\ding{54} (0.9650 )	&\ding{54} (0.9792)	&\ding{54} (0.9824)	\\
8	&0.0094 	&0.0110 	&0.0109 	&0.0109 	&0.0108 	&0.0107 	&0.0105 	&	&\ding{52} (1.1128)	&\ding{54} (0.9470)	&\ding{54} (0.9558)	&\ding{54} (0.9603 )	&\ding{54} (0.9707)	&\ding{54} (0.9736)	\\
10	&0.0116 	&0.0135 	&0.0137 	&0.0133 	&0.0133 	&0.0132 	&0.0129 	&	&\ding{52} (1.1129)	&\ding{54} (0.9550)	&\ding{54} (0.9429)	&\ding{54} (0.9732 )	&\ding{54} (0.9733)	&\ding{54} (0.9773)	\\
15	&0.0169 	&0.0195 	&0.0196 	&0.0193 	&0.0194 	&0.0192 	&0.0188 	&	&\ding{52} (1.1128)	&\ding{54} (0.9637)	&\ding{54} (0.9598)	&\ding{54} (0.9745 )	&\ding{54} (0.9708)	&\ding{54} (0.9786)	\\
\hline

\end{tabular}}																
\end{table*}	

\begin{table*}																
\centering																
\footnotesize																
\caption{\textbf{RRT Version:} Mean \textbf{P-measure} Results and Statistical Pairwise Comparisons of LSH against other Methods for \textbf{Real-life Programs}}																\label{TAB:RRT-real-p}															
\resizebox{\textwidth}{!}{
\begin{tabular}{@{}crrrrrrrrrrrrrr@{}}																
\hline																
\textbf{Test Set Size}  &\multicolumn{7}{c}{\textbf{Methods}} & &\multicolumn{6}{c}{\textit{LSH}} \\\cline{2-8}\cline{10-15}																
($|T|$)  &\textit{RT} &\textit{ART} &\textit{RF} &\textit{CR} &\textit{DF} &\textit{KD} &\textit{LSH} & &\textit{vs. RT} &\textit{vs. ART} &\textit{vs. RF} &\textit{vs. CR} &\textit{vs. DF} &\textit{vs. KD}  \\\hline	
2	&0.0022 	&0.0022 	&0.0024 	&0.0022 	&0.0024 	&0.0025 	&0.0023 	&	&\ding{52} (1.0304)	&\ding{52} (1.0450)	&\ding{54} (0.9718)	&\ding{52} (1.0369)	&\ding{54} (0.9646)	&\ding{54} (0.9165)	\\
4	&0.0048 	&0.0046 	&0.0048 	&0.0046 	&0.0047 	&0.0051 	&0.0047 	&	&\ding{54} (0.9780)	&\ding{109} (1.0186)	&\ding{109} (0.9880)	&\ding{52} (1.0260)	&\ding{109} (1.0139)	&\ding{54} (0.9344)	\\
6	&0.0072 	&0.0070 	&0.0071 	&0.0071 	&0.0071 	&0.0075 	&0.0071 	&	&\ding{109} (0.9851)	&\ding{109} (1.0111)	&\ding{109} (0.9886)	&\ding{109} (0.9925)	&\ding{109} (0.9999)	&\ding{54} (0.9433)	\\
8	&0.0094 	&0.0095 	&0.0094 	&0.0095 	&0.0094 	&0.0098 	&0.0095 	&	&\ding{109} (1.0097)	&\ding{109} (1.0007)	&\ding{109} (1.0055)	&\ding{109} (1.0012)	&\ding{109} (1.0112)	&\ding{54} (0.9642)	\\
10	&0.0116 	&0.0117 	&0.0119 	&0.0118 	&0.0117 	&0.0122 	&0.0119 	&	&\ding{52} (1.0284)	&\ding{52} (1.0168)	&\ding{109} (1.0039)	&\ding{52} (1.0153)	&\ding{52} (1.0197)	&\ding{54} (0.9823)	\\
15	&0.0169 	&0.0174 	&0.0176 	&0.0175 	&0.0179 	&0.0178 	&0.0176 	&	&\ding{52} (1.0425)	&\ding{52} (1.0131)	&\ding{109} (1.0005)	&\ding{109} (1.0087)	&\ding{54} (0.9874)	&\ding{109} (0.9915)	\\
\hline

\end{tabular}}																
\end{table*}	

\begin{table*}
\parbox{\textwidth}{
\subsection{Test Case Generation Time Results}
\vspace{10pt}
}
\parbox{\textwidth}{
Table \ref{TAB:executionTime} presents the test-case generation time for each dimension, when the number of test cases, $n$, was fixed at four representative values: 
$500$;
$1000$;
$5000$; and
$10,000$.
}
\end{table*}

\begin{table*}[!ht]
\centering
\footnotesize
\caption{\textbf{Test-case generation time} (ms) for various input domain dimensions}
\label{TAB:executionTime}
\resizebox{\textwidth}{!}{
\begin{tabular}{@{}c|c|r|rrrrrr|rrrrrr@{}}
\hline
\textbf{No. of Test}  &\textbf{Dimension} &\multirow{2}*{\textbf{RT}}  &\multicolumn{6}{c|}{\textbf{FSCS}} &\multicolumn{6}{c}{\textbf{RRT}}\\\cline{4-15}
\textbf{Cases} ($n$)   &($d$)  & &\textit{ART} &\textit{RF} &\textit{CR} &\textit{DF} &\textit{KD} &\textit{LSH} &\textit{ART} &\textit{RF} &\textit{CR} &\textit{DF} &\textit{KD} &\textit{LSH}\\\hline
\multirow{6}*{500}	&1	&0.04 	&5.89 	&1.62 	&1.55 	&4.41 	&1.74 	&3.27 	&1.22 	&0.35 	&0.30 	&1.02 	&0.75 	&0.59 	\\
	&2	&0.07 	&118.05 	&26.20 	&26.13 	&22.52 	&4.41 	&15.62 	&20.32 	&4.39 	&4.45 	&6.56 	&1.27 	&3.05 	\\
	&3	&0.07 	&115.62 	&26.76 	&26.64 	&49.46 	&8.84 	&16.13 	&18.33 	&4.02 	&4.02 	&11.71 	&1.98 	&2.64 	\\
	&4	&0.08 	&117.85 	&27.66 	&27.43 	&118.91 	&16.55 	&16.65 	&16.68 	&3.80 	&3.75 	&16.69 	&2.85 	&2.44 	\\
	&5	&0.11 	&118.55 	&28.35 	&28.05 	&242.99 	&27.43 	&17.04 	&15.69 	&3.57 	&3.58 	&29.78 	&4.40 	&2.33 	\\
	&10	&0.21 	&129.15 	&31.23 	&30.84 	&25,482.86 	&123.89 	&19.43 	&13.80 	&3.33 	&3.31 	&3309.81 	&14.79 	&2.25 	\\
\hline
 
 \end{tabular}}																
\end{table*}
\clearpage
\newpage
\begin{table*}
\parbox{\textwidth}{
\subsection{F-time Results}
\vspace{10pt}
}
\parbox{\textwidth}{
Tables \ref{TAB:FSCS-Ftime} and \ref{TAB:RRT-Ftime} present the F-time results for the 23 subject programs.
}
\end{table*}

\begin{table*}[!ht]
\centering
\footnotesize
\caption{\textbf{FSCS Version:} Mean \textbf{F-time} Results (ms) and Statistical Pairwise Comparisons of LSH-FSCS against Other Methods for \textbf{{Real-life Programs}}}
\label{TAB:FSCS-Ftime}
\resizebox{\textwidth}{!}{
\begin{tabular}{@{}ccrrrrrrrrrrrrrr@{}}
\hline
\multirow{2}*{\textbf{ID}} &\multirow{2}*{\textbf{Program}}  &\multicolumn{6}{c}{\textbf{Methods}} & &\multicolumn{5}{c}{\textit{LSH}} \\\cline{3-9}\cline{11-16}
  & &\textit{RT} &\textit{ART} &\textit{RF} &\textit{CR} &\textit{DF} &\textit{KD} &\textit{LSH} & &\textit{vs. RT} &\textit{vs. ART} &\textit{vs. RF} &\textit{vs. CR} &\textit{vs. DF} &\textit{vs. KD}  \\\hline
P1	&airy	&0.44 	&23.93 	&4.02 	&4.07 	&10.54 	&7.17 	&3.44 	&	&\ding{54} (0.08)	&\ding{52} (0.74) 	&\ding{109} (0.49) 	&\ding{109} (0.50) 	&\ding{52} (0.81) 	&\ding{52} (0.70) 	\\
P2	&bessj0	&0.21 	&6.95 	&2.11 	&2.06 	&5.82 	&3.70 	&1.94 	&	&\ding{54} (0.10)	&\ding{52} (0.64) 	&\ding{54} (0.47) 	&\ding{54} (0.46) 	&\ding{52} (0.79) 	&\ding{52} (0.67) 	\\
P3	&erfcc	&0.62 	&43.85 	&5.60 	&5.58 	&13.81 	&10.00 	&4.50 	&	&\ding{54} (0.09)	&\ding{52} (0.77) 	&\ding{109} (0.51) 	&\ding{109} (0.51) 	&\ding{52} (0.81) 	&\ding{52} (0.71) 	\\
P4	&probks	&22.67 	&109.97 	&29.00 	&30.52 	&32.35 	&26.85 	&19.31 	&	&\ding{109} (0.49)	&\ding{52} (0.74) 	&\ding{52} (0.56) 	&\ding{52} (0.58) 	&\ding{52} (0.67) 	&\ding{52} (0.60) 	\\
P5	&tanh	&0.13 	&3.08 	&1.52 	&1.46 	&3.92 	&2.41 	&1.47 	&	&\ding{54} (0.10)	&\ding{52} (0.57) 	&\ding{54} (0.45) 	&\ding{54} (0.45) 	&\ding{52} (0.78) 	&\ding{52} (0.62) 	\\
P6	&bessj	&0.94 	&77.80 	&12.63 	&12.13 	&23.95 	&5.37 	&8.90 	&	&\ding{54} (0.12)	&\ding{52} (0.70) 	&\ding{52} (0.57) 	&\ding{52} (0.57) 	&\ding{52} (0.71) 	&\ding{54} (0.38) 	\\
\hline
\end{tabular}}
\end{table*}

\begin{table*}[!ht]
\centering
\footnotesize
\caption{\textbf{RRT Version:} Mean \textbf{F-time} Results (ms) and Statistical Pairwise Comparisons of LSH-RRT against Other Methods for \textbf{{Real-life Programs}}}
\label{TAB:RRT-Ftime}
\resizebox{\textwidth}{!}{
\begin{tabular}{@{}ccrrrrrrrrrrrrrr@{}}
\hline
\multirow{2}*{\textbf{ID}} &\multirow{2}*{\textbf{Program}}  &\multicolumn{6}{c}{\textbf{Methods}} & &\multicolumn{5}{c}{\textit{LSH}} \\\cline{3-9}\cline{11-16}
  & &\textit{RT} &\textit{ART} &\textit{RF} &\textit{CR} &\textit{DF} &\textit{KD} &\textit{LSH} & &\textit{vs. RT} &\textit{vs. ART} &\textit{vs. RF} &\textit{vs. CR} &\textit{vs. DF} &\textit{vs. KD}  \\\hline
P1	&airy	&0.44 	&6.59 	&1.29 	&1.20 	&4.97 	&2.05 	&1.19 	&	&\ding{54} (0.29)	&\ding{52} (0.70) 	&\ding{109} (0.50) 	&\ding{109} (0.49) 	&\ding{52} (0.74) 	&\ding{52} (0.64) 	\\
P2	&bessj0	&0.21 	&1.75 	&0.66 	&0.58 	&2.11 	&1.01 	&0.63 	&	&\ding{54} (0.34)	&\ding{52} (0.61) 	&\ding{109} (0.51) 	&\ding{109} (0.49) 	&\ding{52} (0.73) 	&\ding{52} (0.61) 	\\
P3	&erfcc	&0.62 	&11.15 	&1.77 	&1.69 	&7.56 	&2.73 	&1.68 	&	&\ding{54} (0.28)	&\ding{52} (0.72) 	&\ding{109} (0.50) 	&\ding{109} (0.48) 	&\ding{52} (0.74) 	&\ding{52} (0.63) 	\\
P4	&probks	&22.67 	&36.89 	&24.54 	&24.16 	&34.68 	&17.18 	&15.35 	&	&\ding{52} (0.57) 	&\ding{52} (0.63) 	&\ding{52} (0.59) 	&\ding{52} (0.59) 	&\ding{52} (0.64) 	&\ding{52} (0.54)	\\
P5	&tanh	&0.13 	&0.89 	&0.45 	&0.42 	&1.52 	&0.70 	&0.39 	&	&\ding{54} (0.38)	&\ding{52} (0.59) 	&\ding{52} (0.52) 	&\ding{109} (0.51) 	&\ding{52} (0.74) 	&\ding{52} (0.61) 	\\
P6	&bessj	&0.94 	&20.62 	&3.18 	&3.10 	&7.40 	&1.84 	&2.33 	&	&\ding{54} (0.30)	&\ding{52} (0.73) 	&\ding{52} (0.57) 	&\ding{52} (0.56) 	&\ding{52} (0.74) 	&\ding{54} (0.44) 	\\
\hline
\end{tabular}}
\end{table*}
\clearpage
\newpage

\begin{table*}
\parbox{\textwidth}{

\section{Detailed Experiment Results for Non-Numerical Input Domains}
\addtocounter{table}{-19}
\vspace{10pt}
}
\parbox{\textwidth}{
Tables \ref{TAB:fscs-real-drupal} to \ref{TAB:rrt-real-linux} show the mean number of test cases required to identify each fault in the three configurable SUTs 
---
the $\textrm{F}^{\textrm{f}.x}\textrm{-measure}$
of 
DRUPAL,
BUSYBOX, and
LINUX KERNEL.
}
\end{table*}

\begin{table*}[!t]
\centering
\scriptsize
\caption{\textbf{FSCS version:} Mean Number of test cases to detect each fault and Statistical Pairwise Comparisons of LSH for program \\\textbf{DRUPAL}}
\label{TAB:fscs-real-drupal}
\resizebox{\textwidth}{!}{
\begin{tabular}{@{}crrrrrrrrrrrrrrrrrrrrrrrr@{}}
\hline
\multirow{2}{*}{\textbf{Fault ID}} & \multicolumn{12}{c}{\textbf{Methods}} && \multicolumn{1}{c}{} && \multicolumn{9}{c}{\textit{LSH}}\\\cline{3-13}\cline{16-25}
 && \textit{RT} && \textit{ART} && \textit{RF} && \textit{CR} && \textit{KD} && \textit{LSH} && && \textit{vs. RT} && \textit{vs. ART} && \textit{vs. RF} && \textit{vs. CR} && \textit{vs. KD}\\\hline
 f.1   &  & 4.0630  &  & 2.6957  &  & 2.6987  &  & 2.7120  &  & 2.7667  &  & 2.7713  &  &  &  & \ding{52} (0.58) &  & \ding{109} (0.49) &  & \ding{109} (0.49) &  & \ding{109} (0.49) &  & \ding{109} (0.50) \\
f.2   &  & 4.0983  &  & 3.0850  &  & 3.0617  &  & 3.2053  &  & 3.2220  &  & 3.1693  &  &  &  & \ding{52} (0.55) &  & \ding{109} (0.49) &  & \ding{109} (0.48) &  & \ding{109} (0.50) &  & \ding{109} (0.51) \\
f.3   &  & 4.0723  &  & 3.1357  &  & 3.1010  &  & 3.1890  &  & 3.1040  &  & 3.1457  &  &  &  & \ding{52} (0.55) &  & \ding{109} (0.50) &  & \ding{109} (0.49) &  & \ding{109} (0.50) &  & \ding{109} (0.49) \\
f.4   &  & 4.0557  &  & 3.3333  &  & 3.3960  &  & 3.3847  &  & 3.4353  &  & 3.3833  &  &  &  & \ding{52} (0.53) &  & \ding{109} (0.50) &  & \ding{109} (0.50) &  & \ding{109} (0.50) &  & \ding{109} (0.51) \\
f.5   &  & 4.0497  &  & 3.0613  &  & 3.1570  &  & 3.1957  &  & 3.1107  &  & 3.1483  &  &  &  & \ding{52} (0.55) &  & \ding{109} (0.49) &  & \ding{109} (0.51) &  & \ding{109} (0.50) &  & \ding{109} (0.50) \\
f.6   &  & 3.9743  &  & 3.2927  &  & 3.3377  &  & 3.4180  &  & 3.4183  &  & 3.4160  &  &  &  & \ding{52} (0.53) &  & \ding{109} (0.49) &  & \ding{109} (0.50) &  & \ding{109} (0.50) &  & \ding{109} (0.50) \\
\hline
\end{tabular}}
\end{table*}

\begin{table*}[!t]
\parbox{\textwidth}{
\centering{(\textit{Continue.})}
}
\end{table*}
\begin{table*}[!t]
\centering
\scriptsize
\resizebox{\textwidth}{!}{
\begin{tabular}{@{}crrrrrrrrrrrrrrrrrrrrrrrr@{}}
\hline
\multirow{2}{*}{\textbf{Fault ID}} & \multicolumn{12}{c}{\textbf{Methods}} && \multicolumn{1}{c}{} && \multicolumn{9}{c}{\textit{LSH}}\\\cline{3-13}\cline{16-25}
 && \textit{RT} && \textit{ART} && \textit{RF} && \textit{CR} && \textit{KD} && \textit{LSH} && && \textit{vs. RT} && \textit{vs. ART} && \textit{vs. RF} && \textit{vs. CR} && \textit{vs. KD}\\\hline
f.81  &  & 6.3033  &  & 3.4457  &  & 3.4670  &  & 3.4460  &  & 3.5267  &  & 3.4990  &  &  &  & \ding{52} (0.64) &  & \ding{109} (0.49) &  & \ding{109} (0.50) &  & \ding{109} (0.49) &  & \ding{109} (0.50) \\
f.82  &  & 12.7533 &  & 5.9493  &  & 5.8247  &  & 6.1170  &  & 5.9733  &  & 5.8427  &  &  &  & \ding{52} (0.68) &  & \ding{109} (0.51) &  & \ding{109} (0.50) &  & \ding{109} (0.50) &  & \ding{109} (0.51) \\
f.83  &  & 25.1877 &  & 11.6820 &  & 12.0427 &  & 13.1077 &  & 12.2490 &  & 12.1843 &  &  &  & \ding{52} (0.67) &  & \ding{109} (0.49) &  & \ding{109} (0.50) &  & \ding{109} (0.50) &  & \ding{109} (0.50) \\
f.84  &  & 4.0487  &  & 2.9773  &  & 2.9697  &  & 3.0460  &  & 3.0133  &  & 3.0767  &  &  &  & \ding{52} (0.56) &  & \ding{109} (0.49) &  & \ding{109} (0.49) &  & \ding{109} (0.50) &  & \ding{109} (0.49) \\
f.85  &  & 31.7293 &  & 11.4743 &  & 11.2567 &  & 12.2433 &  & 11.4383 &  & 11.2607 &  &  &  & \ding{52} (0.74) &  & \ding{109} (0.50) &  & \ding{109} (0.50) &  & \ding{109} (0.51) &  & \ding{109} (0.50) \\
f.86  &  & 6.3033  &  & 3.4457  &  & 3.4670  &  & 3.4460  &  & 3.5267  &  & 3.4990  &  &  &  & \ding{52} (0.64) &  & \ding{109} (0.49) &  & \ding{109} (0.50) &  & \ding{109} (0.49) &  & \ding{109} (0.50) \\
\hline
\end{tabular}}
\end{table*}

\begin{table*}[!ht]
\centering
\scriptsize
\caption{\textbf{RRT version:} Mean Number of test cases to detect each fault and Statistical Pairwise Comparisons of LSH for program \\\textbf{DRUPAL}}
\label{TAB:rrt-real-drupal}
\resizebox{\textwidth}{!}{
\begin{tabular}{@{}crrrrrrrrrrrrrrrrrrrrrrrr@{}}
\hline
\multirow{2}{*}{\textbf{Fault ID}} & \multicolumn{12}{c}{\textbf{Methods}} && \multicolumn{1}{c}{} && \multicolumn{9}{c}{\textit{LSH}}\\\cline{3-13}\cline{16-25}
 && \textit{RT} && \textit{ART} && \textit{RF} && \textit{CR} && \textit{KD} && \textit{LSH} && && \textit{vs. RT} && \textit{vs. ART} && \textit{vs. RF} && \textit{vs. CR} && \textit{vs. KD}\\\hline
f.1   &  & 4.0630  &  & 4.0320  &  & 3.9583  &  & 4.0690  &  & 4.0557  &  & 3.9763  &  &  &  & \ding{109} (0.50) &  & \ding{109} (0.50) &  & \ding{109} (0.49) &  & \ding{109} (0.50) &  & \ding{109} (0.50) \\
f.2   &  & 4.0983  &  & 4.0167  &  & 4.0217  &  & 3.9367  &  & 3.9373  &  & 4.0013  &  &  &  & \ding{109} (0.51) &  & \ding{109} (0.50) &  & \ding{109} (0.50) &  & \ding{109} (0.50) &  & \ding{109} (0.50) \\
f.3   &  & 4.0723  &  & 3.9450  &  & 3.8730  &  & 3.9027  &  & 4.1197  &  & 4.0387  &  &  &  & \ding{109} (0.50) &  & \ding{109} (0.50) &  & \ding{109} (0.49) &  & \ding{109} (0.49) &  & \ding{109} (0.50) \\
f.4   &  & 4.0557  &  & 3.9613  &  & 4.0040  &  & 4.0407  &  & 3.9963  &  & 3.8907  &  &  &  & \ding{109} (0.51) &  & \ding{109} (0.50) &  & \ding{109} (0.51) &  & \ding{109} (0.51) &  & \ding{109} (0.51) \\
f.5   &  & 4.0497  &  & 3.9790  &  & 4.0310  &  & 3.9713  &  & 3.9770  &  & 4.0587  &  &  &  & \ding{109} (0.50) &  & \ding{109} (0.49) &  & \ding{109} (0.50) &  & \ding{109} (0.49) &  & \ding{109} (0.49) \\
f.6   &  & 3.9743  &  & 3.9660  &  & 4.0340  &  & 4.0057  &  & 3.9957  &  & 3.9330  &  &  &  & \ding{109} (0.50) &  & \ding{109} (0.50) &  & \ding{109} (0.51) &  & \ding{109} (0.51) &  & \ding{109} (0.50) \\
\hline

\end{tabular}}
\end{table*}
\clearpage
\newpage
\begin{table*}[!t]
\parbox{\textwidth}{
\centering{(\textit{Continue.})}
}
\end{table*}
\begin{table*}[!ht]
\centering
\scriptsize
\resizebox{\textwidth}{!}{
\begin{tabular}{@{}crrrrrrrrrrrrrrrrrrrrrrrr@{}}
\hline
\multirow{2}{*}{\textbf{Fault ID}} & \multicolumn{12}{c}{\textbf{Methods}} && \multicolumn{1}{c}{} && \multicolumn{9}{c}{\textit{LSH}}\\\cline{3-13}\cline{16-25}
 && \textit{RT} && \textit{ART} && \textit{RF} && \textit{CR} && \textit{KD} && \textit{LSH} && && \textit{vs. RT} && \textit{vs. ART} && \textit{vs. RF} && \textit{vs. CR} && \textit{vs. KD}\\\hline
f.81  &  & 6.3033  &  & 6.5150  &  & 6.3100  &  & 6.1980  &  & 6.4203  &  & 6.2060  &  &  &  & \ding{109} (0.50) &  & \ding{109} (0.51) &  & \ding{109} (0.50) &  & \ding{109} (0.50) &  & \ding{109} (0.50) \\
f.82  &  & 12.7533 &  & 13.0633 &  & 12.6553 &  & 12.1040 &  & 13.1467 &  & 12.4887 &  &  &  & \ding{109} (0.50) &  & \ding{109} (0.51) &  & \ding{109} (0.51) &  & \ding{109} (0.49) &  & \ding{109} (0.51) \\
f.83  &  & 25.1877 &  & 25.7820 &  & 25.8513 &  & 25.7507 &  & 26.1200 &  & 25.5387 &  &  &  & \ding{109} (0.50) &  & \ding{109} (0.51) &  & \ding{109} (0.50) &  & \ding{109} (0.50) &  & \ding{109} (0.51) \\
f.84  &  & 4.0487  &  & 4.0170  &  & 4.0630  &  & 4.0160  &  & 3.9217  &  & 4.0177  &  &  &  & \ding{109} (0.50) &  & \ding{109} (0.50) &  & \ding{109} (0.50) &  & \ding{109} (0.50) &  & \ding{109} (0.49) \\
f.85  &  & 31.7293 &  & 30.5097 &  & 31.7427 &  & 32.7957 &  & 31.5767 &  & 30.7853 &  &  &  & \ding{109} (0.51) &  & \ding{109} (0.50) &  & \ding{109} (0.51) &  & \ding{109} (0.51) &  & \ding{109} (0.50) \\
f.86  &  & 6.3033  &  & 6.5150  &  & 6.3100  &  & 6.1980  &  & 6.4203  &  & 6.2060  &  &  &  & \ding{109} (0.50) &  & \ding{109} (0.51) &  & \ding{109} (0.50) &  & \ding{109} (0.50) &  & \ding{109} (0.50) \\
\hline

\end{tabular}}
\end{table*}

\begin{table*}[!ht]
\centering
\footnotesize
\caption{\textbf{FSCS version:} Mean Number of test cases to detect each fault and Statistical Pairwise Comparisons of LSH for program \\\textbf{BUSYBOX}}
\label{TAB:fscs-real-busybox}
\resizebox{\textwidth}{!}{
\begin{tabular}{@{}crrrrrrrrrrrrrrrrrrrrrrrr@{}}
\hline
\multirow{2}{*}{\textbf{Fault ID}} & \multicolumn{12}{c}{\textbf{Methods}} && \multicolumn{1}{c}{} && \multicolumn{9}{c}{\textit{LSH}}\\\cline{3-13}\cline{16-25}
 && \textit{RT} && \textit{ART} && \textit{RF} && \textit{CR} && \textit{KD} && \textit{LSH} && && \textit{vs. RT} && \textit{vs. ART} && \textit{vs. RF} && \textit{vs. CR} && \textit{vs. KD}\\\hline
f.1 && 6.0553 && 5.6820 && 5.6550 && 5.4823 && 5.5247 && 5.6737 &&  && \ding{109} (0.51) && \ding{109} (0.50) && \ding{109} (0.50) && \ding{109} (0.50) && \ding{109} (0.50)\\
f.2 && 1.9880 && 1.7880 && 1.8187 && 1.8323 && 1.8283 && 1.8280 &&  && \ding{109} (0.51) && \ding{109} (0.49) && \ding{109} (0.49) && \ding{109} (0.49) && \ding{109} (0.50)\\
f.3 && 1.9913 && 1.8333 && 1.8447 && 1.8363 && 1.8393 && 1.8167 &&  && \ding{109} (0.51) && \ding{109} (0.50) && \ding{109} (0.51) && \ding{109} (0.51) && \ding{109} (0.51)\\
f.4 && 1.3427 && 1.2923 && 1.2957 && 1.2660 && 1.2787 && 1.2970 &&  && \ding{109} (0.51) && \ding{109} (0.50) && \ding{109} (0.50) && \ding{109} (0.49) && \ding{109} (0.49)\\
f.5 && 32.4917 && 28.0937 && 27.2970 && 29.0703 && 27.8867 && 27.9310 &&  && \ding{52} (0.53) && \ding{109} (0.50) && \ding{109} (0.49) && \ding{109} (0.50) && \ding{109} (0.50)\\
f.6 && 5.1470 && 5.4730 && 5.5533 && 5.5517 && 5.5660 && 5.4917 &&  && \ding{109} (0.48) && \ding{109} (0.50) && \ding{109} (0.51) && \ding{109} (0.50) && \ding{109} (0.51)\\
\hline

\end{tabular}}
\end{table*}

\begin{table*}[!ht]
\centering
\footnotesize
\caption{\textbf{RRT version:} Mean Number of test cases to detect each fault and Statistical Pairwise Comparisons of LSH for program \\\textbf{BUSYBOX}}
\label{TAB:rrt-real-busybox}
\resizebox{\textwidth}{!}{
\begin{tabular}{@{}crrrrrrrrrrrrrrrrrrrrrrrr@{}}
\hline
\multirow{2}{*}{\textbf{Fault ID}} & \multicolumn{12}{c}{\textbf{Methods}} && \multicolumn{1}{c}{} && \multicolumn{9}{c}{\textit{LSH}}\\\cline{3-13}\cline{16-25}
&& \textit{RT} && \textit{ART} && \textit{RF} && \textit{CR} && \textit{KD} && \textit{LSH} &&  && \textit{vs. RT} && \textit{vs. ART} && \textit{vs. RF} && \textit{vs. CR} && \textit{vs. KD}\\\hline
f.1 && 6.0553 && 6.0173 && 6.1587 && 6.1270 && 5.9973 && 6.1610 &&  && \ding{109} (0.49) && \ding{109} (0.49) && \ding{109} (0.50) && \ding{109} (0.50) && \ding{109} (0.49)\\
f.2 && 1.9880 && 2.0413 && 2.0420 && 1.9833 && 2.0420 && 2.0293 &&  && \ding{109} (0.49) && \ding{109} (0.51) && \ding{109} (0.50) && \ding{109} (0.49) && \ding{109} (0.50)\\
f.3 && 1.9913 && 1.9643 && 2.0553 && 2.0257 && 1.9967 && 2.0090 &&  && \ding{109} (0.50) && \ding{109} (0.49) && \ding{109} (0.51) && \ding{109} (0.51) && \ding{109} (0.50)\\
f.4 && 1.3427 && 1.3463 && 1.3323 && 1.3217 && 1.3287 && 1.3290 &&  && \ding{109} (0.50) && \ding{109} (0.51) && \ding{109} (0.50) && \ding{109} (0.50) && \ding{109} (0.50)\\
f.5 && 32.4917 && 31.9430 && 33.0497 && 31.9413 && 31.5933 && 31.2703 &&  && \ding{109} (0.51) && \ding{109} (0.51) && \ding{109} (0.52) && \ding{109} (0.51) && \ding{109} (0.50)\\
f.6 && 5.1470 && 5.3900 && 5.3397 && 5.2187 && 5.3497 && 5.3493 &&  && \ding{109} (0.48) && \ding{109} (0.49) && \ding{109} (0.50) && \ding{109} (0.49) && \ding{109} (0.49)\\
\hline
\end{tabular}}
\end{table*}

\begin{table*}[!ht]
    \centering
    \footnotesize
    \caption{\textbf{FSCS version:} Mean Number of test cases to detect each fault and Statistical Pairwise Comparisons of LSH for program \\\textbf{LINUX KERNEL}}
    \label{TAB:fscs-real-linux}
\resizebox{\textwidth}{!}{
    \begin{tabular}{@{}crrrrrrrrrrrrrrrrrrrrrrrr@{}}
    \hline
\multirow{2}{*}{\textbf{Fault ID}} & \multicolumn{12}{c}{\textbf{Methods}} && \multicolumn{1}{c}{} && \multicolumn{9}{c}{\textit{LSH}}\\\cline{3-13}\cline{16-25}
 && \textit{RT} && \textit{ART} && \textit{RF} && \textit{CR} && \textit{KD} && \textit{LSH} &&  && \textit{vs. RT} && \textit{vs. ART} && \textit{vs. RF} && \textit{vs. CR} && \textit{vs. KD}\\\hline
f.1 && 1.9743 && 1.8550 && 1.9150 && 1.8340 && 1.8580 && 1.8330 &&  && \ding{52} (0.52) && \ding{109} (0.50) && \ding{52} (0.52) && \ding{109} (0.50) && \ding{109} (0.51)\\
f.2 && 1.9770 && 1.8453 && 1.8427 && 1.9223 && 1.8363 && 1.8597 &&  && \ding{109} (0.51) && \ding{109} (0.50) && \ding{109} (0.50) && \ding{109} (0.52) && \ding{109} (0.50)\\
f.3 && 1.9663 && 1.8007 && 1.8263 && 1.8250 && 1.7930 && 1.7863 &&  && \ding{52} (0.52) && \ding{109} (0.50) && \ding{109} (0.51) && \ding{109} (0.51) && \ding{109} (0.50)\\
f.4 && 16.0240 && 13.3863 && 13.5443 && 13.9347 && 14.0003 && 14.5613 &&  && \ding{52} (0.53) && \ding{54} (0.48) && \ding{109} (0.48) && \ding{109} (0.49) && \ding{109} (0.50)\\
f.5 && 2.0453 && 1.8670 && 1.8223 && 1.8850 && 1.8530 && 1.8757 &&  && \ding{52} (0.52) && \ding{109} (0.50) && \ding{109} (0.49) && \ding{109} (0.51) && \ding{109} (0.50)\\
f.6 && 5.2663 && 3.7253 && 3.6757 && 3.7840 && 3.7677 && 3.8657 &&  && \ding{52} (0.57) && \ding{109} (0.50) && \ding{109} (0.49) && \ding{109} (0.50) && \ding{109} (0.50)\\
\hline

\end{tabular}}
\end{table*}

\begin{table*}[!ht]
    \centering
    \footnotesize
    \caption{\textbf{RRT version:} Mean Number of test cases to detect each fault and Statistical Pairwise Comparisons of LSH for program \\\textbf{LINUX KERNEL}}
    \label{TAB:rrt-real-linux}
\resizebox{\textwidth}{!}{
    \begin{tabular}{@{}crrrrrrrrrrrrrrrrrrrrrrrr@{}}
    \hline
\multirow{2}{*}{\textbf{Fault ID}} & \multicolumn{12}{c}{\textbf{Methods}} && \multicolumn{1}{c}{} && \multicolumn{9}{c}{\textit{LSH}}\\\cline{3-13}\cline{16-25}
 && \textit{RT} && \textit{ART} && \textit{RF} && \textit{CR} && \textit{KD} && \textit{LSH} && && \textit{vs. RT} && \textit{vs. ART} && \textit{vs. RF} && \textit{vs. CR} && \textit{vs. KD}\\\hline
f.1 && 1.9743 && 2.0213 && 2.0083 && 1.9807 && 2.0220 && 1.9997 &&  && \ding{109} (0.50) && \ding{109} (0.50) && \ding{109} (0.50) && \ding{109} (0.50) && \ding{109} (0.50)\\
f.2 && 1.9770 && 2.0373 && 1.9890 && 2.0223 && 1.9643 && 2.0317 &&  && \ding{109} (0.49) && \ding{109} (0.50) && \ding{109} (0.49) && \ding{109} (0.50) && \ding{109} (0.49)\\
f.3 && 1.9663 && 2.0300 && 2.0070 && 2.0503 && 1.9917 && 1.9877 &&  && \ding{109} (0.49) && \ding{109} (0.50) && \ding{109} (0.50) && \ding{109} (0.50) && \ding{109} (0.50)\\
f.4 && 16.0240 && 16.3890 && 16.1507 && 16.1723 && 16.0700 && 15.4287 &&  && \ding{109} (0.51) && \ding{109} (0.51) && \ding{109} (0.51) && \ding{109} (0.51) && \ding{109} (0.50)\\
f.5 && 2.0453 && 1.9953 && 2.0030 && 2.0040 && 1.9710 && 2.0090 &&  && \ding{109} (0.51) && \ding{109} (0.50) && \ding{109} (0.50) && \ding{109} (0.50) && \ding{109} (0.50)\\
f.6 && 5.2663 && 5.3923 && 5.3057 && 5.2893 && 5.4537 && 5.1720 &&  && \ding{109} (0.50) && \ding{109} (0.51) && \ding{109} (0.50) && \ding{109} (0.51) && \ding{109} (0.51)\\
\hline
\end{tabular}}
\end{table*}

\vfill

\end{document}